\documentclass{article}
\usepackage{qic,epsfig}

\input{Qcircuit}

\begin{document}
%%
%% I commented out this redefinition of length
%\setlength{\textheight}{8.0truein}    %FOR 2ND PAGE ONWARDS
%%

\runninghead{Accuracy threshold for postselected quantum computation}
            {P. Aliferis, D. Gottesman, and J. Preskill}

\normalsize\textlineskip
\thispagestyle{empty}
\setcounter{page}{1}

%% I commented out the copyrightheadings
%\copyrightheading{Vol.}{No.}{Year}{Page Nos.}
%\copyrightheading{0}{0}{2004}{000--000}

\vspace*{0.88truein}

\alphfootnote

\fpage{1}

\centerline{\bf
%%%%%%%%%%%%%%%%%%%%%
%Put in titles here
%%%%%%%%%%%%%%%%%%%%%
ACCURACY THRESHOLD FOR POSTSELECTED QUANTUM COMPUTATION}
\vspace*{0.37truein}
\centerline{\footnotesize
%%%%%%%%%%%%%%%%%%%%%%%%%%%%%%%%%%%%
%put authors' name and address here
%%%%%%%%%%%%%%%%%%%%%%%%%%%%%%%%%%%%
PANOS ALIFERIS}
\vspace*{0.015truein}
\centerline{\footnotesize\it Institute for Quantum Information, California Institute of Technology }
\baselineskip=10pt
\centerline{\footnotesize\it Pasadena, CA 91125, USA}
\vspace*{10pt}
\centerline{\footnotesize 
DANIEL GOTTESMAN}
\vspace*{0.015truein}
\centerline{\footnotesize\it Perimeter Institute }
\baselineskip=10pt
\centerline{\footnotesize\it Waterloo ON N2V 1Z3, Canada}
\vspace*{10pt}
\centerline{\footnotesize 
JOHN PRESKILL}
\vspace*{0.015truein}
\centerline{\footnotesize\it Institute for Quantum Information, California Institute of Technology }
\baselineskip=10pt
\centerline{\footnotesize\it Pasadena, CA 91125, USA}
\vspace*{0.225truein}
%% I commented out the \publisher line, and added a line dating this draft
%\publisher{(received date)}{(revised date)}
%\centerline{DRAFT --- 27 March 2007}
%
%
\vspace*{0.21truein}
%
%% \abstracts{first paragraph}{second paragraph}{third paragraph}
%% If there is only one paragraph, just keep the second and third empty 
%% like the following one 
\abstracts{
%%%%%%%%%%%%%%%%%%%%
% put abstract here
%%%%%%%%%%%%%%%%%%%%
We prove an accuracy threshold theorem for fault-tolerant quantum computation based on error detection and postselection. Our proof provides a rigorous foundation for the scheme suggested by Knill, in which preparation circuits for ancilla states are protected by a concatenated error-detecting code and the preparation is aborted if an error is detected. The proof applies to independent stochastic noise but (in contrast to proofs of the quantum accuracy threshold theorem based on concatenated error-correcting codes) not to strongly-correlated adversarial noise. Our rigorously established lower bound on the accuracy threshold, $1.04\times 10^{-3}$, is well below Knill's numerical estimates.
}{}{}
\vspace*{10pt}
%% I commented out the key words, the following space, and the communication line
%\keywords{Quantum error correction, fault tolerance, accuracy threshold}
%\vspace*{3pt}
%\communicate{to be filled by the Editorial}
%%
%%
\vspace*{1pt}\textlineskip    %) USE THIS MEASUREMENT WHEN THERE IS
   %) A SECTION HEADING
%\vspace*{-0.5pt}
%\noindent
%%%%%%%%%%%%%%%%%%%%%%%%%%%%%%%%
%put the text of the paper here
%%%%%%%%%%%%%%%%%%%%%%%%%%%%%%%%%\documentclass[aps,pra,showpacs,preprint]{revtex4}

\section{Introduction}
\label{sec:intro}

The theory of fault-tolerant quantum computation \cite{shor_ft} establishes that a noisy quantum computer can simulate an ideal quantum computer accurately. In particular, the quantum accuracy threshold theorem \cite{ben-or,kitaev_threshold,knill, jp_threshold,gottesman_threshold,AGP,reichardt-threshold} asserts that an arbitrarily long quantum computation can be executed reliably, provided that the noise afflicting the computer's hardware is weaker than a certain critical value, the {\em accuracy threshold}. 

The numerical value of the accuracy threshold is of considerable practical interest, as it provides a target for the prospective quantum hardware designer. For a noise model such that faults in the quantum circuit are independent and identically distributed with fault rate $\varepsilon$ (and also for more general ``adversarial'' local stochastic noise models), a lower bound on the threshold $\varepsilon_{\rm th} > 2.7 \times 10^{-5}$ was rigorously established in \cite{AGP}; this lower bound was improved  to $\varepsilon_{\rm th} > 1.9\times 10^{-4}$ in \cite{aliferis-cross}. 

These lower bounds on the accuracy threshold have been derived by studying fault-tolerant circuits that process quantum information protected by a concatenated quantum error-correcting code (a self-similar hierarchy of codes within codes). But Knill \cite{knill_detect} has suggested a different approach to fault-tolerant quantum computing, and numerical studies of Knill's scheme indicate that a much higher value of the threshold (higher than 1\%) can be achieved. Knill's scheme is based on the {\em quantum software} strategy \cite{steane-software,gottesman-chuang}, in which the execution of fault-tolerant quantum gates is facilitated by the offline preparation of ancilla states, which are permitted to interact with the data only after the fidelity of the preparation has been suitably verified. The novel feature of Knill's scheme is that the ancilla preparation circuit is protected by a concatenated error-{\em detecting} code, and the ancilla is accepted only if no errors are detected during the preparation; we refer to this general approach as {\em postselected} quantum computation. 

The purpose of this paper is to formulate and prove an accuracy threshold theorem for postselected quantum computation. We are motivated by a desire to put Knill's optimistic threshold estimates on a rigorous basis, and also because the theory of postselected quantum computation poses intriguing conceptual questions, due to the subtlety of assessing the reliability of a quantum circuit conditioned on the acceptance of all ancilla states. In particular, we will see that to obtain a threshold theorem for postselected computation, we must posit limits on the correlations in the noise beyond what would be needed if error correction were used instead of error detection. For a noise model with suitable locality properties, we establish a new rigorous lower bound on the quantum accuracy threshold, $1.04\times 10^{-3}$, an improvement by a factor of about 5.5 compared to the best previously established rigorous lower bound, but still an order of magnitude below Knill's estimates based on numerical simulations. We note that a threshold theorem for postselected quantum computation has also been proved recently by Reichardt \cite{reichardt-postselection,reichardt-thesis}, using completely different methods; he too reports evidence for a lower bound close to $10^{-3}$.

The rest of this paper is organized as follows. In Sec.~\ref{sec:knill} we review Knill's proposed strategy for boosting the accuracy threshold using postselected computation. In Sec.~\ref{sec:sketch} we preview some of the ingredients in our analysis, emphasizing the subtlety of estimating the probability of failure in a circuit simulation {\em conditioned on acceptance} by all error-detection gadgets in the circuit, and in Sec.~\ref{sec:noise} we formulate the {\em locally correlated stochastic noise model} that is assumed in our proof. We prove in Sec.~\ref{sec:goodness} a fundamental lemma relating {\em goodness} (sparseness of faults) to {\em correctness} (accurate simulation). In Sec.~\ref{sec:carving} we develop a method for ``carving'' a circuit into nonoverlapping gadgets. This procedure ensures that each fault contributes to the failure of only one gadget, and enables us to derive an upper bound on the failure probability of a gadget conditioned on {\em local acceptance}; that is, under the assumption that no errors are detected by the gadget. In Sec.~\ref{sec:local-to-global} we take the crucial step of relating the failure probability conditioned on local acceptance to the failure probability conditioned on {\em global acceptance} by every error detection in the circuit. We discuss the analysis of open circuits (circuits that prepare quantum states) in Sec.~\ref{sec:level-reduction}, and in Sec.~\ref{sec:threshold} we complete our new proof of the quantum threshold theorem based on postselection.

After that, the remainder of the paper is devoted to obtaining an optimized numerical estimate of the threshold. In Sec.~\ref{sec:malignant} we outline our method, based on counting the ``malignant pairs'' of locations inside a gadget. We describe our gadget constructions in Sec.~\ref{sec:gadgets}, and perform the threshold analysis in Sec.~\ref{sec:estimate}. We make some general remarks about the resource requirements for postselected quantum computation in Sec.~\ref{sec:overhead}, and Sec.~\ref{sec:conclusions} contains our conclusions.

\section{Knill's proposal}
\label{sec:knill}

The goal of fault-tolerant quantum computation is to simulate an ideal quantum circuit using noisy gates. In this simulation, the logical qubits processed by the computer are protected from damage using a quantum error-correcting code \cite{shor_9,steane_7}, and the gates acting on the logical qubits are realized by ``gadgets'' that act on the code blocks. The gadgets exploit the redundancy of the quantum error-correcting code to diagnose and remove errors caused by faults; they are carefully designed to minimize propagation of errors among qubits within the same code block. 

The quantum accuracy threshold theorems proved in \cite{ben-or,kitaev_threshold,knill, jp_threshold,gottesman_threshold,AGP,reichardt-threshold} are based on {\em concatenated} quantum error-correcting codes \cite{knill-concatenated}. The code block of a concatenated code is constructed as a hierarchy of codes within codes --- the code block at level $k$ of this hierarchy is built from logical qubits encoded at level $k{-}1$ of the hierarchy. Likewise, the fault-tolerant gadgets are constructed as a hierarchy of gadgets within gadgets --- the gadgets at level $k$ are built from gate gadgets at level $k{-}1$. 

The key idea underlying the threshold theorem is that if faults are sufficiently rare and not too strongly correlated, then errors are very likely to be corrected at some level of concatenation before they percolate up to cause an encoded error at the top level. The precise formulation of the threshold condition depends on how we choose to model the noise. For example, in a {\em stochastic} model the noise is described probabilistically. We use the term {\em location} to speak of an operation in a quantum circuit that is performed in a single time step; a location may be a single-qubit or multi-qubit gate, a qubit preparation step, a qubit measurement, or the identity operation in the case of a qubit that is idle during the time step. We assume that at each circuit location either the ideal operation is executed perfectly or else a fault occurs; a stochastic noise model assigns a probability to each {\em fault path} --- that is, to each possible set of faulty locations in the circuit. We speak of {\em local stochastic noise} with fault rate $\varepsilon$ if, for any $r$ specified locations in the circuit, the sum of the probabilities of all fault paths with faults at those $r$ locations is no larger than $\varepsilon^r$. In this model no further restrictions are imposed on the noise --- for each fault path the trace-preserving quantum operation applied at the faulty locations is arbitrary and can be chosen adversarially; therefore although $\varepsilon$ quantifies the strength of the noise, the faults can be correlated both temporally and spatially. For this local stochastic noise model, it was shown in \cite{AGP} that an ideal quantum circuit can be simulated accurately and with reasonable overhead provided that $\varepsilon < \varepsilon_{\rm th}= 2.7\times 10^{-5}$, and this rigorous lower bound on the threshold has since been improved to $1.9 \times 10^{-4}$ in \cite{aliferis-cross}. 

Once we have established a lower bound $\varepsilon_{\rm th}$ on the threshold, we can obtain a stronger lower bound $\tilde\varepsilon_0$ by showing that a universal set of gates with fault rate below $\varepsilon_{\rm th}$ can be simulated by noisier gates with fault rate $\varepsilon$, where  $\tilde\varepsilon_0 > \varepsilon > \varepsilon_{\rm th}$.  Knill follows this strategy \cite{knill_detect}, where the simulation is achieved using error detection and postselection.  

The reason to study postselected circuits is that error detection is easier to execute than error correction, and therefore the accuracy threshold for postselected quantum computation is higher than estimates of the quantum accuracy threshold based on quantum error-correcting codes. Our core result in this paper is a proof of a threshold theorem for postselected quantum computation using concatenated error-detecting codes, and an estimate of the threshold based on a particular distance-2 quantum code; however, this proof works not for the local stochastic noise model described above, but only for a more restricted noise model that we will define in Sec.~\ref{sec:noise}. In some ways our analysis of Knill's postselected fault-tolerant quantum computation is similar to the analysis of fault-tolerant quantum computation based on error correction --- here any errors are very likely to be {\em detected} at some level before they reach the top level of a concatenated code to cause an encoded error. Thus, if no errors are detected, there are probably no faults at all (or any errors due to faulty gates have been eliminated by subsequent error detection steps that project the encoded blocks back to the code space before an encoded error can occur). Therefore, the postselected computation is reliable.

The price paid for using postselection is a significant overhead cost, because the large majority of the attempted ancilla preparation circuits are aborted due to detection of an error.  However, the goal of the postselected computation is to achieve simulated gates not with arbitrarily small fault rate, but rather with fault rate below $\varepsilon_{\rm th}$. Therefore, formally the additional overhead cost incurred by preparing ancillas using Knill's method is a (large) multiplicative constant, independent of the size of the quantum computation to be executed.

\begin{figure}
\begin{center}
%\vspace{0.5cm}
\begin{picture}(260,84)
\put(0,12){\line(1,0){10}}
\put(0,24){\line(1,0){10}}
\put(10,0){\framebox(42,36){\shortstack{encode\\$C_1^{\circ k}$}}}
\put(52,12){\line(1,0){10}}
\put(52,24){\line(1,0){10}}
\put(62,0){\framebox(42,36){\shortstack{prepare \\$C_2$\\ancilla}}}
\put(104,12){\line(1,0){10}}
\put(104,24){\line(1,0){10}}
\put(114,0){\framebox(42,36){\shortstack{decode\\$C_1^{\circ k}$}}}
\put(156,12){\line(1,0){20}}
\put(218,6){\makebox(20,12){$C_2$ encoded data out}}
\put(156,24){\line(1,1){24}}

\put(104,54){\makebox(20,12){$C_2$ encoded data in}}
\put(160,60){\line(1,0){20}}
\put(180,36){\framebox(60,36){\shortstack{transversal\\Bell\\measurement}}}
\end{picture}
\vspace{0.3cm}
\end{center}
\fcaption{Knill's scheme for teleporting an encoded quantum gate, shown schematically. Blocks of the level-$k$ concatenated error-detecting code $C_1^{\circ k}$ are encoded, and then a circuit is simulated that prepares the encoded ancillas for the error-correcting code $C_2$, where each simulated gate is preceded by an error detection step. Finally, the $C_1^{\circ k}$ blocks are decoded to obtain the desired $C_2$ ancilla, and a transversal Bell measurement is performed to teleport the gate and extract the $C_2$ error syndrome.}
\label{fig:knill}
\end{figure}

Knill's scheme (see Fig.~\ref{fig:knill}) uses two codes, which we will call $C_1$ and $C_2$. The simulated gates act on the encoded blocks of the code $C_2$; these $C_2$ gates, as well as the error correction, are realized using ``teleportation.'' Once suitable $C_2$-encoded ancilla blocks are prepared, the execution of the gate or error correction is completed by performing a transversal Bell measurement on a data block and one block of the encoded ancilla. (Pauli operators conditioned on the measurement outcomes would then complete the teleportation of the $C_2$ gate or error correction, but it is not actually necessary to perform these Pauli gates; it suffices to keep a classical record of the syndrome inferred from the measurements, a record that is continually updated as the computation proceeds.)

The code $C_2$ will be chosen to be a code with relatively high distance, such that if {\em independent} errors occur in the $C_2$ ancilla blocks at a rate {\em well above} $\varepsilon_{\rm th}$, the probability of an encoded error in the simulated gate is safely below $\varepsilon_{\rm th}$. Thus, to obtain an interesting lower bound on the accuracy threshold, it is sufficient to describe how to prepare suitable ancillas, such that the errors in the ancillas are sufficiently rare and have negligible correlations.

To prepare the $C_2$ ancillas, we need to build encoded $C_2$ blocks.  But encoding circuits typically propagate errors badly, so, unless we use a special trick, the errors in the $C_2$ ancilla blocks will be too strongly correlated. This is where the code $C_1$ comes in. This code will be used to {\em detect} errors, not to correct them, so it can be chosen to have distance 2. In our threshold calculations we (following Knill) will choose $C_1$ to be the $[[4,2,2]]$ quantum code (length $n=4$, $k=2$ protected qubits, distance $d=2$) except that we will make use of only one of the two logical qubits in the block.

We will simulate the $C_2$ ancilla preparation circuits using gadgets that act on the encoded blocks of $C_1^{\circ k}$ --- the code $C_1$ concatenated with itself $k$ times. (The value of $k$ will depend on how close the fault rate is to the threshold value.)  In this simulation, each gate is replaced by what we call a $k$-Rectangle (or $k$-Rec), consisting of a level-$k$ encoded gate ($k$-Ga) {\em preceded} by level-$k$ error-detection steps ($k$-EDs) on all input blocks; see Fig.~\ref{fig:level-1-sim}. If an error is detected in any error-detection step at any level of concatenation, the ancilla preparation is aborted, and begins again from scratch. 

\begin{figure}
\begin{center}
\setlength{\unitlength}{1.2pt}
\vspace{0.5cm}
\begin{picture}(160,54)
\put(0,19){\line(1,0){5}}
\put(0,35){\line(1,0){5}}

\put(5,13){\framebox(20,27){\footnotesize 0-Ga}}
\put(25,19){\line(1,0){5}}
\put(25,35){\line(1,0){5}}
\put(40,21){\makebox(30,12){$\Longrightarrow$}}
\put(78,12){\line(1,0){10}}
\put(78,42){\line(1,0){10}}
\put(88,0){\framebox(24,24){1-ED}}
\put(88,30){\framebox(24,24){1-ED}}
\put(112,12){\line(1,0){10}}
\put(112,42){\line(1,0){10}}
\put(122,0){\framebox(32,54){1-Ga}}
\put(154,12){\line(1,0){10}}
\put(154,42){\line(1,0){10}}
\end{picture}
\vspace{0.3cm}
\end{center}
\fcaption{Level-1 simulation. Each level-0 gate (0-Ga) in the ideal circuit is replaced by a 1-Rectangle, which consists of the level-1 encoded gate (1-Ga) that simulates the 0-Ga, preceded by a level-1 error detection gadget (1-ED) acting on each input block of the 1-Ga.}
\label{fig:level-1-sim}
\end{figure}

If the $C_2$ ancilla preparation is simulated successfully (without any error being detected at any level in any simulated gate), the result is a $C_2$ ancilla encoded using $C_1^{\circ k}$. At this stage $C_1^{\circ k}$ is decoded. This decoding proceeds one level at a time, with an error detection included in each decoding step. Finally, each $C_1^{\circ k}$ block has been mapped to a single qubit, and the preparation of the $C_2$ ancilla is now complete.

The rate of error in this $C_2$ ancilla is dominated by the final decoding step, in which a level-1 $C_1$ block is mapped to a qubit. This final step is unprotected by $C_1$, so that any fault in the decoding circuit could result in an undetected error in the decoded qubit. Thus, if the fault rate is $\varepsilon$, the error rate after decoding will be about $D\varepsilon$, if there are $D$ locations in the decoding circuit. But the key point is that the correlations among errors are negligible. Though the decoding of the $C_1^{\circ k}$ blocks is not well protected, the $C_2$ ancilla encoding circuit itself is well protected by the $C_1^{\circ k}$ code, so that undetected logical errors in the $C_1^{\circ k}$ blocks are highly unlikely. Instead, any undetected errors in the ancilla are likely to arise during the decoding of $C_1^{\circ k}$ blocks, while these blocks are out of contact with one another, and therefore these errors are uncorrelated. 

Thus the whole purpose of the concatenated error-detecting code $C_1^{\circ k}$ is to ensure that the errors in the encoded $C_2$ ancillas are very nearly independent. Then if the error-correcting code $C_2$ is chosen appropriately, the (nearly independent) errors arising from faults that occur at rate $\varepsilon > \varepsilon_{\rm th}$ can be corrected with reasonably high probability, so that encoded errors in the $C_2$ gadgets occur at a rate below $\varepsilon_{\rm th}$. This allows us to establish an improved lower bound on the threshold $\tilde\varepsilon_0>\varepsilon_{\rm th}$.

To avoid potential confusion, we remark that our notation differs from Knill's in \cite{knill_detect}. Knill uses $C_e$ to denote the high-distance quantum error-correcting code that we call $C_2$, and he uses $C_4/C_6$ to denote the quantum error-detecting code that we call $C_1$. 

\section{Sketch of our analysis}
\label{sec:sketch}
To analyze this scheme, we need to verify that the $C_1^{\circ k}$ gates are highly reliable for $k$ sufficiently large and noise that is sufficiently weak. By ``highly reliable'' we mean that the {\em conditional} probability of an encoded error is extremely small {\em given that no errors are detected at any level and in any $k$-ED}. That is, we are interested in how accurately the {\em postselected} circuit simulates the ideal $C_2$ ancilla encoding circuit. 

For the analysis of postselected circuits, we can borrow some of the same methods that were used in \cite{AGP} (and earlier in \cite{ben-or}) to analyze fault-tolerance based on concatenated error-correcting codes. There the key lemma established that a {\em good} gadget (one with sparse faults) is also {\em correct} (simulates the corresponding ideal gadget accurately). The threshold theorem was proved by showing that, for sufficiently weak noise, all gadgets are very likely to be good when the level of concatenation is high. Another central observation in \cite{AGP} was that it is convenient to characterize the performance of gadgets {\em syntactically} rather than {\em semantically} --- that is, to speak of the properties of operators rather than the properties of states --- and we will again adopt a syntactic approach here. 

The connection between goodness and correctness will be analyzed ``level by level.'' That is, we will show that a circuit built from level-1 gadgets is essentially equivalent to a corresponding circuit built from level-0 gates, where each good level-1 gadget is replaced by an ideal gate, and each bad level-1 gadget is replaced by a faulty gate. Using this reasoning $k$ times in succession, we can reduce a level-$k$ circuit to an equivalent level-0 circuit. If we are using a distance-2 code that detects one error, we may regard a good gadget as one that contains no more than one fault, and a bad gadget as one that contains two or more faults. Goodness implies correctness because the error due to a single fault in a good gadget is sure to be detected before it is joined by a second error that restores the block to the code space, causing an undetectable encoded error.

To prove the threshold theorem for postselected quantum computation, we need to consider how the effective fault rate evolves under ``level reduction.'' Naively, if we replace each level-1 gadget by an equivalent level-0 gate (which is what we mean by ``level reduction''), then if the fault rate is $\varepsilon$ before level reduction it becomes $\varepsilon^{(1)}$ after level reduction, where $\varepsilon^{(1)} = O(\varepsilon^2)$ because two faults in a gadget are required for the gadget to be bad. For the case of fault-tolerant quantum computation without postselection, the level reduction procedure was formalized in \cite{AGP,aliferis-thesis}. However, for postselected quantum computation the estimate of the fault rate after level reduction is actually somewhat subtle, because we wish to quantify the fault rate in the reduced circuit conditioned on ``global acceptance'' by the level-1 circuit --- that is, under the assumption that no level-1 error detection anywhere in the circuit detects an error. It is not hard to see that the probability is $O(\varepsilon^2$) for a given gadget to be bad  conditioned on ``local acceptance'' (that is, under the assumption that no error is detected by the 1-EDs that immediately precede and follow the level-1 gate). But correlations in the noise might cause the probability of badness conditioned on global acceptance to be substantially higher than the probability conditioned on local acceptance. 

In fact, to obtain a threshold condition for postselected quantum computation, we must specify a noise model that limits the correlations in the noise. While we may allow {\em local} adversaries at each faulty circuit location to choose the particular operation performed by the faulty gate, communication among the adversaries must be restricted. Otherwise, conspiring adversaries could boost substantially the probability (conditioned on global acceptance) of a particular fault path by turning off faults at strategically chosen locations.

Details of the noise model assumed in our proof will be specified in Sec.~\ref{sec:noise}. For now we just point out two criteria that the noise model should satisfy for the proof to work. First, the noise model should limit correlations in the noise to the extent that the probability of badness for any set of level-1 gadgets, conditioned on global acceptance, does not differ drastically from the probability of badness conditioned on local acceptance. Second, these limitations on the correlations should be preserved under level reduction, so that the level reduction can be applied repeatedly, with the probability of badness evolving in a similar way each time. 

To obtain useful lower bounds on the threshold, we will also need to face two technicalities that were encountered previously in \cite{AGP}. We will need to consider gadgets that overlap with one another, and therefore must adjust our notion of badness to ensure that two overlapping bad gadgets really fail independently of one another. And not all pairs of fault locations in a gadget are comparable; we will distinguish between ``malignant'' pairs of locations such that faults at the pair of locations can cause the gadget to be incorrect, and ``benign'' pairs of locations such that the gadget is correct even if arbitrary faults occur at that pair of locations. 

We remark that our method of analysis applies to 1-Recs constructed as in Fig.~\ref{fig:level-1-sim}, where gate gadgets are preceded by error-detecting gadgets applied to all input blocks. A different formulation of the level-1 simulation uses teleportation to execute the level-1 encoded gates, thereby integrating the error detection into the gate implementation \cite{knill-schemes}, and our analysis does not apply to such gadgets. Our methods, appropriately adapted, can be applied to teleported gates that have suitable properties, but we will not discuss that extension in this paper.

The rest of this paper will fill in the details of the above proof sketch.

\section{Locally correlated stochastic noise}
\label{sec:noise}

\subsection{The trouble with local stochastic noise}
For the threshold estimate in \cite{AGP} based on quantum error {\em correction}, a ``local stochastic noise model'' was assumed. In this model, a quantum circuit is expressed as a sum over ``fault paths,'' each of which is assigned a probability. A fault path specifies a subset of all the level-0 locations in the quantum circuit where the faults occur. For the 0-locations that are not faulty, the quantum gates are assumed to be ideal, and for the faulty 0-locations the quantum gates are arbitrary. 

%[{\em Give a formal definition?}]

In the local stochastic noise model with noise strength $\varepsilon$, the sum of the probabilities of all the fault paths that are faulty at $r$ specified 0-locations is assumed to be no larger than $\varepsilon^r$. (Here no condition is imposed on the circuit at the 0-locations outside the specified set of $r$ 0-locations --- these may or may not by faulty.) Another way to formulate this model is to suppose that faults are distributed in the quantum circuit by an independent and identically distributed (i.i.d.) process: for each set of $s$ level-0 locations in the circuit, the probability that these locations, and only these locations, are faulty is  $(1-\varepsilon)^{L_0-s}\varepsilon^s$, where $L_0$ is the total number of 0-locations in the circuit. Again, for each fault path, the gates are arbitrary at the faulty locations and ideal elsewhere. 

An adversary who decides how to assign an operation to each fault path has the power to impose strong correlations (both temporal and spatial) among the faults at different locations. An adversarial noise model allowing such correlations may seem overly pessimistic, but is nevertheless a natural choice for the analysis of fault-tolerant circuits based on error correction, because its structure is preserved under level reduction. Even if we assume that faults in level-0 gates are completely uncorrelated, bad level-1 gadgets can be correlated, and these become correlated level-0 faults after one level reduction step. The correlations could become more and more complex to analyze as further level reduction steps are performed. On the other hand, the adversarial local stochastic noise model is already so pessimistic that level reduction does not make the correlations any worse, and each level reduction step can be analyzed in the same way. Furthermore, allowing adversarial noise does not much affect the rigorous lower bound on the threshold derived using concatenated error-correcting codes in \cite{AGP}. However, if the same local stochastic noise model is assumed for postselected quantum computation based on error {\em detection}, then the adversary can exploit the correlations in the noise to foil the simulation. To prove a threshold theorem, we must refine the noise model to limit the adversary's power.

To understand how the correlations can be exploited by the adversary, consider a level-1 simulation. Suppose we are interested in estimating the probability that a certain level-1 location fails (i.e., is incorrect), given that no errors are detected by any level-1 error detection (1-ED) in the circuit. Fix a pair of 0-locations in the 1-location such that faults at that pair of 0-locations can be chosen so that the 1-location is incorrect and no errors are detected by any of the leading or trailing 1-EDs associated with that 1-location. Let us call these two 0-locations $A$ and $B$. Faults at the pair of 0-locations $AB$ occur with probability $\varepsilon^2$. 

Now divide all the fault paths into two classes -- those with faults at $AB$ and the rest. The adversary can arrange for every fault path with faults at $AB$ to be accepted, by choosing the faults at $AB$ appropriately and ``turning off'' all other faults (i.e., setting the operation at all the other ``faulty'' locations to be ideal). The adversary can also arrange for nearly every fault path that does not have faults at $AB$ to be rejected. For any fault path with at least one fault, the adversary can turn off all faults except one, and choose that one fault so that it triggers a subsequent 1-ED. The only fault path that the adversary cannot choose to be rejected is the trivial one with no faults anywhere. The trivial fault path occurs with probability $(1-\varepsilon)^{L_0}$, where $L_0$ is the total number of 0-locations.

Therefore (using Bayes's rule), the local stochastic noise model is compatible with the conditional probability of failure at a specified 1-location
\begin{equation}
P_{\rm fail}^{(1)}= {\rm Prob(faults~at~} AB~|~{\rm global ~acceptance)}=\frac{\varepsilon^2}{\varepsilon^2 + (1-\varepsilon)^{L_0}}~.
\end{equation}
This failure probability is at least $1/2$ for 
\begin{equation}
\varepsilon^2\ge e^{-\varepsilon L_0} \ge (1-\varepsilon)^{L_0}~,
\end{equation}
or
\begin{equation}
L_0 \ge \left(2 \varepsilon^{-1}\right)\ln(1/\varepsilon)~.
\end{equation}
If the minimal number of 0-locations contained in any 1-Rec is $C$, then the probability of failure is greater than $1/2$ for 
\begin{equation}
L_1 \ge (2/C)\varepsilon^{-1}\ln(1/\varepsilon)~,
\end{equation}
where $L_1$ is the number of 1-Recs. 
The simulation fails (there is an undetected encoded error) with probability greater than $1/2$ for a number of 1-locations that scales almost like $\varepsilon^{-1}$ (with only a logarithmic correction). In contrast, for a level-1 simulation based on a code that corrects one error, the circuit size that can be simulated accurately scales like $\varepsilon^{-2}$.

Similarly, at level $k$, we can choose faults at $2^k$ 0-locations that cause a particular $k$-location to fail, and the adversary can achieve
\begin{equation}
P_{\rm fail}^{(k)}= \frac{\varepsilon^{2^k}}{\varepsilon^{2^k} + (1-\varepsilon)^{L_0}}~,
\end{equation}
which is greater than $1/2$ for 
\begin{equation}
L_0 \ge \left(2^k \varepsilon^{-1}\right)\ln(1/\varepsilon)~,
\end{equation}
or
\begin{equation}
L_k \ge (2/C)^k\varepsilon^{-1}\ln(1/\varepsilon)~,
\end{equation}
where $L_k$ is the number of $k$-locations (the number of 0-locations inside each $k$-location is at least $C^k$). The scaling with $\varepsilon$ is the same as for $k=1$, but the factor $(2/C)^k$ further suppresses the size of a circuit that can be reliably simulated. In contrast, for a level-$k$ simulation based on a concatenated error-correcting code, the circuit size that can be simulated accurately scales like $\varepsilon^{-2^k}$.

\subsection{A more local noise model}
\label{sec:more-local-noise}

A postselected computation is vulnerable to highly adversarial correlated noise because most fault paths are rejected due to detection of an error. Therefore an adversary with global control over the correlations has unreasonable power --- by arranging for the acceptance of fault paths that would otherwise be rejected, she can wield substantial influence over how the faults are distributed when the circuit is accepted. To formulate and prove a threshold theorem for postselected quantum computation, we must choose a noise model that, on the one hand, disallows overly conspiratorial noise correlations and that, on the other hand, propagates simply under level reduction. 

An independent noise model with no correlations at all meets the first criterion, but not the second one. The trouble is that, in a level-1 circuit, there are two types of quantum information propagating through the circuit. In addition to the encoded qubits that are being processed by the circuit, there is also ``syndrome information'' characterizing the deviation from the code space. Undetected errors can carry information about faults that occurred in earlier gadgets forward to later gadgets. Therefore, even if the noise in the level-0 gates were uncorrelated, there would be correlations among the bad level-1 gadgets, which would become correlations among level-0 gates after one level reduction step.

Fortunately, though, there are limitations on the flow of syndrome information through the level-1 circuit and corresponding limitations on the correlations among bad level-1 gadgets. A 1-ED that has no faults and detects no errors necessarily projects the block to the code space, destroying any evidence of preceding faults, and preventing information about these faults from propagating forward. This observation provides guidance regarding how to formulate a noise model whose properties are preserved by level reduction.

With each level-1 simulated gate we associate an extended rectangle (1-exRec), consisting of the encoded gate together with the 1-EDs immediately preceding the gate acting on all input blocks (the gate's ``leading 1-EDs''), and the 1-EDs immediately following the gate acting on all output blocks (the gate's ``trailing 1-EDs''); see Fig.~\ref{fig:exRec}. Let us say that a 1-exRec is good if it contains no more than one fault, and that a 1-ED is ``very good'' if it contains no faults. Then it is impossible for syndrome information about faults that occured prior to a good 1-exRec to propagate through the good 1-exRec and reach a later 1-exRec. The reason is that if there is at most one fault in the 1-exRec, then either all of the leading 1-EDs are very good, or else all of the trailing 1-EDs are very good. Either way (after postselection), the blocks are restored to the code space by the very good 1-EDs, and any evidence of previous faults is erased.

\begin{figure}
\begin{center}
\vspace{0.5cm}
\begin{picture}(201,96)
\put(5,12){\line(1,0){5}}
\put(10,0){\framebox(24,24){1-ED}}
\put(34,12){\line(1,0){14}}
\put(5,44){\line(1,0){5}}
\put(10,32){\framebox(24,24){1-ED}}
\put(34,42){\line(1,0){14}}
\put(48,0){\framebox(28,56){1-Ga}}
\put(76,12){\line(1,0){14}}
\put(76,44){\line(1,0){5}}
\put(85,44){\line(1,0){5}}
\put(90,0){\framebox(24,24){1-ED}}
\put(114,12){\line(1,0){5}}
\put(90,32){\framebox(24,24){1-ED}}
\put(114,44){\line(1,0){5}}
\put(123,44){\line(1,0){5}}
\put(85,76){\line(1,0){5}}
\put(90,64){\framebox(24,24){1-ED}}
\put(114,76){\line(1,0){14}}
\put(128,32){\framebox(28,56){1-Ga}}
\put(156,44){\line(1,0){14}}
\put(156,76){\line(1,0){14}}
\put(170,32){\framebox(24,24){1-ED}}
\put(194,44){\line(1,0){5}}
\put(170,64){\framebox(24,24){1-ED}}
\put(194,76){\line(1,0){5}}
\put(3,-4){\dashbox(118,64){}}
\put(83,28){\dashbox(118,64){}}
\end{picture}
\vspace{0.3cm}
\end{center}
\fcaption{Two consecutive level-1 extended rectangles (indicated by dashed lines). Each 1-exRec consists of a gate gadget (denoted 1-Ga) that realizes an encoded two-qubit gate, preceded by ``leading'' error detection gadgets (1-EDs) applied to each input block and followed by ``trailing'' error detection gadgets applied to each output block. The 1-exRecs overlap --- a trailing 1-ED of the earlier 1-exRec is also a leading 1-ED of the later 1-exRec}
\label{fig:exRec}
\end{figure}

Now, the story is not quite so simple. As will be explained in more detail in Sec.~\ref{sec:carving}, a good 1-exRec that precedes a bad 1-exRec may be regarded as ``truncated'' --- one or more of its trailing 1-EDs is amputated, so that the above argument may not apply as such. Nevertheless, we will be able to formulate a definition of goodness for truncated 1-exRecs such that syndrome information is unable to propagate through a good 1-exRec to another good 1-exRec that immediately follows. After level reduction, two successive good 1-exRecs become consecutive ideal level-0 gates, where there is no communication from the first ideal gate to the one that follows. 

Thus, if the noise were uncorrelated at level 0, then the effective noise model that we would obtain after one level reduction step is correlated, but the correlations are of a special kind. ``Syndrome information'' can be passed from one gate to the next, and the bad gates can consult this information when they decide how to fail. {\em However, wherever two good gates occur in succession, communication between the two gates is not allowed.} We will regard this feature as the defining property of a new noise model that we call {\em locally correlated stochastic noise}. 

That is, let us imagine that a local adversary resides at each location of a level-0 circuit, and suppose that the bad (i.e. faulty) 0-locations are chosen by an i.i.d. process with fault probability $\varepsilon$.  The ideal gate is applied at the good 0-locations, but at a bad 0-location the local adversary may choose an arbitrary operation to apply in place of the ideal gate. Each local adversary has a quantum memory of unbounded size; she may receive quantum messages from the adversaries at the immediately preceding or immediately following locations, and she may send quantum messages to the adversaries at the immediately preceding or immediately following locations. (Though it may seem perverse to allow messages to be passed both forward and backward in time, this feature is required to ensure the stability of the noise model under level reduction.) A local adversary at a bad location may perform operations that act jointly on the message system that she receives and on the data that is being processed at her 0-location. In addition, local adversaries at two bad 0-locations that both immediately follow the same good 0-location are permitted to ``communicate,'' so that they can apply collective operations to their data. But no communication from a good gate to an immediately following or immediately preceding good gate is permitted. See Fig.~\ref{fig:locally-correlated} and Fig.~\ref{fig:locally-correlated2}.

\begin{figure}
\begin{center}
\vspace{0.3cm}
\begin{picture}(330,46)
\put(0,12){\line(1,0){10}}
\put(10,0){\framebox(30,24){\shortstack{bad\\$0$-Ga}}}
\put(40,12){\line(1,0){10}}
\put(50,0){\framebox(30,24){\shortstack{good\\$0$-Ga}}}
\put(80,12){\line(1,0){10}}
\put(90,0){\framebox(30,24){\shortstack{bad\\$0$-Ga}}}

\put(120,12){\line(1,0){10}}
\put(25,24){\line(0,1){10}}
\put(65,34){\vector(-1,0){25}}
\put(25,34){\line(1,0){15}}
\put(90,34){\line(1,0){15}}
\put(65,34){\vector(1,0){25}}
\put(105,24){\line(0,1){10}}
\put(54,34){\makebox(20,12){message}}
\put(200,12){\line(1,0){10}}
\put(210,0){\framebox(30,24){\shortstack{bad\\$0$-Ga}}}
\put(240,12){\line(1,0){10}}
\put(250,0){\framebox(30,24){\shortstack{good\\$0$-Ga}}}
\put(280,12){\line(1,0){10}}
\put(290,0){\framebox(30,24){\shortstack{good\\$0$-Ga}}}

\put(320,12){\line(1,0){10}}
\put(225,24){\line(0,1){10}}
\put(225,34){\vector(1,0){55}}
\put(235,34){\makebox(20,12){message}}
\put(275,28){\makebox(20,12){$\times$}}
\end{picture}
\vspace{0.4cm}
\end{center}
\fcaption{Communication among local adversaries allowed by the locally correlated stochastic noise model. If a good gate is both preceded and followed by bad gates, then a message can pass through the good gate from one bad gate to the other. But if two good gates occur in succession, communication through the pair of good gates is not permitted.}
\label{fig:locally-correlated}
\end{figure}

\begin{figure}
\begin{center}
\vspace{0.2cm}
\begin{picture}(100,72)
\put(0,12){\line(1,0){10}}
\put(0,60){\line(1,0){10}}
\put(10,0){\framebox(30,72){\shortstack{good\\$0$-Ga}}}
\put(40,12){\line(1,0){10}}
\put(40,60){\line(1,0){10}}
\put(50,0){\framebox(30,24){\shortstack{bad\\$0$-Ga}}}
\put(50,48){\framebox(30,24){\shortstack{bad\\$0$-Ga}}}
\put(80,12){\line(1,0){10}}
\put(80,60){\line(1,0){10}}

\put(65,36){\vector(0,1){12}}
\put(65,36){\vector(0,-1){12}}
\put(80,24){\makebox(20,24){\shortstack{{\small collective}\\{\small operation}}}}

\end{picture}
\vspace{0.3cm}
\end{center}
\fcaption{Another type of ``communication'' allowed by the locally correlated stochastic noise model. Local adversaries at two bad 0-locations that both immediately follow the same good 0-location are permitted to apply a collective operation to their shared data.}
\label{fig:locally-correlated2}
\end{figure}

The locally correlated stochastic noise model has been constructed so that the effective noise after level reduction can also be described as locally correlated stochastic noise (though with a different value of the fault rate). This works because, as noted above, a very good 1-ED that detects no error erases the syndrome, but that property by itself is not quite sufficient. If communication is already allowed at level 0, there are actually (at least in principle) two ways to induce correlations among bad 1-gadgets: information about past faults can be carried forward by the syndrome {\em or} information about past or future faults can be carried by the messages that are passed between local adversaries. Fortunately, though, both kinds of communication will be blocked by a very good 1-ED if the 1-ED circuit is suitably constructed. For example, if the 1-ED has depth at least two, then there is no path through a very good 1-ED that can evade passing through two consecutive good 0-locations. Therefore, neither syndrome information nor level-0 messages can penetrate a very good 1-ED. 

Note that if a good 0-location is both preceded and followed by a bad 0-location, then communication between the two bad 0-locations that passes through the good 0-location is allowed by the locally correlated stochastic noise model. This feature is included because, as discussed in Sec.~\ref{sec:carving}, uncorrected errors might propagate through a good truncated 1-exRec, which after level reduction provides an open communication channel through the resulting ideal 0-gate. As will also be further explained in Sec.~\ref{sec:carving}, operations acting collectively at two bad 0-locations that both immediately follow the same good 0-location are included as well, to ensure that the noise model is preserved by level reduction. 

To improve the threshold estimate, it is desirable to allow a good 1-exRec to contain faults at a ``benign'' pair of locations. For this purpose we must be sure that the definition of benign is compatible with the requirement that the noise model is stable under level reduction. One possible approach is to design 1-ED gadgets of sufficiently high depth. For our threshold estimate in this paper we will instead formulate a variant of the locally correlated stochastic noise model for which the analysis proceeds smoothly --- this reformulated noise model will be explained in Sec.~\ref{sec:seal-cluster-Knill}. 

The threshold theorem based on concatenated error correcting codes also applies to non-stochastic noise models in which fault paths are added in superposition \cite{ben-or,terhal,AGP,aharonov-kitaev-preskill}. It may be possible to prove a threshold theorem for postselected quantum computation for non-stochastic noise, as long as the noise correlations are suitably restricted, but we will not attempt an analysis of non-stochastic noise in this paper. The task of estimating the probability that a circuit fails, conditioned on acceptance by all error detection steps, is simplified if we assign probabilities rather than amplitudes to the fault paths.

\section{Goodness and correctness}
\label{sec:goodness}

To prove the threshold theorem for postselected quantum computing, we need to analyze how the noise in a postselected circuit is affected by level reduction. That analysis has two parts. In the first part, discussed here and in Sec.~\ref{sec:carving}, we consider the probability for 1-exRecs to be bad conditioned on local acceptance; that is, assuming that the leading and trailing 1-EDs in the 1-exRec detect no errors. In the second part we will relate the probability of badness conditioned on local acceptance to the probability of badness conditioned on global acceptance; that is, assuming that no 1-ED anywhere in the circuit detects any errors. The restrictions on correlations in the noise assumed in our noise model come into play only in the second part of the analysis, which we will postpone until Sec.~\ref{sec:local-to-global}. 

A mathematical concept that turns out to be useful for analyzing the fault-tolerant circuits is the {\em ideal decoder} that maps an encoded 1-block to the corresponding state of a single qubit. We will actually make a distinction between two types of decoders, which we call an ideal 1-decoder and an ideal 1-$^*$decoder. The 1-decoder measures the error syndrome, uses the syndrome information to perform a canonical mapping to the code space, decodes the 1-block to a qubit, and then discards the syndrome. Of course, a nontrivial syndrome for a distance-2 code is ambiguous because it could arise from more than one weight-1 Pauli error. But we may adopt an arbitrary rule that associates each nontrivial syndrome with a particular recovery operation. For example, for the [[4,2,2]] code that will be discussed in detail in Sec.~\ref{sec:gadgets}, if the syndrome measurement reveals that $Z^{\otimes 4}=-1$, the value $Z^{\otimes 4}=1$ could be restored by applying $X$ to any one of the four qubits in the block. Our standard choice might be to apply $X$ to the first qubit. The 1-$^*$decoder (denoted ${\cal D}$) is just like the 1-decoder, except that the syndrome is extracted coherently rather than measured and retained rather than discarded. The 1-decoder and the 1-$^*$decoder map an input block that is in the code space to the same output qubit state. Since we are using only one of the two logical qubits in the [[4,2,2]] code block, the ``syndrome'' also includes the value of the ``spectator'' (or ``gauge'' \cite{bacon,poulin}) qubit in the block. That is, the action of the 1-$^*$decoder can be expressed as
\begin{equation}
\label{one-decoder-action}
\bar X_S^{s} X_1^x Z_1^z |\bar\psi,\bar 0\rangle \mapsto |\psi\rangle\otimes |s\rangle\otimes |x\rangle\otimes |z\rangle~,\
\end{equation}
where $|\bar \psi,\bar 0\rangle$ denotes the state of the code block with the logical qubit in the state $|\psi\rangle$ and the spectator qubit in the state $|0\rangle$, $\bar{X}_S$ denotes the Pauli $X$ acting on the spectator, and $X_1$, $Z_1$ are Pauli operators acting on the first qubit in the block. 

We make a similar distinction between the ideal 1-encoder and the ideal 1-$^*$encoder. The 1-encoder maps an input qubit state to the corresponding logical state of the encoded block. The 1-$^*$encoder's input consists of both a qubit and a syndrome, and it maps the input qubit to the state of the encoded block that deviates from the corresponding logical state by the standard Pauli operator associated with the syndrome. Thus the 1-$^*$encoder ${\cal D}^{-1}$ is the inverse of the  1-$^*$decoder ${\cal D}$: ${\cal D}{\cal D}^{-1}=I$. Also note that the action of the 1-encoder on an input qubit coincides with the action of the 1-$^*$encoder on an input qubit if the input syndrome is trivial.

Now recall that a level-1 simulation of an ideal circuit is constructed by replacing each gate by the corresponding level-1 rectangle (1-Rec), which consists of a level-1 gate gadget (1-Ga) preceded by level-1 error detection steps (1-EDs) acting on all input blocks, as in Fig.~\ref{fig:level-1-sim}. The 1-Rec combined with the following 1-EDs acting on all output blocks is called a level-1 extended rectangle (1-exRec), as shown in Fig.~\ref{fig:exRec}.

For a specified fault path, we will classify level-1 gadgets as very good, good, or pre-bad according to the following rules:

\medskip
\noindent {\bf Classification of noisy gadgets}
\begin{description}
\item[$1.$] A 1-ED, 1-Ga, or 1-Rec is {\em very good} if it contains no faults. 
\item[$2.$] A 1-ED, 1-Ga, 1-Rec, or 1-exRec is {\em good} if it contains no more than one fault. 
\item[$3.$] A 1-exRec is {\em pre-bad} if it contains two or more faults. 
\end{description}

\noindent To begin our analysis of the level-1 simulation (with a specified fault path), we examine each 1-exRec in the circuit, and designate it as either good or pre-bad. Since the 1-exRecs can overlap with one another (each 1-ED is contained in two 1-exRecs), some faults may contribute to the pre-badness of two distinct 1-exRecs, but ignore that for now --- we will return to this issue in Sec.~\ref{sec:carving}. We use the term ``pre-bad'' here because, after taking the overlaps into account, some of the pre-bad 1-exRecs will be reclassified as good, according to rules that will be formulated in Sec.~\ref{sec:good-prime}. 

The ideal quantum circuit can be regarded as an acyclic directed graph, and the location of each pre-bad 1-exRec can be identified as a vertex in this graph. These pre-bad locations form a subgraph of the circuit, and we refer to each connected component of this subgraph as a ``bad cluster.'' Thus, for each fault path there is a corresponding archipelago consisting of disjoint connected clusters of pre-bad 1-exRecs, each embedded in a surrounding sea of good 1-exRecs. Let us imagine
enclosing each bad cluster with insertions of decoder-encoder pairs ${\cal D}{\cal D}^{-1}=I$. We insert ${\cal D}{\cal D}^{-1}$ after the trailing 1-EDs of the 1-exRecs at the rear of the connected bad cluster, and before the leading 1-EDs of the 1-exRecs at the front of the cluster.

Of course, these insertions are equivalent to doing nothing. The reason for the insertions is that we may now imagine moving the encoders forward through the sea of good 1-exRecs. The encoders start at either the beginning of the circuit or at the rear end of a cluster, and sweep forward until they meet up with the decoders and annihilate them, either at the end of the circuit or at the front end of a cluster. We will argue that the sweeping encoders convert each good 1-exRec in the sea into an ideal level-0 gate. Then in Sec.~\ref{sec:carving} we will consider how to reduce the pre-bad 1-exRecs in the clusters into level-0 gates.  

We wish to show that the 1-Rec contained in a good 1-exRec becomes an ideal gate under level reduction, i.e., as the ideal encoders sweep to the right; hence we say that the good 1-exRec is {\em correct}. To show that a good 1-exRec is correct, we will need to use some properties of fault-tolerant error detection and gate gadgets for a distance-two code. Before we specify these properties in detail, let us state more carefully what these properties imply. In fact, we will need to distinguish two different notions of correctness for single-qubit gates (and three different notions for two-qubit gates) --- which notion applies depends on whether the preceding 1-exRec is good or pre-bad. 

If the 1-exRec preceding the good 1-exRec is also good, then the relevant notion of correctness is what we will call A-correctness:

\vspace{0.5cm}
\begin{picture}(328,24)
\setlength{\unitlength}{0.9pt}
\put(0,12){\line(1,0){10}}
\put(10,0){\framebox(24,24){${\cal D}^{-1}$}}
\put(34,12){\line(1,0){10}}
\put(44,0){\framebox(30,24){1-ED}}
\put(74,12){\line(1,0){10}}
\put(84,0){\framebox(30,24){1-Ga}}
\put(114,12){\line(1,0){10}}
\put(124,0){\framebox(30,24){1-ED}}
\put(154,12){\line(1,0){10}}
\put(164,6){\makebox(20,12){=}}
\put(184,12){\line(1,0){10}}
\put(194,0){\framebox(30,24){\shortstack{ideal\\$0$-Ga}}}
\put(224,12){\line(1,0){10}}
\put(234,0){\framebox(24,24){${\cal D}^{-1}$}}
\put(258,12){\line(1,0){10}}
\put(268,0){\framebox(30,24){1-ED}}
\put(298,12){\line(1,0){10}}
\put(308,6){\makebox(20,12){.}}
\end{picture}
\vspace{0.3cm}

\noindent
A-correctness says that we can move a 1-encoder forward past the 1-Rec in a good 1-exRec, thereby converting the 1-Rec to an ideal level-0 gate (0-Ga). This equality means that for any single-qubit input, {\em after postselection} (i.e., conditioning on acceptance by both the leading and trailing 1-EDs) the output 1-block from the 1-Ga on the left-hand side agrees with the output of the 1-encoder on the right-hand side. 

Our level-1 gadgets include a  1-preparation (which has no input) and a 1-measurement (which has a classical bit as output, and includes error detection). A-correctness properties can also be defined for 1-preparation and 1-measurement:

\vspace{0.5cm}
\begin{picture}(266,24)
\setlength{\unitlength}{0.9pt}
%
%\put(0,12){\line(1,0){10}}
\put(10,0){\framebox(36,24){1-prep}}
\put(46,12){\line(1,0){10}}
\put(56,0){\framebox(30,24){1-ED}}
\put(86,12){\line(1,0){10}}
\put(96,6){\makebox(20,12){=}}
%
%\put(116,12){\line(1,0){10}}
\put(126,0){\framebox(36,24){\shortstack{ideal\\$0$-prep}}}
\put(162,12){\line(1,0){10}}
\put(172,0){\framebox(24,24){${\cal D}^{-1}$}}
\put(196,12){\line(1,0){10}}
\put(206,0){\framebox(30,24){1-ED}}
\put(236,12){\line(1,0){10}}
\put(246,6){\makebox(20,12){,}}
\end{picture}

\vspace{0.3cm}
%
%\vspace{0.5cm}
\begin{picture}(216,24)
\setlength{\unitlength}{0.9pt}
\put(0,12){\line(1,0){10}}
\put(10,0){\framebox(24,24){${\cal D}^{-1}$}}
\put(34,12){\line(1,0){10}}
\put(44,0){\framebox(30,24){1-ED}}
\put(74,12){\line(1,0){10}}
\put(84,0){\framebox(36,24){1-meas}}
\put(130,6){\makebox(20,12){=}}
\put(150,12){\line(1,0){10}}
\put(160,0){\framebox(36,24){\shortstack{ideal\\$0$-meas}}}
\put(196,6){\makebox(20,12){.}}
\end{picture}
\vspace{0.3cm}

If the 1-exRec preceding the good 1-exRec is pre-bad, then the relevant notion of correctness is what we will call B-correctness:

\vspace{0.4cm}
\begin{picture}(396,34)
\setlength{\unitlength}{0.9pt}
\put(0,12){\line(1,0){10}}
\put(10,0){\framebox(30,24){1-ED}}
\put(40,12){\line(1,0){10}}
\put(50,0){\framebox(24,24){${\cal D}$}}
\put(74,12){\line(1,0){10}}
\put(84,0){\framebox(24,24){${\cal D}^{-1}$}}
\put(108,12){\line(1,0){10}}
\put(62,24){\line(0,1){10}}
\put(62,34){\line(1,0){34}}
\put(96,24){\line(0,1){10}}
\put(118,0){\framebox(30,24){1-Ga}}
\put(148,12){\line(1,0){10}}
\put(158,0){\framebox(30,24){1-ED}}
\put(188,12){\line(1,0){10}}
\put(198,6){\makebox(20,12){=}}
\put(218,12){\line(1,0){10}}
\put(228,0){\framebox(30,24){1-ED}}
\put(258,12){\line(1,0){10}}
\put(268,0){\framebox(24,24){${\cal D}$}}
\put(292,12){\line(1,0){10}}
\put(280,24){\line(0,1){10}}
\put(280,34){\line(1,0){15}}
\put(290,28){\makebox(20,12){$\times$}}
\put(302,0){\framebox(30,24){\shortstack{ideal\\$0$-Ga}}}
\put(332,12){\line(1,0){10}}
\put(342,0){\framebox(24,24){${\cal D}^{-1}$}}
\put(366,12){\line(1,0){10}}
\put(376,0){\framebox(30,24){1-ED}}
\put(406,12){\line(1,0){10}}
\put(416,6){\makebox(20,12){.}}
\end{picture}
\vspace{0.1cm}

\noindent
B-correctness says that if a decoder-encoder pair is inserted after the leading 1-ED in a good 1-exRec, then we can move the 1-$^*$encoder forward past the 1-Ga, thereby converting it to an ideal 0-Ga. This equality means that for any state of the input 1-block, {\em after postselection} the output 1-block from the 1-Ga on the left-hand side agrees with the output of the 1-encoder on the right-hand side. (In our diagrammatic notation, the horizontal input and output to a 1-$^*$encoder or 1-$^*$decoder represents the data, while the ``hooked'' line represents the input or output syndrome; if the hooked line is missing, that indicates that the syndrome is trivial, in which case the 1-$^*$encoder is equivalent to a 1-encoder.) Note that on the right-hand side, the 1-encoder has no input syndrome; that is, B-correctness includes the assertion that the output of the 1-Ga on the left-hand side (after postselection) is a codeword. An analogous B-correctness property can also be defined for  1-measurement:

\vspace{0.4cm}
\begin{picture}(396,34)
\setlength{\unitlength}{0.9pt}
\put(0,12){\line(1,0){10}}
\put(10,0){\framebox(30,24){1-ED}}
\put(40,12){\line(1,0){10}}
\put(50,0){\framebox(24,24){${\cal D}$}}
\put(74,12){\line(1,0){10}}
\put(84,0){\framebox(24,24){${\cal D}^{-1}$}}
\put(108,12){\line(1,0){10}}
\put(62,24){\line(0,1){10}}
\put(62,34){\line(1,0){34}}
\put(96,24){\line(0,1){10}}
\put(118,0){\framebox(36,24){1-meas}}

\put(164,6){\makebox(20,12){=}}
\put(184,12){\line(1,0){10}}
\put(194,0){\framebox(30,24){1-ED}}
\put(224,12){\line(1,0){10}}
\put(234,0){\framebox(24,24){${\cal D}$}}
\put(258,12){\line(1,0){10}}
\put(246,24){\line(0,1){10}}
\put(246,34){\line(1,0){15}}
\put(256,28){\makebox(20,12){$\times$}}
\put(268,0){\framebox(36,24){\shortstack{ideal\\$0$-meas}}}
\put(306,6){\makebox(20,12){.}}
\end{picture}
\vspace{0.1cm}

\noindent
For two-qubit gates, we will distinguish AA-correctness (both encoders initially in front of the leading 1-EDs), AB-correctness (one decoder-encoder pair inserted after a leading 1-ED) and BB-correctness (two decoder-encoder pairs inserted after the leading 1-EDs).

In the equations we have written to express A-correctness and B-correctness, a noisy operation on the left-hand side of the equation is replaced by an ideal operation on the right-hand side. Therefore, these equations are not fully satisfactory as written, because the two sides of the equation seem to tell conflicting stories concerning what actions by the adversary will be accepted. This shortcoming can be repaired by including on the right-hand side a noisy circuit that acts on an arbitrary dummy input, a point that will be clarified below.  But for now we have suppressed the noisy processing of the dummy input on the right-hand side in order to emphasize what is really most important --- that, after postselection, the processing of the encoded data is ideal. 

Now we are ready to specify the properties of fault-tolerant 1-ED and 1-Ga gadgets that can be invoked to show that a good 1-exRec is correct. The properties can be expressed as equalities that hold when all 1-EDs accept (detect no errors).

\medskip
\noindent {\bf Properties of fault-tolerant level-1 gadgets}. 

\medskip
\noindent {\em Property ED1}:

\vspace{0.2cm}
\begin{picture}(322,40)
\put(0,12){\line(1,0){10}}
\put(10,0){\framebox(24,24){${\cal D}$}}
\put(22,24){\line(0,1){10}}
\put(22,34){\line(1,0){34}}
\put(34,12){\line(1,0){10}}
\put(44,0){\framebox(24,24){${\cal D}^{-1}$}}
\put(56,24){\line(0,1){10}}
\put(68,12){\line(1,0){10}}
\put(78,0){\framebox(48,24){\small\shortstack{very good\\1-ED}}}
\put(126,12){\line(1,0){10}}
\put(136,0){\framebox(24,24){${\cal D}$}}
\put(148,24){\line(0,1){10}}
\put(148,34){\line(1,0){34}}
\put(160,12){\line(1,0){10}}
\put(170,0){\framebox(24,24){${\cal D}^{-1}$}}
\put(182,24){\line(0,1){10}}
\put(194,12){\line(1,0){10}}
\put(204,6){\makebox(20,12){=}}
\put(224,12){\line(1,0){10}}
\put(234,0){\framebox(24,24){${\cal D}$}}
\put(246,24){\line(0,1){10}}
\put(246,34){\line(1,0){10}}
\put(256,28){\makebox(12,12){0}}
\put(258,12){\line(1,0){30}}
\put(288,0){\framebox(24,24){${\cal D}^{-1}$}}
\put(300,24){\line(0,1){10}}
\put(300,34){\line(-1,0){10}}
\put(278,28){\makebox(12,12){0}}
\put(312,12){\line(1,0){10}}
\end{picture}
\vspace{0.3cm}

\noindent We say that a 1-ED is ``very good'' if it contains no faults at all. Property ED1 just says that a 1-ED with no faults does what it is supposed to do --- it detects a deviation from the code space. The ``0'' on the right-hand side indicates that the syndrome is projected to its trivial value, and thus that the very good 1-ED on the left-hand side projects its input to the code space if it does not detect an error. Furthermore, property ED1 expresses that the operation applied to the data qubit on the right-hand side is the identity; thus the output of the very good 1-ED on the left-hand side matches its postselected input.

\medskip
\noindent {\em Property ED2}:

\vspace{0.2cm}
\begin{picture}(322,40)
\put(0,12){\line(1,0){10}}
\put(10,0){\framebox(24,24){${\cal D}$}}
\put(22,24){\line(0,1){10}}
\put(22,34){\line(1,0){11}}
\put(33,28){\makebox(12,12){0}}
\put(45,34){\line(1,0){11}}
\put(34,12){\line(1,0){10}}
\put(44,0){\framebox(24,24){${\cal D}^{-1}$}}
\put(56,24){\line(0,1){10}}
\put(68,12){\line(1,0){10}}
\put(78,0){\framebox(48,24){\small\shortstack{good\\1-ED}}}
\put(126,12){\line(1,0){10}}
\put(136,0){\framebox(24,24){${\cal D}$}}
\put(148,24){\line(0,1){10}}
\put(148,34){\line(1,0){11}}
\put(159,28){\makebox(12,12){0}}
\put(171,34){\line(1,0){11}}
\put(160,12){\line(1,0){10}}
\put(170,0){\framebox(24,24){${\cal D}^{-1}$}}
\put(182,24){\line(0,1){10}}
\put(194,12){\line(1,0){10}}
\put(204,6){\makebox(20,12){=}}
\put(224,12){\line(1,0){10}}
\put(234,0){\framebox(24,24){${\cal D}$}}
\put(246,24){\line(0,1){10}}
\put(246,34){\line(1,0){10}}
\put(256,28){\makebox(12,12){0}}
\put(258,12){\line(1,0){30}}
\put(288,0){\framebox(24,24){${\cal D}^{-1}$}}
\put(300,24){\line(0,1){10}}
\put(300,34){\line(-1,0){10}}
\put(278,28){\makebox(12,12){0}}
\put(312,12){\line(1,0){10}}
\end{picture}
\vspace{0.3cm}

\noindent We say that a 1-ED is ``good'' if it contains no more than one fault. Property ED2 says that the 1-ED is fault tolerant in the sense that errors resulting from a fault do not propagate badly and cause an encoded error. Here ``0'' on the left-hand side indicates that the syndrome is projected to its trivial value, and thus that both the input and the output to the good 1-ED is in the code space. The right-hand side of property ED2 indicates that the output codeword matches the input codeword --- thus the good 1-ED, if it accepts and its output is in the code space, simulates the identity operation.

In property ED2, the adversary applies a faulty operation on the left-hand side but not on the right-hand side. To ensure that the same actions by the adversary are accepted on both sides of the equation, we need to augment the right-hand side with a noisy computation acting on a dummy input:

\vspace{0.2cm}
\begin{picture}(196,40)
\put(0,6){\makebox(50,12){dummy input}}
\put(59,28){\makebox(12,12){0}}
\put(71,34){\line(1,0){11}}
\put(60,12){\line(1,0){10}}
\put(70,0){\framebox(24,24){${\cal D}^{-1}$}}
\put(82,24){\line(0,1){10}}
\put(94,12){\line(1,0){10}}
\put(104,0){\framebox(48,24){\small\shortstack{good\\1-ED}}}
\put(152,12){\line(1,0){10}}
\put(162,0){\framebox(24,24){${\cal D}$}}
\put(174,24){\line(0,1){10}}
\put(174,34){\line(1,0){11}}
\put(185,28){\makebox(12,12){0}}
\put(186,12){\line(1,0){10}}
\end{picture}
\vspace{0.3cm}

\noindent We don't need to specify the dummy input to this computation because, if the adversary can attack only one location in the 1-ED circuit, and if the input to the 1-ED is a codeword, then whether the noisy 1-ED accepts or not does not depend on {\em which} codeword is chosen as the input.

\medskip
\noindent {\em Property ED3}:

\vspace{0.2cm}
\begin{picture}(224,40)
\put(0,12){\line(1,0){10}}
\put(10,0){\framebox(48,24){\small\shortstack{good\\1-ED}}}
\put(58,12){\line(1,0){10}}
\put(68,0){\framebox(24,24){${\cal D}$}}
\put(80,24){\line(0,1){10}}
\put(80,34){\line(1,0){11}}
%\put(91,28){\makebox(12,12){0}}
\put(92,12){\line(1,0){10}}
\put(102,6){\makebox(20,12){=}}
\put(122,12){\line(1,0){10}}
\put(132,0){\framebox(48,24){\small\shortstack{good\\1-ED}}}
\put(180,12){\line(1,0){10}}
\put(190,0){\framebox(24,24){${\cal D}$}}
\put(202,24){\line(0,1){10}}
\put(202,34){\line(1,0){11}}
\put(213,28){\makebox(12,12){1}}
\put(214,12){\line(1,0){10}}
\end{picture}
\vspace{0.3cm}

\noindent Here the ``1'' labeling the output syndrome on the right-hand side indicates a syndrome that points to a state deviating from the code space due to the action of a weight-one operator. That is, the decoder with a ``1'' syndrome is a projector onto the space spanned by all states that can be obtained by acting on a codeword with an operator that acts on a single qubit in the code block. Property ED3 expresses another fault-tolerance property --- for {\em any} input, the output of a 1-ED with at most one fault is no more than 1-deviated from the code space. In the case of the [[4,2,2]] code, which is a perfect distance-2 quantum code, property ED3 is automatic, because the 1-deviated states span the entire Hilbert space of the code block (and because we don't care about the state of the ``gauge'' qubit). But nevertheless it is useful to include ED3 in our list of specified properties, both for the sake of logical clarity and to emphasize that our analysis can apply to non-perfect codes.

\medskip
\noindent {\em Property Ga1}:

\vspace{0.2cm}
\begin{picture}(360,40)
\put(0,12){\line(1,0){10}}
\put(10,0){\framebox(24,24){${\cal D}$}}
\put(22,24){\line(0,1){10}}
\put(22,34){\line(1,0){11}}
\put(33,28){\makebox(12,12){0}}
\put(45,34){\line(1,0){11}}
\put(34,12){\line(1,0){10}}
\put(44,0){\framebox(24,24){${\cal D}^{-1}$}}
\put(56,24){\line(0,1){10}}
\put(68,12){\line(1,0){10}}
\put(78,0){\framebox(48,24){\small\shortstack{very good\\1-Ga}}}
\put(126,12){\line(1,0){10}}
\put(136,0){\framebox(24,24){${\cal D}$}}
\put(148,24){\line(0,1){10}}
\put(148,34){\line(1,0){34}}
\put(160,12){\line(1,0){10}}
\put(170,0){\framebox(24,24){${\cal D}^{-1}$}}
\put(182,24){\line(0,1){10}}
\put(194,12){\line(1,0){10}}
\put(204,6){\makebox(20,12){=}}
\put(224,12){\line(1,0){10}}
\put(234,0){\framebox(24,24){${\cal D}$}}
\put(246,24){\line(0,1){10}}
\put(246,34){\line(1,0){10}}
\put(256,28){\makebox(12,12){0}}
\put(258,12){\line(1,0){10}}
\put(268,0){\framebox(48,24){\small\shortstack{ideal\\0-Ga}}}
\put(316,12){\line(1,0){10}}
\put(326,0){\framebox(24,24){${\cal D}^{-1}$}}
\put(338,24){\line(0,1){10}}
\put(338,34){\line(-1,0){10}}
\put(316,28){\makebox(12,12){0}}
\put(350,12){\line(1,0){10}}
\end{picture}
\vspace{0.3cm}

\noindent Property Ga1 says that a 1-Ga with no faults does what it is supposed to do --- its action on the code space simulates the desired ideal operation.

\medskip
\noindent {\em Property Ga2}:

\vspace{0.2cm}
\begin{picture}(360,40)
\put(0,12){\line(1,0){10}}
\put(10,0){\framebox(24,24){${\cal D}$}}
\put(22,24){\line(0,1){10}}
\put(22,34){\line(1,0){11}}
\put(33,28){\makebox(12,12){0}}
\put(45,34){\line(1,0){11}}
\put(34,12){\line(1,0){10}}
\put(44,0){\framebox(24,24){${\cal D}^{-1}$}}
\put(56,24){\line(0,1){10}}
\put(68,12){\line(1,0){10}}
\put(78,0){\framebox(48,24){\small\shortstack{good\\1-Ga}}}
\put(126,12){\line(1,0){10}}
\put(136,0){\framebox(24,24){${\cal D}$}}
\put(148,24){\line(0,1){10}}
\put(148,34){\line(1,0){11}}
\put(159,28){\makebox(12,12){0}}
\put(171,34){\line(1,0){11}}
\put(160,12){\line(1,0){10}}
\put(170,0){\framebox(24,24){${\cal D}^{-1}$}}
\put(182,24){\line(0,1){10}}
\put(194,12){\line(1,0){10}}
\put(204,6){\makebox(20,12){=}}
\put(224,12){\line(1,0){10}}
\put(234,0){\framebox(24,24){${\cal D}$}}
\put(246,24){\line(0,1){10}}
\put(246,34){\line(1,0){10}}
\put(256,28){\makebox(12,12){0}}
\put(258,12){\line(1,0){10}}
\put(268,0){\framebox(48,24){\small\shortstack{ideal\\0-Ga}}}
\put(316,12){\line(1,0){10}}
\put(326,0){\framebox(24,24){${\cal D}^{-1}$}}
\put(338,24){\line(0,1){10}}
\put(338,34){\line(-1,0){10}}
\put(316,28){\makebox(12,12){0}}
\put(350,12){\line(1,0){10}}
\end{picture}
\vspace{0.3cm}

\noindent Property Ga2 says that the 1-Ga is fault tolerant in the sense that errors resulting from a fault do not propagate badly and cause an encoded error. If the input to a 1-Ga with at most one fault is a codeword, then the postselected operation that arises from projecting the output to the code space reproduces the action of the ideal gate. Again, to ensure that the same actions by the adversary are accepted on both sides of this equation, we must augment the right-hand side with a noisy computation acting on a dummy input:

\vspace{0.2cm}
\begin{picture}(196,40)
\put(0,6){\makebox(50,12){dummy input}}
\put(59,28){\makebox(12,12){0}}
\put(71,34){\line(1,0){11}}
\put(60,12){\line(1,0){10}}
\put(70,0){\framebox(24,24){${\cal D}^{-1}$}}
\put(82,24){\line(0,1){10}}
\put(94,12){\line(1,0){10}}
\put(104,0){\framebox(48,24){\small\shortstack{good\\1-Ga}}}
\put(152,12){\line(1,0){10}}
\put(162,0){\framebox(24,24){${\cal D}$}}
\put(174,24){\line(0,1){10}}
\put(174,34){\line(1,0){11}}
\put(185,28){\makebox(12,12){0}}
\put(186,12){\line(1,0){10}}
\end{picture}
\vspace{0.3cm}

\medskip
\noindent {\em Property Ga3}:

\vspace{0.2cm}
\begin{picture}(360,40)
\put(0,12){\line(1,0){10}}
\put(10,0){\framebox(24,24){${\cal D}$}}
\put(22,24){\line(0,1){10}}
\put(22,34){\line(1,0){11}}
\put(33,28){\makebox(12,12){1}}
\put(45,34){\line(1,0){11}}
\put(34,12){\line(1,0){10}}
\put(44,0){\framebox(24,24){${\cal D}^{-1}$}}
\put(56,24){\line(0,1){10}}
\put(68,12){\line(1,0){10}}
\put(78,0){\framebox(48,24){\small\shortstack{very good\\1-Ga}}}
\put(126,12){\line(1,0){10}}
\put(136,0){\framebox(24,24){${\cal D}$}}
\put(148,24){\line(0,1){10}}
\put(148,34){\line(1,0){11}}
\put(159,28){\makebox(12,12){0}}
\put(171,34){\line(1,0){11}}
\put(160,12){\line(1,0){10}}
\put(170,0){\framebox(24,24){${\cal D}^{-1}$}}
\put(182,24){\line(0,1){10}}
\put(194,12){\line(1,0){10}}
\put(204,6){\makebox(20,12){=}}
\put(224,12){\line(1,0){10}}
\put(234,0){\framebox(24,24){${\cal D}$}}
\put(246,24){\line(0,1){10}}
\put(246,34){\line(1,0){10}}
\put(256,28){\makebox(12,12){0}}
\put(258,12){\line(1,0){10}}
\put(268,0){\framebox(48,24){\small\shortstack{ideal\\0-Ga}}}
\put(316,12){\line(1,0){10}}
\put(326,0){\framebox(24,24){${\cal D}^{-1}$}}
\put(338,24){\line(0,1){10}}
\put(338,34){\line(-1,0){10}}
\put(316,28){\makebox(12,12){0}}
\put(350,12){\line(1,0){10}}
\end{picture}
\vspace{0.3cm}

\noindent Property Ga3 says that a 1-Ga with no faults is fault tolerant in the sense that an error in an input block does not propagate badly and cause an encoded error in an output block. In the case of a multi-qubit gate, only one of the input blocks is 1-deviated on the left-hand side, while the other blocks have a trivial syndrome.

Our analysis of postselected circuits in this paper applies to fault-tolerant simulations using gadgets that satisfy the six properties listed above. (We also note that minor variants of these properties apply to the 1-preparation and 1-measurement gadgets.) In particular, using these properties we can prove the crucial fact that goodness implies correctness:

\medskip
\noindent {\bf Lemma 1. Good 1-exRecs are correct}. {\em Suppose that the level-1 error detection and gate gadgets obey properties ED1, ED2, ED3, Ga1, Ga2, Ga3. Then a 1-exRec containing no more than one fault (a good 1-exRec) is correct. That is, a good 1-exRec that simulates a single-qubit gate obeys A-correctness and B-correctness, and a good 1-exRec that simulates a two-qubit gate obeys AA-correctness, AB-correctness, and BB-correctness.} 
\medskip 

\noindent {\bf Proof}: We consider the case of a single-qubit gate (the proof for two-qubit gates is similar). First consider A correctness. Since there is at most one fault in the 1-exRec, there are three cases:
\begin{description}
\item[Case 1.] The leading 1-ED and the 1-Ga are both very good. Then property ED1 implies that the leading 1-ED simulates the identity and that the input to the 1-Ga is a codeword; property Ga1 implies that the output of the 1-Ga is a codeword and that the 1-Ga simulates the ideal 0-Ga.

\item[Case 2.] The leading 1-ED and the trailing 1-ED are both very good. Then property ED1 implies that the leading 1-ED simulates the identity and that the input and the output of the 1-Ga are both codewords; property Ga2 implies that the 1-Ga simulates the ideal 0-Ga.

\item[Case 3.] The 1-Ga and the trailing 1-ED are both very good. Then property ED1 implies that the output of the 1-Ga is a codeword, and property ED3 implies that the input to the 1-Ga is 1-deviated. Property Ga3 implies that the 1-Ga simulates the ideal 0-Ga and that the output of the 1-ED is actually a codeword; therefore the 1-ED simulates the identity by property ED2.

\end{description}

\noindent For B-correctness, again there are three cases, and in all three cases the proof of B-correctness is similar to the proof of A-correctness. In case 3 in particular, properties ED1, ED3, and Ga3 imply that the input and output of the 1-Ga are actually codewords, and that the 1-Ga simulates the ideal 0-Ga.

\rightline{$\square$}

By linking together the noisy circuits acting on dummy inputs appearing on the right-hand side of properties ED2 and Ga2, we obtain the circuit that augments the right-hand side of the A-correctness equation:

\vspace{0.2cm}
\begin{picture}(196,40)
\put(0,6){\makebox(50,12){dummy input}}
\put(59,28){\makebox(12,12){0}}
\put(71,34){\line(1,0){11}}
\put(60,12){\line(1,0){10}}
\put(70,0){\framebox(24,24){${\cal D}^{-1}$}}
\put(82,24){\line(0,1){10}}
\put(94,12){\line(1,0){10}}
\put(104,0){\framebox(48,24){\small\shortstack{1-ED}}}
\put(152,12){\line(1,0){10}}
\put(162,0){\framebox(24,24){${\cal D}$}}
\put(174,24){\line(0,1){10}}
\put(174,34){\line(1,0){11}}
\put(185,28){\makebox(12,12){0}}
\put(197,34){\line(1,0){11}}
\put(186,12){\line(1,0){10}}
\put(196,0){\framebox(24,24){${\cal D}^{-1}$}}
\put(208,24){\line(0,1){10}}
\put(220,12){\line(1,0){10}}
\put(230,0){\framebox(48,24){\small\shortstack{1-Ga}}}
\put(278,12){\line(1,0){10}}
\put(288,0){\framebox(24,24){${\cal D}$}}
\put(300,24){\line(0,1){10}}
\put(300,34){\line(1,0){11}}
\put(311,28){\makebox(12,12){0}}
\put(312,12){\line(1,0){10}}
\end{picture}
\vspace{0.3cm}

\noindent In fact, the projection onto the code space inserted between the 1-ED and the 1-Ga is superfluous and can be removed; when both 1-EDs in a good 1-exRec accept, then the output of the leading 1-ED is guaranteed to be a codeword, as our proof of Lemma 1 has shown. For B-correctness, the noisy computation on the right-hand side is just as in property Ga2.

Our explanation of why goodness implies correctness has been precise, but perhaps as a result it has also been pedantic. Actually, the idea is quite simple --- if our gadgets are fault tolerant and the 1-exRec contains only one fault, then either the 1-Rec is very good, or else any error caused by a fault in the 1-Rec will be detected by the very good trailing 1-ED. Therefore, after postselection, there really is no fault in the 1-Rec after all and the 1-Rec acts as though it were ideal.

Lemma 1 establishes, as desired, that good 1-exRecs are converted to the corresponding ideal 0-Ga's as the 1-encoders sweep forward through the circuit. Each 1-encoder originates at either a good 1-preparation or at the rear end of a bad cluster. For a good 1-exRec that is adjacent to the rear end of a cluster, we may use B-correctness for a one qubit gate, or one of AB-correctness or BB-correctness for a two-qubit gate, to justify moving the 1-encoder one step forward, and converting the 1-Rec to an ideal 0-Ga. For a good 1-exRec that immediately follows other good 1-exRecs, we can use A-correctness or AA-correctness to justify moving the 1-encoder forward. Under repeated application of Lemma 1, the 1-encoders march forward until they meet either the 1-decoders at the front end of another bad cluster or they meet a good 1-measurement. In either case the 1-encoders disappear. 

At this stage, the original noisy level-1 circuit has been partially reduced to an equivalent level-0 circuit. All of the good 1-exRecs have become ideal 0-Ga's. And the bad clusters are surrounded by 1-encoders in front (i.e., just before the bad cluster) and 1-decoders behind (i.e., just after the bad cluster). To proceed further we must study the level-0 circuit simulated by a bad cluster.

\section{Carving bad clusters}
\label{sec:carving}

\subsection{Reclassification of pre-bad 1-exRecs}

To complete the level reduction, we are to map each of the 1-exRecs in the bad cluster to a corresponding level-0 gate, obtaining a level-0 circuit that is equivalent to the noisy level-1 circuit. This mapping will be constructed by inserting decoder-encoder pairs at properly chosen positions inside the bad cluster. Specifically, a decoder-encoder pair ${\cal DD}^{-1}$ is inserted either immediately  before or immediately after each 1-ED in the bad cluster. For each 1-ED, whether ${\cal DD}^{-1}$ is inserted before or after the 1-ED depends on the fault path, and is determined by an algorithm that we will describe. We refer to the placement of the decoder-encoder pairs as the ``carving'' of the bad cluster.

Once the decoder-encoder pairs are placed, the bad cluster has been expressed as a circuit of level-0 operations, where each level-0 operation is a level-1 circuit preceded by an encoder (or two encoders in the case of a two-qubit gate) and followed by a decoder (or two decoders). The level-1 circuit is a subset of the corresponding 1-exRec; let us call it the carved 1-exRec. For example, for the case of a single-qubit gate, there are four possibilities for the carved 1-exRec, depending on the placement of the decoder-encoder pairs: (1) a 1-Ga, (2) a 1-Ga followed by a 1-ED, (3) a 1-Ga preceded by a 1-ED, and (4) a 1-Ga preceded and followed by a 1-ED. Thus each 1-exRec in the circuit is mapped to a corresponding level-0 gate: the carved 1-exRec preceded by encoder(s) and followed by decoder(s). This mapping almost completes the level reduction but not quite. In addition some faults in the resulting level-0 circuit are propagated forward to the following gate, by a procedure that we will explain.

We will say that a 1-exRec is good* if it is mapped by level reduction to the ideal level-0 gate. Thus, a good 1-exRec is good*, but in addition some of the pre-bad 1-exRecs in a bad cluster will be designated as good*. (Actually, we will distinguished two kinds of good* 1-exRecs in the bad cluster, which we will refer to as {\em good$'$} and {\em good$''$}; both kinds are mapped to ideal gates.) Otherwise the pre-bad 1-exRec is mapped to a level-0 fault, and we say that the 1-exRec is bad. Of course, whether each 1-exRec is designated good* or bad depends on the fault path.

The goal of our carving algorithm is to assure that if the fault rate is $\varepsilon$ in the original circuit, then the fault rate is $O(\varepsilon^2)$ in the reduced circuit. This may seem obvious at first, since two faults are needed to cause a 1-exRec to be incorrect. But it is not quite so obvious because the 1-exRecs overlap. For example, consider the two overlapping pre-bad 1-exRecs shown in Fig.~\ref{fig:bad-overlapping-exRecs}. Each 1-exRec contains two faults, but because one fault is shared, the probability of the fault-path shown is $O(\varepsilon^3)$. The level reduction procedure may map one of these 1-exRecs to a fault, but not both, since the probability of two faults in the reduced circuit should be $O(\varepsilon^4)$. On the other hand, the carved 1-exRecs do not overlap. Our carving algorithm should be designed so that, for any fault path, the carved 1-exRec contained in each bad 1-exRec contains at least two faults.

\begin{figure}
\begin{center}
\vspace{0.5cm}
\begin{picture}(250,48)
\put(2,22){\line(1,0){5}}
\put(7,4){\framebox(36,36){1-ED}}
\put(43,22){\line(1,0){14}}
\put(57,4){\framebox(36,36){1-Ga}}
\put(85,26){$\times$}
\put(93,22){\line(1,0){5}}
\put(102,22){\line(1,0){5}}
\put(107,4){\framebox(36,36){1-ED}}
\put(113,6){$\times$}
\put(143,22){\line(1,0){5}}
\put(152,22){\line(1,0){5}}
\put(157,4){\framebox(36,36){1-Ga}}
\put(193,22){\line(1,0){14}}
\put(207,4){\framebox(36,36){1-ED}}
\put(231,32){$\times$}
\put(243,22){\line(1,0){5}}
\put(0,0){\dashbox(150,44){}}
\put(100,-4){\dashbox(150,52){}}
\end{picture}
\vspace{0.3cm}
\end{center}
\fcaption{Two overlapping pre-bad 1-exRecs indicated by dashed lines, with fault locations indicated by $\times$. Because one of the three faults is contained in the shared 1-ED, the pre-bad 1-exRecs are not independent events. }
\label{fig:bad-overlapping-exRecs}
\end{figure}

\subsection{Carving rules}
\label{sec:good-prime}
The carving procedure consists of two steps. In the first step each pre-bad 1-exRec in the cluster is classified as bad, good$'$, or good$''$. In the second step, we determine for each pair of consecutive 1-exRecs whether the ${\cal DD}^{-1}$ is inserted just before or just after the 1-ED shared by the two 1-exRecs.

To classify the 1-exRecs, we start at the rear edge of the bad cluster --- with pre-bad 1-exRecs that are followed by only good 1-exRecs --- and work forward from there. The classification of each 1-exRec is determined by the classification of the 1-exRecs that immediately follow it, according to these rules:

\medskip
\noindent {\bf Rules for classification of pre-bad 1-exRecs}
\begin{description}
\item[$0.$] We say that a 1-exRec is good* if it is good, good$'$, or good$''$.
\item[$1.$] A pre-bad 1-exRec is bad if all of the following 1-exRecs are good*.
\item[$2.$] A pre-bad 1-exRec is good$'$ if all of the following 1-exRecs are bad.
\item[$3.$] If the 1-exRecs following a pre-bad 1-exRec are neither all bad nor all good*, then:
\begin{description} 
\item[$i.$] the pre-bad 1-exRec is good$''$ if the {\em truncated} 1-exRec contains no more than one fault.  
\item[$ii.$] the pre-bad 1-exRec is bad if the {\em truncated} 1-exRec contains more than one fault.  
\end{description}
\end{description}

\noindent In the statement of Rule 3, we say that a 1-exRec is ``truncated'' if all 1-EDs that it shares with following {\em bad} 1-exRecs have been amputated. When we count the faults in the pre-bad 1-exRec to determine whether it is bad, we should exclude from the count any faults that are shared with the bad 1-exRecs that follow, because these faults may have already been held responsible for the badness of the following 1-exRecs. However, faults shared with following good* 1-exRecs should be included in the count.

For the gadgets that we will analyze in detail, all quantum gates will be either single-qubit or two-qubit gates; therefore, Rule 3 will be applied to two-qubit gates that are followed by one bad 1-exRec and one good* 1-exRec. However, we have stated Rule 3 in a more general form to emphasize that it can also be applied to quantum gates that act on more than two qubits. We also remark that Rule 1 is interpreted so that  a pre-bad measurement 1-exRec (which is not followed by another 1-exRec) is always declared bad.

The next step is to decide on the placement of the decoder-encoder pairs. Here the rules are quite simple:

\medskip
\noindent {\bf Rules for placement of decoder-encoder pairs}
\begin{description}
\item[$1.$] If a bad 1-exRec is followed by a good* 1-exRec, then ${\cal DD}^{-1}$ is inserted immediately {\em after} the shared 1-ED.
\item[$2.$] For any other consecutive pair of 1-exRecs, ${\cal DD}^{-1}$ is inserted immediately {\em before} the shared 1-ED.
\end{description}

\noindent For a specified fault path, the classification rules and placement rules unambiguously define how a cluster of pre-bad 1-exRecs is divided into carved 1-exRecs. Now we must explain how each carved 1-exRec is mapped to a level-0 gate under level reduction.

The explanation is simplest in the case of a bad 1-exRec. If the bad 1-exRec is not followed by other bad 1-exRecs, then decoder-encoder pairs are placed in front of the leading 1-EDs and behind the trailing 1-EDs; the carved 1-exRec is the full 1-exRec and contains at least two faults. Then the bad 1-exRec, together with the 1-$^*$encoder(s) in front and the 1-$^*$decoder(s) in back, becomes a faulty level-0 gate, denoted ${\cal N}$; see Fig.~\ref{fig:bad}. The input to the 1-$^*$encoder may include a nontrivial syndrome; in that case the 1-exRec is transformed to a faulty 0-Ga that acts jointly on the level-0 data and the syndrome. Similarly, if the bad 1-exRec is followed by another bad 1-exRec, then the carved 1-exRec is the truncated 1-exRec, containing at least two faults; the truncated 1-exRec, together with the 1-$^*$encoder(s) in front and the 1-$^*$decoder(s) in back, becomes a faulty level-0 gate.

\begin{figure}
\begin{center}
\vspace{0.5cm}
\begin{picture}(352,34)
\put(0,12){\line(1,0){10}}
\put(10,0){\framebox(32,24){${\cal D}{\cal D}^{-1}$}}
\put(42,12){\line(1,0){10}}
\put(18,24){\line(0,1){10}}
\put(34,24){\line(0,1){10}}
\put(18,34){\line(1,0){16}}
\put(52,0){\framebox(32,24){\small 1-ED}}
\put(84,12){\line(1,0){10}}
\put(94,0){\framebox(32,24){\small 1-Ga}}
\put(126,12){\line(1,0){10}}
\put(136,0){\framebox(32,24){\small 1-ED}}
\put(168,12){\line(1,0){10}}
\put(178,0){\framebox(32,24){${\cal D}{\cal D}^{-1}$}}
\put(210,12){\line(1,0){10}}
\put(186,24){\line(0,1){10}}
\put(202,24){\line(0,1){10}}
\put(186,34){\line(1,0){16}}
\put(220,6){\makebox(20,12){=}}
%%%
%%%
\put(240,12){\line(1,0){10}}
\put(250,0){\framebox(24,24){${\cal D}$}}
\put(274,12){\line(1,0){10}}
\put(284,0){\framebox(24,24){${\cal N}$}}
\put(308,12){\line(1,0){10}}
\put(318,0){\framebox(24,24){${\cal D}^{-1}$}}
\put(342,12){\line(1,0){10}}
\put(262,24){\line(0,1){10}}
\put(292,24){\line(0,1){10}}
\put(262,34){\line(1,0){30}}
\put(300,24){\line(0,1){10}}
\put(330,24){\line(0,1){10}}
\put(300,34){\line(1,0){30}}
\end{picture}
\vspace{0.3cm}
\end{center}
\fcaption{Level reduction applied to a bad 1-exRec. Sandwiched between an ideal encoder in front and an ideal decoder behind, the bad 1-exRec becomes a faulty level-0 operation ${\cal N}$ that acts jointly on the data and the syndrome.}
\label{fig:bad}
\end{figure}

\subsection{Good$\, '$ 1-exRecs}

Consider a good$'$ 1-exRec that simulates a single-qubit gate. By definition (i.e., by classification Rule 2), the good$'$ 1-exRec is followed by a bad 1-exRec, and (by placement Rule 2) a decoder-encoder pair is inserted before the trailing 1-ED. If the good$'$ 1-exRec is preceded by a good* (i.e., good or good$''$) 1-exRec, then a decoder-encoder pair is inserted before the leading 1-ED, and if the 1-exRec is preceded by a bad 1-exRec, then a decoder-encoder pair is inserted after the leading 1-ED. Either way, the carved 1-exRec, preceded by an encoder and followed by a decoder, becomes a faulty 0-Ga that acts on the data and the input syndrome. But this faulty gate can be written as $U\circ {\cal N}$, where $U$ is the ideal 0-Ga and ${\cal N}$ is a noisy gate acting on data and syndrome. By absorbing ${\cal N}$ into the following bad location, we may associate the good$'$ 1-exRec with the ideal 0-Ga $U$. (Here, in deference to the standard convention in which circuits are drawn with input on the left and output on the right,  we use  $A\circ B$ to denote the composite operator obtained by applying first $A$ and then $B$.) Thus, level-reduction maps the good$'$ 1-exRec to the corresponding ideal gate. We note that the declaration that the pre-bad 1-exRec is good$'$ does not depend on the fault path in the 1-exRec --- all that matters is that the following 1-exRec is bad. 

Similarly, for a good$'$ two-qubit gate, {\em both} of the following 1-exRecs have been declared bad and associated with faulty 0-Ga's, and decoder-encoder pairs are placed before both trailing 1-EDs. Furthermore decoder-encoder pairs are placed before any leading 1-EDs that are shared with preceding good* 1-exRecs, and after any leading 1-EDs that are shared with preceding bad 1-exRecs.  Then the carved 1-exRec, together with its preceding encoders and following decoders, becomes the ideal gate followed by a faulty operation ${\cal N}$ that can be absorbed into the bad locations that follow; see Fig.~\ref{fig:good-prime}. Now, ${\cal N}$ might be an entangling operation acting collectively on the two output qubits. But recall that in the locally correlated stochastic noise model, local adversaries at two bad 0-locations that both immediately follow the same good 0-location are permitted to apply a collective operation to their shared data. Therefore, the collective faulty operation ${\cal N}$ is compatible with locally correlated stochastic noise. Indeed, we incorporated this feature into the noise model in order to accommodate the situation shown in Fig.~\ref{fig:good-prime}.

\begin{figure}
\begin{center}
\setlength{\unitlength}{0.85pt}
%\vspace{0.5cm}
\begin{picture}(434,76)
\put(0,6){\makebox(20,12){\shortstack{{\small good or} \\{\small good$''$}}}}
\put(0,38){\makebox(20,12){\small bad}}
\put(30,12){\line(1,0){10}}
\put(40,0){\framebox(32,24){\small ${\cal D}{\cal D}^{-1}$}}
\put(72,12){\line(1,0){10}}
\put(30,44){\line(1,0){10}}
\put(40,32){\framebox(32,24){\small 1-ED}}
\put(72,44){\line(1,0){10}}
\put(82,0){\framebox(32,24){\small 1-ED}}
\put(114,12){\line(1,0){10}}
\put(82,32){\framebox(32,24){\small ${\cal D}{\cal D}^{-1}$}}
\put(114,44){\line(1,0){10}}
\put(90,56){\line(0,1){10}}
\put(106,56){\line(0,1){10}}
\put(90,66){\line(1,0){16}}
\put(124,0){\framebox(32,56){1-Ga}}
\put(156,12){\line(1,0){10}}
\put(156,44){\line(1,0){10}}
\put(166,32){\framebox(32,24){\small ${\cal D}{\cal D}^{-1}$}}
\put(198,44){\line(1,0){10}}
\put(174,56){\line(0,1){10}}
\put(190,56){\line(0,1){10}}
\put(174,66){\line(1,0){16}}
\put(212,38){\makebox(20,12){\small bad}}
\put(166,0){\framebox(32,24){\small ${\cal D}{\cal D}^{-1}$}}
\put(198,12){\line(1,0){10}}
\put(174,0){\line(0,-1){10}}
\put(190,0){\line(0,-1){10}}
\put(174,-10){\line(1,0){16}}
\put(212,6){\makebox(20,12){\small bad}}
\put(232,22){\makebox(20,12){=}}
%%%
%%%
\put(250,44){\line(1,0){10}}
\put(262,32){\framebox(32,24){\small 1-ED}}
\put(294,42){\line(1,0){10}}
\put(294,12){\line(1,0){10}}
\put(304,0){\framebox(24,24){${\cal D}$}}
\put(328,12){\line(1,0){10}}
\put(304,32){\framebox(24,24){${\cal D}$}}
\put(328,42){\line(1,0){10}}
\put(338,0){\framebox(32,56){\shortstack{ideal\\$0$-Ga}}}
\put(370,12){\line(1,0){10}}
\put(370,44){\line(1,0){10}}
\put(380,0){\framebox(32,56){${\cal N}$}}
\put(412,12){\line(1,0){10}}
\put(412,44){\line(1,0){10}}
\put(422,32){\framebox(24,24){${\cal D}^{-1}$}}
\put(446,44){\line(1,0){10}}
\put(422,0){\framebox(24,24){${\cal D}^{-1}$}}
\put(446,12){\line(1,0){10}}
\put(316,56){\line(0,1){10}}
\put(388,56){\line(0,1){10}}
\put(316,66){\line(1,0){72}}
\put(316,56){\line(0,1){10}}
\put(388,56){\line(0,1){10}}
\put(316,66){\line(1,0){72}}
\put(404,56){\line(0,1){10}}
\put(434,56){\line(0,1){10}}
\put(404,66){\line(1,0){30}}
\put(404,0){\line(0,-1){10}}
\put(434,0){\line(0,-1){10}}
\put(404,-10){\line(1,0){30}}
\end{picture}
\vspace{0.5cm}
\end{center}
\fcaption{Level reduction applied to a good$'$ 1-exRec. A pre-bad 1-exRec followed by two bad 1-exRecs is mapped to an ideal gate followed by a noisy operation ${\cal N}$ that can be absorbed into the following faulty gates. Decoder-encoder pairs are inserted before a leading 1-ED that is shared with a preceding good* 1-exRec, and after a leading 1-ED that is shared with a preceding bad 1-exRec. }
\label{fig:good-prime}
\end{figure}

Note that a pre-bad single-qubit 1-exRec that precedes a good$'$ 1-exRec (or a pre-bad two-qubit 1-exRec that precedes two good$'$ 1-exRecs) is declared bad.  We might have argued that this 1-exRec faithfully simulates the ideal 0-Ga, simply by propagating any faults through the following good$'$ location to the  bad location(s) further downstream. However, that prescription would not be compatible with the requirement that the locally correlated stochastic noise model is preserved by level reduction. 

\subsection{Good$\,''$ 1-exRecs and weak correctness}

Now we wish to argue that a good$''$ 1-exRec is mapped by level reduction to the corresponding ideal 0-Ga. For this purpose we observe that the good$''$ truncated 1-exRec has a property that we will call {\em weak correctness}: the carved 1-exRec, together with the encoders in front and the decoders in back can be expressed as $U\circ ({\cal N}\otimes I)$, where $U$ is the ideal two-qubit 0-Ga and ${\cal N}$ is a noisy operation that acts on the syndrome and on the one output qubit that enters the following bad 1-exRec. We can absorb ${\cal N}$ into the subsequent fault, thereby associating the  carved good$''$ 1-exRec with the ideal level-0 gate. Of course, the nontrivial content of weak correctness is that the other output qubit, the one that enters the subsequent good* location, agrees with the output that would be produced by the ideal 0-Ga. Fig.~\ref{fig:weak-correct2} illustrates the case where one of the  preceding 1-exRecs is bad (weak AB-correctness).

\begin{figure}
\begin{center}
\setlength{\unitlength}{0.7pt}
\vspace{0.5cm}
\begin{picture}(514,66)
\put(0,6){\makebox(20,12){\tiny {\shortstack{good or\\good$''$}}}}
\put(0,38){\makebox(20,12){\tiny bad}}

\put(30,12){\line(1,0){10}}
\put(40,0){\framebox(32,24){\tiny ${\cal D}{\cal D}^{-1}$}}
\put(72,12){\line(1,0){10}}
\put(30,44){\line(1,0){10}}
\put(40,32){\framebox(32,24){\tiny 1-ED}}
\put(72,42){\line(1,0){10}}
\put(82,0){\framebox(32,24){\tiny 1-ED}}
\put(114,12){\line(1,0){10}}
\put(82,32){\framebox(32,24){\tiny ${\cal D}{\cal D}^{-1}$}}
\put(114,42){\line(1,0){10}}
\put(90,56){\line(0,1){10}}
\put(106,56){\line(0,1){10}}
\put(90,66){\line(1,0){16}}
\put(124,0){\framebox(32,56){\tiny 1-Ga}}
\put(156,12){\line(1,0){10}}
\put(156,44){\line(1,0){10}}
\put(166,0){\framebox(32,24){\tiny ${\cal D}{\cal D}^{-1}$}}
\put(198,12){\line(1,0){10}}
\put(166,32){\framebox(32,24){\tiny ${\cal D}{\cal D}^{-1}$}}
\put(198,44){\line(1,0){10}}
\put(176,56){\line(0,1){10}}
\put(192,56){\line(0,1){10}}
\put(176,66){\line(1,0){16}}
\put(208,38){\makebox(20,12){\tiny bad}}
\put(208,0){\framebox(32,24){\tiny 1-ED}}
\put(240,12){\line(1,0){10}}
\put(264,6){\makebox(20,12){\tiny {\shortstack{good, \\good$'$ or\\good$''$}}}}
%
%\put(270,22){\makebox(20,12){=}}
\put(290,22){\makebox(20,12){=}}
%%%
%%%
\put(310,44){\line(1,0){10}}
\put(320,32){\framebox(32,24){\tiny 1-ED}}
\put(352,42){\line(1,0){10}}

\put(352,12){\line(1,0){10}}
\put(362,0){\framebox(24,24){\tiny ${\cal D}$}}
\put(386,12){\line(1,0){10}}
\put(362,32){\framebox(24,24){\tiny ${\cal D}$}}
\put(386,42){\line(1,0){10}}
\put(374,56){\line(0,1){10}}
\put(396,0){\framebox(32,56){\shortstack{\tiny ideal\\\tiny $0$-Ga}}}
\put(428,12){\line(1,0){10}}
\put(428,44){\line(1,0){10}}
\put(438,0){\framebox(24,24){\tiny ${\cal D}^{-1}$}}
\put(462,12){\line(1,0){10}}
\put(472,0){\framebox(32,24){\tiny 1-ED}}
\put(504,12){\line(1,0){10}}
\put(438,32){\framebox(32,24){\tiny ${\cal N}$}}
\put(470,44){\line(1,0){10}}
\put(446,56){\line(0,1){10}}
\put(462,56){\line(0,1){10}}
\put(446,66){\line(-1,0){72}}
\put(462,66){\line(1,0){30}}
\put(480,32){\framebox(24,24){\tiny ${\cal D}^{-1}$}}
\put(504,44){\line(1,0){10}}
\put(492,56){\line(0,1){10}}
\end{picture}
\vspace{0.3cm}
\end{center}
\fcaption{Weak correctness, in the case where one of the preceding 1-exRecs is bad and the other is good* (i.e., weak AB-correctness). If a 1-exRec is followed by one bad 1-exRec and one good* 1-exRec, and if the truncated 1-exRec has no more than one fault, then it satisfies the identity shown (after postselection). Decoder-encoder pairs are inserted after a leading 1-ED that is shared with a preceding bad 1-exRec, and before a leading 1-ED that is shared with a preceding good* 1-exRec. Weak correctness means that the carved 1-exRec, combined with the preceding encoders and following decoders, becomes the ideal level-0 gate followed by a noisy operation that acts on the output qubit that enters the following bad 1-exRec. Syndrome information can propagate to the following bad 1-exRec, but not to the following good* 1-exRec.}
\label{fig:weak-correct2}
\end{figure}

Weak correctness follows from some further properties of our fault-tolerant gate gadgets which are weaker versions of properties Ga2 and Ga3. In place of Ga2 we have 

\medskip
\noindent {\em Property wGa2}:

\vspace{0.2cm}
\begin{picture}(382,84)
\put(0,22){\line(1,0){10}}
\put(10,10){\framebox(24,24){${\cal D}$}}
\put(22,10){\line(0,-1){10}}
\put(22,0){\line(1,0){11}}
\put(33,-6){\makebox(12,12){0}}
\put(45,0){\line(1,0){11}}
\put(34,22){\line(1,0){10}}
\put(0,56){\line(1,0){10}}
\put(10,44){\framebox(24,24){${\cal D}$}}
\put(22,68){\line(0,1){10}}
\put(22,78){\line(1,0){11}}
\put(33,72){\makebox(12,12){0}}
\put(45,78){\line(1,0){11}}
\put(34,54){\line(1,0){10}}
\put(44,10){\framebox(24,24){${\cal D}^{-1}$}}
\put(56,10){\line(0,-1){10}}
\put(68,22){\line(1,0){10}}
\put(44,44){\framebox(24,24){${\cal D}^{-1}$}}
\put(56,68){\line(0,1){10}}
\put(68,54){\line(1,0){10}}
\put(78,10){\framebox(36,60){\small\shortstack{good\\1-Ga}}}
\put(114,22){\line(1,0){10}}
\put(124,10){\framebox(24,24){${\cal D}$}}
\put(136,10){\line(0,-1){10}}
\put(136,0){\line(1,0){11}}
\put(147,-6){\makebox(12,12){0}}
\put(159,0){\line(1,0){11}}
\put(148,22){\line(1,0){10}}
\put(114,56){\line(1,0){10}}
\put(124,44){\framebox(24,24){${\cal D}$}}
\put(136,68){\line(0,1){10}}
\put(136,78){\line(1,0){34}}
\put(148,54){\line(1,0){10}}
\put(158,10){\framebox(24,24){${\cal D}^{-1}$}}
\put(170,10){\line(0,-1){10}}
\put(182,22){\line(1,0){10}}
\put(158,44){\framebox(24,24){${\cal D}^{-1}$}}
\put(170,68){\line(0,1){10}}
\put(182,54){\line(1,0){10}}
\put(198,34){\makebox(20,12){=}}
\put(224,22){\line(1,0){10}}
\put(234,10){\framebox(24,24){${\cal D}$}}
\put(246,10){\line(0,-1){10}}
\put(246,0){\line(1,0){11}}
\put(257,-6){\makebox(12,12){0}}
\put(258,22){\line(1,0){10}}
\put(224,56){\line(1,0){10}}
\put(234,44){\framebox(24,24){${\cal D}$}}
\put(246,68){\line(0,1){10}}
\put(246,78){\line(1,0){11}}
\put(257,72){\makebox(12,12){0}}
\put(258,56){\line(1,0){10}}
\put(268,10){\framebox(36,60){\small\shortstack{ideal\\0-Ga}}}
\put(304,22){\line(1,0){10}}
\put(304,56){\line(1,0){10}}
\put(314,10){\framebox(24,24){${\cal D}^{-1}$}}
\put(326,10){\line(0,-1){10}}
\put(303,-6){\makebox(12,12){0}}
\put(315,0){\line(1,0){11}}
\put(338,22){\line(1,0){10}}
\put(314,44){\framebox(24,24){${\cal N}$}}
\put(326,68){\line(0,1){10}}
\put(338,56){\line(1,0){10}}
\put(348,44){\framebox(24,24){${\cal D}^{-1}$}}
\put(360,68){\line(0,1){10}}
\put(372,56){\line(1,0){10}}
\put(326,78){\line(1,0){34}}
\end{picture}
\vspace{0.3cm}

\noindent Property wGa2 says that if there is a single fault in the 1-Ga, and the second output block is projected to a codeword, then the second output block has no encoded error. Here the action of the adversary determines the faulty operation that acts on the first output on the right side of the equation. Another way to express property wGa2 would be to say that if the inverse $U^{-1}$ of the ideal 0-Ga is inserted on the left-hand side after the input is decoded, then (after postselection) on the right-hand side the identity operation is applied to the second input codeword. 

The weakened version of property Ga3 is

\medskip
\noindent {\em Property wGa3}:

\vspace{0.2cm}
\begin{picture}(382,84)
\put(0,22){\line(1,0){10}}
\put(10,10){\framebox(24,24){${\cal D}$}}
\put(22,10){\line(0,-1){10}}
\put(22,0){\line(1,0){11}}
\put(33,-6){\makebox(12,12){0}}
\put(45,0){\line(1,0){11}}
\put(34,22){\line(1,0){10}}
\put(0,56){\line(1,0){10}}
\put(10,44){\framebox(24,24){${\cal D}$}}
\put(22,68){\line(0,1){10}}
\put(22,78){\line(1,0){11}}
\put(33,72){\makebox(12,12){1}}
\put(45,78){\line(1,0){11}}
\put(34,54){\line(1,0){10}}
\put(44,10){\framebox(24,24){${\cal D}^{-1}$}}
\put(56,10){\line(0,-1){10}}
\put(68,22){\line(1,0){10}}
\put(44,44){\framebox(24,24){${\cal D}^{-1}$}}
\put(56,68){\line(0,1){10}}
\put(68,54){\line(1,0){10}}
\put(78,10){\framebox(36,60){\small\shortstack{very\\good\\1-Ga}}}
\put(114,22){\line(1,0){10}}
\put(124,10){\framebox(24,24){${\cal D}$}}
\put(136,10){\line(0,-1){10}}
\put(136,0){\line(1,0){11}}
\put(147,-6){\makebox(12,12){0}}
\put(159,0){\line(1,0){11}}
\put(148,22){\line(1,0){10}}
\put(114,56){\line(1,0){10}}
\put(124,44){\framebox(24,24){${\cal D}$}}
\put(136,68){\line(0,1){10}}
\put(136,78){\line(1,0){34}}
\put(148,54){\line(1,0){10}}
\put(158,10){\framebox(24,24){${\cal D}^{-1}$}}
\put(170,10){\line(0,-1){10}}
\put(182,22){\line(1,0){10}}
\put(158,44){\framebox(24,24){${\cal D}^{-1}$}}
\put(170,68){\line(0,1){10}}
\put(182,54){\line(1,0){10}}
\put(198,34){\makebox(20,12){=}}
\put(224,22){\line(1,0){10}}
\put(234,10){\framebox(24,24){${\cal D}$}}
\put(246,10){\line(0,-1){10}}
\put(246,0){\line(1,0){11}}
\put(257,-6){\makebox(12,12){0}}
\put(258,22){\line(1,0){10}}
\put(224,56){\line(1,0){10}}
\put(234,44){\framebox(24,24){${\cal D}$}}
\put(246,68){\line(0,1){10}}
\put(246,78){\line(1,0){32}}
\put(278,72){\makebox(12,12){1}}
\put(290,78){\line(1,0){32}}
\put(258,56){\line(1,0){10}}
\put(268,10){\framebox(36,60){\small\shortstack{ideal\\0-Ga}}}
\put(304,22){\line(1,0){10}}
\put(304,56){\line(1,0){10}}
\put(314,10){\framebox(24,24){${\cal D}^{-1}$}}
\put(326,10){\line(0,-1){10}}
\put(303,-6){\makebox(12,12){0}}
\put(315,0){\line(1,0){11}}
\put(338,22){\line(1,0){10}}
\put(314,44){\framebox(24,24){${\cal N}$}}
\put(322,68){\line(0,1){10}}
\put(330,68){\line(0,1){10}}
\put(338,56){\line(1,0){10}}
\put(348,44){\framebox(24,24){${\cal D}^{-1}$}}
\put(360,68){\line(0,1){10}}
\put(372,56){\line(1,0){10}}
\put(330,78){\line(1,0){30}}
\end{picture}
\vspace{0.3cm}

\noindent Property wGa3 says that, for a 1-Ga containing no faults, if the second output block is projected to the code space, then one error in the first input block does not propagate badly through the 1-Ga and cause an encoded error in the second output block. Similarly (though this is not illustrated by the diagram), one error in the second input block does not cause an encoded error in the second output block.

We note that, for a distance-two code, a nontrivial syndrome does not point to a unique weight-one error; therefore, the decomposition of the input state into codeword and syndrome depends on the particular conventions we have used in defining the 1-$^*$decoder. Property wGa3 holds no matter how these conventions are chosen. Consider, for example, the transversal {\sc cnot} gate for the [[4,2,2]] code (see Sec.~\ref{sec:gadgets}). If the input control block deviates from the code space by one $X$ error, that error will propagate to the target block and be detected. Thus, if the output target block is projected to the code space, we may infer that after postselection the input block does not have an $X$ error after all; only a $Z$ error is possible. That single $Z$ error could cause a logical $Z_L$ error in the input control block, but the {\sc cnot} gate does not propagate this logical $Z_L$ to the target block. On the other hand, one error in the input target block is sure to be detected in the output target block. 

Using these properties we can prove:

\medskip
\noindent {\bf Lemma 2. Good$''$ truncated 1-exRecs are weakly correct}. {\em Suppose that the level-1 error detection and gate gadgets obey properties ED1, ED2, ED3, Ga1, wGa2, wGa3. Then a truncated two-qubit 1-exRec containing no more than one fault (a good$''$ truncated 1-exRec) is weakly correct. That is, a good$''$ truncated 1-exRec obeys weak AA-correctness, weak AB-correctness, and weak BB-correctness.} 
\medskip 

\noindent {\bf Proof}: We consider weak AB-correctness as illustrated in Fig.~\ref{fig:weak-correct2} (the proofs of weak AA-correctness and weak BB-correctness are similar). Since there is at most one fault in the truncated 1-exRec, there are four cases:
\begin{description}
\item[Case 1.] The leading 1-EDs and the 1-Ga are very good. Then property ED1 implies that both inputs to the 1-Ga are codewords, and that the leading 1-ED acting on the second input block simulates the identity. Property Ga1 implies that both output blocks of the 1-Ga are in the code space and that the 1-Ga simulates the ideal 0-Ga.

\item[Case 2.] The leading 1-EDs and the untruncated trailing 1-ED are very good. Then property ED1 implies that the input blocks and the second output block of the 1-Ga are codewords, and that the leading 1-ED acting on the second input block simulates the identity. Property wGa2 implies that there is no encoded error in the second output block.

\item[Case 3.] The 1-Ga, the untruncated trailing 1-ED, and the 1-ED acting on the second input block are very good.  Then property ED1 implies that the second input block and the second output block of the 1-Ga are in the code space, and that the 1-ED acting on the second input block simulates the identity; furthermore, property ED3 implies that first input block to the 1-Ga is 1-deviated. Property wGa3 implies that there is no encoded error in the second output block.

\item[Case 4.] The 1-Ga, the untruncated trailing 1-ED, and the 1-ED acting on the first input block are very good.  Then property ED1 implies that the first input block and the second output block of the 1-Ga are in the code space;  furthermore, property ED3 implies that the second input block to the 1-Ga is 1-deviated. Property wGa3 implies that there is no encoded error in the second output block. 

However, in contrast to Case 3, the 1-ED acting on the second input
block does not necessarily simulate the identity, at least for the
standard convention for decomposing a 1-deviated state into an encoded
qubit and a nontrivial syndrome --- the state of the input qubit to the
1-encoder that precedes the 1-ED may differ from the state of the output
qubit from the 1-*decoder that follows the 1-ED.  However, we may choose
a different syndrome decoding convention for this particular 1-*decoder,
allowing the decoded qubit to be in the same state as the input qubit.  Since
property wGa3 holds irrespective of the syndrome decoding convention,
there is still no encoded error in the second output block.  The choice
of a different 1-*decoder may cause an additional error when it meets
the standard 1-*encoder in the first output block, but that error can be
absorbed into the noisy operation ${\cal N}$.

\end{description}

\rightline{$\square$}

The level reduction of a good$''$ 1-exRec is carried out by inserting $U\circ U^{-1}$ immediately before the 1-$^*$encoders that precede the 1-exRec, where $U$ is the ideal 0-Ga. Then we group $U^{-1}$ with the carved good$''$ 1-exRec, obtaining, according to Lemma 2, a noisy level-0 operation that acts only on the input to the following bad 1-exRec. This noisy operation is absorbed into the level-0 fault that results from applying level reduction to the bad 1-exRec;  thus the good$''$ 1-exRec has been replaced by the ideal 0-Ga $U$.

Therefore we have shown, as desired, that our level reduction procedure maps a two-qubit good$''$ 1-exRec to the corresponding ideal gate. Note that to reach this conclusion we assumed that the error detection steps in the bad cluster detect no errors, as postselection is a key ingredient in the proof of weak correctness. It was also necessary to propagate faults forward from the level-0 gates associated with carved good$''$ 1-exRecs to the faulty gates that immediately follow.

\subsection{Probability of bad 1-exRecs}
\label{sec:prob-bad-exRec}

To summarize, we have now articulated a procedure, depending on the fault path, for dividing the bad cluster into nonoverlapping carved 1-exRecs, and designating each pre-bad 1-exRec in the cluster as bad, good$'$, or good$''$. Our level reduction algorithm maps the bad cluster, together with the encoders that immediately precede it and the decoders that immediately follow it, to an equivalent circuit of level-0 gates, where each good$'$ or good$''$ 1-exRec is mapped to an ideal 0-Ga, and each bad 1-exRec is mapped to a faulty 1-Ga. Furthermore, each bad carved 1-exRec contains at least two faults. 

Suppose that we specify $r$ 1-exRecs in a level-1 simulation and we consider the set of all fault paths such that these $r$ specified 1-exRecs are bad. In every one of these fault paths, each of the $r$ bad (possibly truncated) 1-exRecs contains at least two faults. Therefore, for each fault path, we may identify a pair of level-0 locations in the 1-exRec that are faulty; the pair may be chosen in no more than $A$ ways, where $A$ is the number of pairs of locations contained in the largest 1-exRec used in the simulation. Since the truncated bad 1-exRecs do not overlap, the pairs are disjoint. Furthermore, once a pair of 0-locations is chosen in each of the $r$ 1-exRecs, the probability of the sum of all fault paths that have faults at the $2r$ specified 0-locations is not more than $\varepsilon^{2r}$, in the local stochastic noise model. Our conclusion is:

\medskip
\noindent {\bf Lemma 3. Probability of bad 1-exRecs}. {\em Suppose that an ideal quantum circuit is simulated using level-1 fault-tolerant error-detecting gadgets, satisfying the properties specified in Lemmas 1 and 2, subject to local stochastic noise with strength $\varepsilon$. Let us say that a 1-exRec is bad if and only if it is mapped by level reduction to a faulty gate (under the assumption that all 1-EDs in the 1-exRec accept). Consider a specified set ${\cal I}_r$ of $r$ (possibly overlapping) 1-exRecs, and consider the union of all fault paths such that all of the 1-exRecs in the set ${\cal I}_r$ are bad.  The probability ${\rm Prob}({\cal I}_r{\rm ~ bad})$ of this set of fault paths obeys
\begin{equation}
\label{eye-are-bad}
{\rm Prob}({\cal I}_r{\rm ~ bad}) \le \left(A\varepsilon^2\right)^r~,
\end{equation}
where $A$ is the number of pairs of locations contained in the largest 1-exRec used in the simulation. }
\medskip

Note that in the statement of Lemma 3 we assume the local stochastic noise model. It will be necessary to specialize to locally correlated stochastic noise only when we consider the probability of badness conditioned on {\em global} acceptance by the circuit in Sec.~\ref{sec:local-to-global}. The content of the lemma is just that with our rules for defining badness, a pair of fault locations can be associated with each bad 1-exRec, that the pairs are disjoint, and that each pair can be chosen in no more than A ways. 

However, since we will adopt the locally correlated stochastic noise model for our arguments in Sec.~\ref{sec:local-to-global}, it will be important to examine whether this noise model applies to the reduced circuit, so that the analysis of the level reduction can be repeated in a self-similar manner. Let us therefore observe that a good$''$ 1-exRec does not allow syndrome information from earlier 1-exRecs to propagate to the good, good$'$ or good$''$ 1-exRec that follows --- in Fig.~\ref{fig:weak-correct2} the decoder acting on the second output block finds a trivial syndrome. On the other hand, syndrome information can propagate through a good$''$ 1-exRec to reach the following bad 1-exRec. A good$'$ 1-exRec might propagate syndrome information forward, but it is always followed by bad 1-exRecs. Thus we see that communication inside a bad cluster has the property assumed in the locally correlated noise model: after level reduction, no syndrome information can propagate from one good 0-Ga to another good 0-Ga that immediately follows. 

We note that, because our carving procedure starts at the rear edge of a bad cluster and works forward from there, whether a given pre-bad 1-exRec is classified as bad or good* can depend on what happens in the future of the 1-exRec. For example, consider the case of a single-qubit 1-exRec with one fault in its 1-Ga and one fault in its trailing 1-ED.  This 1-exRec will be good$'$ if the following 1-exRec is declared bad (because the following 1-exRec contains additional faults), but the 1-exRec will be bad if the following 1-exRec is declared good (because the following 1-exRec contains only the one fault in its leading 1-ED). Furthermore, when a bad 1-exRec is mapped to a level-0 fault under level reduction, the type of fault can depend on what happens at subsequent locations. After level reduction, then, the local adversaries are empowered to pass messages both forward and backward in time. Thus we have incorporated message passing in both directions in our formulation of the locally correlated stochastic noise model, to ensure its stability under level reduction. Messages may be passed backward from a bad 0-Ga to a preceding bad or good 0-Ga, but not from a good 0-Ga to a preceding good 0-Ga.

\section{From local acceptance to global acceptance}
\label{sec:local-to-global}
Our next goal is to obtain an upper bound on the probability that $r$ specified 1-exRecs in the circuit are bad, conditioned on {\em global} acceptance by every 1-ED in the circuit. To obtain this bound we must exploit a crucial feature of the locally correlated stochastic noise model --- that if a 1-ED is very good (contains no faults), then no information about previous faults can propagate through the 1-ED and reach later 0-locations in the circuit.

To be concrete, consider the case $r=1$, and contemplate the set of all fault paths such that one specified 1-exRec is bad. This set of fault paths can be sorted into subsets, with each subset corresponding to a particular ``minimal sealed cluster'' that contains the specified 1-exRec. Here a ``cluster'' means a connected subgraph of the quantum circuit, comprised of 1-Ga's and 1-EDs, whose boundary contains only 1-EDs. We say that the cluster is ``sealed'' if every one of these bounding 1-EDs is very good. The sealed cluster is ``minimal'' if there is no smaller sealed cluster contained within it. We will call the set of very good 1-EDs at the beginning of the sealed cluster the ``leading edge'' of the cluster, and the set of very good 1-EDs at the end of the sealed cluster the ``trailing edge.'' The sealed cluster excluding its leading and trailing edges is the ``interior'' of the sealed cluster. For a simple example, see Fig.~\ref{fig:sealed}.

\begin{figure}
\begin{center}
\vspace{0.5cm}
\begin{picture}(346,48)
\put(2,22){\line(1,0){5}}
\put(7,4){\framebox(36,36){1-ED}}
\put(43,22){\line(1,0){14}}
\put(57,4){\framebox(36,36){1-Ga}}
\put(93,22){\line(1,0){5}}
\put(102,22){\line(1,0){5}}
\put(107,4){\framebox(36,36){1-ED}}
\put(113,6){$\times$}
\put(143,22){\line(1,0){10}}
\put(153,4){\framebox(36,36){1-Ga}}
\put(189,22){\line(1,0){14}}
\put(203,4){\framebox(36,36){1-ED}}
\put(227,32){$\times$}
\put(239,22){\line(1,0){5}}
\put(248,22){\line(1,0){5}}
\put(253,4){\framebox(36,36){1-Ga}}
\put(255,10){$\times$}
\put(289,22){\line(1,0){14}}
\put(303,4){\framebox(36,36){1-ED}}
\put(339,22){\line(1,0){5}}
\put(0,-4){\dashbox(346,52){}}
\put(100,0){\dashbox(146,44){}}
\end{picture}
\vspace{0.3cm}
\end{center}
\fcaption{A bad 1-exRec (inner box) and the sealed cluster (outer box) that contains it.  Fault locations are indicated by $\times$. Note that the leading and trailing 1-EDs of the sealed cluster are very good (contain no faults).}
\label{fig:sealed}
\end{figure}

The very good 1-EDs at the edge isolate the sealed cluster from the surrounding gates; that is, because the output of a very good 1-ED is always a codeword, the error syndrome carries no information about previous faults through the leading edge of the sealed cluster to its interior, and no information about faults in its interior through the trailing edge to subsequent gates. Furthermore, the messages passed by the level-0 local adversaries are blocked by a very good 1-ED (if, for example, the 1-ED has depth at least two). In the locally correlated stochastic noise model, messages can be passed from one level-0 gate to an immediately following or preceding gate, but only if at least one of the two gates is bad. Conceivably, then, a message could pass from a final good gate in a very good 1-ED that is in the leading edge of the sealed cluster to a bad gate immediately following in the interior, or from the final good gate inside a very good 1-ED that is contained in the trailing edge of the sealed cluster to a bad gate immediately following that is outside the sealed cluster. But in both cases, the message does not carry any information about the fault path inside the sealed cluster to the region outside, or any information about the fault path outside the sealed cluster to the inside. 

The concept of a minimal sealed cluster may also be applied to the case where $r>1$ specified 1-exRecs are bad, except that now the sealed cluster might have more than one connected component, where each component contains at least one of the specified 1-exRecs. With the sealed cluster and the fault path inside the sealed cluster fixed, consider the set of fault paths outside the sealed cluster such that all 1-EDs outside the sealed cluster and all 1-EDs in its leading edge accept (detect no errors). Because in the locally correlated stochastic noise model, the local adversaries in the interior of the sealed cluster know nothing about the fault path outside the sealed cluster, and the local adversaries outside the sealed cluster know nothing about the fault path inside, this set of fault paths outside does not depend at all on the fault path inside. That is, with the sealed cluster fixed, the inside and outside fault paths are uncorrelated. Indeed, the set of outside fault paths such that all 1-EDs outside and all 1-EDs in the leading edge of the sealed cluster accept would be exactly the same even if we replaced each fault in the interior of the sealed cluster by an ideal gate, that is, if the sealed cluster were replaced by the corresponding ``very good'' cluster.

Now consider the set of all fault paths in a level-1 circuit such that all 1-exRecs in a specified set ${\cal I}_r$ of $r$ 1-exRecs are bad, and such that the minimal sealed cluster containing these $r$ 1-exRecs is a particular fixed cluster ${\cal K}$. We may use the ``impermeability of the boundary'' of the sealed cluster to obtain a useful upper bound on the probability of this set of fault paths, {\em conditioned on global acceptance by all 1-EDs in the circuit}. The following identities hold:
\begin{eqnarray}
\label{conditional-bounds}
&&\frac{{\rm Prob}({\cal I}_r ~{\rm bad} ~\wedge~ {\cal K}~{\rm sealed~| ~ global ~acceptance})}{{\rm Prob}({\cal K}~{\rm  very~good}~|~ {\rm global~acceptance})}\nonumber\\
&=& \frac{{\rm Prob}({\cal I}_r ~{\rm bad} ~\wedge~{\cal K}~{\rm sealed~\wedge ~ global ~acceptance})}{{\rm Prob}({\cal K}~ {\rm very~good}~\wedge~ {\rm global~acceptance})}\nonumber\\
&=& \frac{{\rm Prob}({\cal I}_r ~{\rm bad} ~\wedge~ {\cal K}~{\rm sealed~\wedge ~ local ~acceptance})}{{\rm Prob}({\cal K}~{\rm very~good}~\wedge~ {\rm local~acceptance})}~.
\end{eqnarray}
Here ``${\cal I}_r ~{\rm bad} ~\wedge~ {\cal K}~{\rm sealed}$'' refers to the union of all fault paths such that the 1-exRecs in ${\cal I}_r$ are bad and ${\cal K}$ is the minimal sealed cluster containing ${\cal I}_r$, while ``${\cal K}$ very good'' refers to the sum over fault paths such that ${\cal K}$ contains no fault at all. The first equality follows from Bayes's rule, and the second is true because of the isolation of the sealed cluster --- the probability of acceptance for 1-EDs outside the sealed cluster does not depend on the fault path inside the sealed cluster. Actually ``local acceptance'' means acceptance by every 1-ED contained in the interior of the sealed cluster and also by every 1-ED in its trailing edge. ``Global acceptance'' includes in addition acceptance by every 1-ED that is outside the sealed cluster and also by every 1-ED in its leading edge.

It is easy to estimate the denominator: if there are no faults at all in the cluster ${\cal K}$, then local acceptance is assured; therefore, 
\begin{equation}
{\rm Prob}({\cal K}~{\rm very~good}~\wedge~ {\rm local~acceptance}) \ge (1- \varepsilon)^{|{\cal K}|}~, 
\end{equation}
where $|{\cal K}|$ is the number of 0-locations in the {\em interior} of the sealed cluster. (When the sealed cluster ${\cal K}$ is specified, we are already requiring the fault path to be very good at the edge of ${\cal K}$; to ensure local acceptance and that the sealed cluster is very good, it suffices to demand that there are no faults in the interior. Therefore, our estimate of ${\rm Prob}({\cal K}~{\rm very~good}~\wedge~ {\rm local~acceptance})$ includes a factor of $(1-\varepsilon)$ for each 0-location in the interior.)

Now let ${\rm Prob}({\cal K} ~|~{\cal I}_r)$ denote the probability that ${\cal K}$ is the minimal sealed cluster containing ${\cal I}_r$, and consider summing over ${\cal K}$ with ${\cal I}_r$ fixed. We observe that
\begin{eqnarray}
{\rm Prob}({\cal I}_r ~{\rm bad} ~\wedge~ {\cal K}~{\rm sealed~\wedge ~ local ~acceptance})
&\le& {\rm Prob}({\cal I}_r ~{\rm bad} ~\wedge~ {\cal K}~{\rm sealed})\nonumber\\
&=&{\rm Prob}({\cal I}_r ~{\rm bad})\cdot {\rm Prob}({\cal K} ~|~{\cal I}_r)~,
\end{eqnarray}
and we recall that ${\rm Prob}({\cal I}_r ~{\rm bad})$ can be bounded using Lemma 3. Since 
\begin{equation} 
{{\rm Prob}({\cal K}~{\rm very~good}~|~ {\rm global~acceptance})}\le 1~,
\end{equation} eq.~(\ref{conditional-bounds}) implies:

\medskip
\noindent {\bf Lemma 4. Probability of a bad cluster conditioned on global acceptance}. {\em Suppose that an ideal quantum circuit is simulated using level-1 fault-tolerant error-detecting gadgets, satisfying the properties specified in Lemmas 1 and 2, subject to locally correlated stochastic noise with strength $\varepsilon$. Consider the union of all fault paths such that a specified set ${\cal I}_r$ of $r$ 1-exRecs in the level-1 circuit are bad. The probability ${\rm Prob}({\cal I}_r~{\rm bad ~| ~ global ~acceptance})$ of this set of fault paths, conditioned on global acceptance by every 1-ED in the circuit, obeys
\begin{eqnarray}
\label{prob-bad-global}
{\rm Prob}({\cal I}_r~ {\rm bad ~| ~ global ~acceptance}) ~\le ~\left(A\varepsilon^2\right)^r \sum_{\cal K} \frac{{\rm Prob}({\cal K} ~|~ {\cal I}_r)}{(1-\varepsilon)^{|{\cal K}|} }~.
\end{eqnarray}
Here  $A$ is the number of pairs of locations contained in the largest 1-exRec, ${\cal K}$ is summed over all minimal sealed clusters that contain ${\cal I}_r$, and $|{\cal K}|$ is the number of level-0 locations contained in the interior of ${\cal K}$.}
\medskip

To go further, we will need an estimate of ${\rm Prob}({\cal K} ~|~{\cal I}_r)$ and a method for enumerating the sealed clusters. For both purposes, it is convenient to regard the minimal sealed cluster ${\cal K}$ as a graph whose nodes are 1-Ga's, with maximal degree $d=2m$ if each quantum gate in our universal set acts on no more than $m$ qubits. We pessimistically assume that every 1-ED in each of the $r$ specified 1-exRecs in the set ${\cal I}_r$ contains a fault, so that all of these 1-EDs are in the interior of ${\cal K}$. Under this assumption, the smallest possible sealed cluster consists of the 1-exRecs in ${\cal I}_r$ plus all neighboring 1-exRecs. This smallest cluster contains no more than $r(d +1)$ 1-Ga's.  Now imagine traversing ${\cal K}$ in a ``breadth-first'' manner. Any new 1-Ga that we encounter must be connected to the graph that has already been traversed by a 1-ED that is not very good (since if the 1-ED is very good it is contained in the cluster's edge).  Therefore, for any ${\cal K}$ larger than the smallest cluster, a fault must occur in each 1-ED in a specified set determined by ${\cal K}$, where the number of these faulty 1-EDs is at least $|G({\cal K})| - r(d+1)$; here $|G({\cal K})|$ denotes the number of 1-Ga's contained in ${\cal K}$. The probability that a specified 1-ED contains at least one fault is no larger than 
\begin{equation}
{\rm Prob(ED~not~very~good)} \le 1-(1-\varepsilon)^{|{\rm ED}|}~\le~ \varepsilon \cdot |{\rm ED}|~,
\end{equation}
 where $|{\rm ED}|$ denotes the number of 0-locations contained in the 1-ED circuit. Thus we obtain the estimate
\begin{equation}
\label{prob-kay-eye}
{\rm Prob}({\cal K} ~|~ {\cal I}_r)~ \le ~ (\varepsilon\cdot |{\rm ED}|)^{|G({\cal K})|- r(d+1)}~.
\end{equation}

If each of the designated 1-exRecs in ${\cal I}_r$ has the maximal degree $d=2m$, then the interior of the smallest possible sealed cluster has size 
\begin{equation}
|{\cal K}_{\rm smallest}| \le r(d+1)\cdot |{\rm Ga}| + rd \cdot |{\rm ED}|= r(d+1)\cdot |{\rm exRec}| - rd^2 \cdot|{\rm ED}|~,
\end{equation}
where $|{\rm Ga}|$ denotes the maximal number of 0-locations in an $m$-qubit 1-Ga, and $|{\rm exRec}|$ denotes the maximal number of 0-locations in an $m$-qubit 1-exRec. Each time we add an additional 1-Ga to the minimal sealed cluster, the size $|{\cal K}|$ of the interior grows by
\begin{equation}
|{\rm ED}| + |{\rm Ga}| = |{\rm exRec}| - (d-1)\cdot |{\rm ED}|~,
\end{equation}
and the probability ${\rm Prob}({\cal K}~|~{\cal I}_r)$ is suppressed by another factor of $\varepsilon\cdot |ED|$. If we sum over the number (denoted by $t$) of 1-Ga's added to the smallest cluster, eq.~(\ref{prob-bad-global}) yields
\begin{eqnarray}
\label{prob-bad-global2}
&&{\rm Prob}({\cal I}_r~ {\rm bad ~| ~ global ~acceptance}) \nonumber\\
&&~\le ~\left(\frac{A\varepsilon^2}{(1-\varepsilon)^{(d+1)\cdot |{\rm exRec}| - d^2 \cdot|{\rm ED}|}}\right)^r 
\sum_{t=0}^{\infty} N({\cal I}_r,t)\left(\frac{\varepsilon\cdot |{\rm ED}|}{(1-\varepsilon)^{|{\rm exRec}| - (d-1)\cdot |{\rm ED}|} }\right)^t~,
\end{eqnarray}
where $N({\cal I}_r,t)$ denotes the number of clusters with $t$ 1-Ga's added to the smallest cluster containing ${\cal I}_r$.

Now we need to obtain an upper bound on the sum over $t$. One convenient way to proceed is to consider, for each of the $d$ 1-Ga's that surround each 1-exRec in ${\cal I}_r$, enumerating all of the connected clusters that contain that particular 1-Ga. Then summing independently over the possible configurations for each of these $rd$ clusters will account for each possible ${\cal K}$, though some of the minimal sealed clusters will be counted more than once. On a graph with maximal degree $d$, let $M_d(s)$ denote the maximal number of connected clusters of $s$ nodes that contain a particular specified site. Then by summing over the number of nodes in each of $rd$ clusters, we obtain from eq.~(\ref{prob-bad-global2}) 
\begin{eqnarray}
\label{prob-bad-global3}
{\rm Prob}({\cal I}_r~ {\rm bad ~| ~ global ~acceptance})
~\le ~\left(\frac{A\gamma(\varepsilon)^d \varepsilon^2}{D(\varepsilon)}\right)^r 
~,
\end{eqnarray}
where
\begin{equation}
\label{gamma-define}
\gamma(\varepsilon) = \sum_{s=1}^\infty M_d(s) \left(\frac{\varepsilon\cdot |{\rm ED}|}{(1-\varepsilon)^{|{\rm exRec}| - (d-1)\cdot |{\rm ED}|} }\right)^{s-1}~,
\end{equation}
and
\begin{equation}
\label{D-define}
D(\varepsilon)= (1-\varepsilon)^{(d+1)\cdot |{\rm exRec}| - d^2 \cdot|{\rm ED}|}~.
\end{equation}

We can derive an upper bound on $\gamma(\varepsilon)$ from an upper bound on $M_d(s)$, which is provided by

\medskip
\noindent {\bf Lemma 5. Cluster counting}. {\em Consider any specified set ${\cal J}_t$ of $t$ nodes in a graph of maximal degree $d$. Let $M_d(s,{\cal J}_t)$ denote the number of connected sets of $s \ge t$ nodes that contain all of the nodes in ${\cal J}_t$. Then $M_d(s,{\cal J}_t)\le e^{t-1}(ed)^{s-t}$ (where $e=2.71828\dots$).} 
\medskip 

\noindent {\bf Proof}: First consider the case $t=1$. We may associate with each connected set containing the specified marked node a directed tree that spans the set, so that it suffices to count the trees. Imagine traversing the tree in a breadth-first manner, starting at the marked node (the root). In the first step we add directed edges that point from the root to neighboring nodes that are in the set. In subsequent steps, we visit the nodes of the tree one at a time, and we add new directed edges that point from each node to its neighbors in the set that are not already contained in the tree. This procedure continues until all $s$ nodes in the set have been exhausted. 

The root of the tree has in-degree 0 and an out-degree that can take any value from 0 to $d$; all other nodes have in-degree 1 and an out-degree that can range from 0 to $d-1$. The final node visited has out-degree 0. Let $i_1,i_2, \dots, i_{s-1}$ denote the out-degrees of the remaining $s-1$ nodes (those whose out-degrees do not necessarily vanish). These out-degrees must sum to $s-1$.

To count the trees, note that once the out-degree of node $j$ is assumed to be $i_j$, there are at most ${d\choose i_j}$ ways to choose the $i_j$ outgoing edges. Thus the number of trees is bounded above by
\begin{eqnarray}
M_d(s,{\cal I}_1)&\le& \sum_{i_1,i_2,\dots, i_{s-1}} {d\choose i_1}{d\choose i_2}\cdots {d\choose i_{s-1}}
\le \sum_{i_1,i_2,\dots, i_{s-1}}\frac{d^{i_1} d^{i_2} \dots d^{i_{s-1}}}{i_1! i_2! \dots i_{s-1}!}\nonumber\\
&\le& ~d^{s-1}\left(\sum_{i=0}^\infty \frac{1}{i!}\right)^{s-1} = ~(ed)^{s-1}~.
\end{eqnarray}

If there are $t$ marked nodes, then the tree has $t$ roots.  Assign an arbitrary ordering to the roots. First we traverse the tree whose root is the first marked node, avoiding the other marked nodes and avoiding nodes that have been visited in previous steps. Then we traverse the tree whose root is the second marked node, again avoiding other marked nodes and nodes that have already been visited, etc., until every one of the $s$ nodes has been reached. Now the out-degrees of the nodes sum to $s-t$, and as before there are $s-1$ potentially nonvanishing out-degrees. (Actually, $t$ of the out-degrees vanish, but because we don't know which ones will vanish we are obligated to sum over all $s-1$ of them.) Therefore counting as above gives $M_d(s,{\cal J}_t)\le e^{s-1}d^{s-t}$.

\rightline{$\square$}

\noindent In fact, the same argument (and therefore the same upper bound) applies if the subgraph is not connected, as long as each unmarked node is connected to at least one of the marked nodes. We will only need the case $t=1$ for our argument in this paper, but we have considered general $t$ in Lemma 5 for completeness.

Using $M_d(s)=(ed)^{s-1}$ and defining $s'=s-1$, we can evaluate the sum in eq.~(\ref{gamma-define}), finding
\begin{eqnarray}
\label{gamma-evaluate}
\gamma(\varepsilon) = \sum_{s'=0}^\infty \left(\frac{ed\varepsilon\cdot |{\rm ED}|}{(1-\varepsilon)^{|{\rm exRec}| - (d-1)\cdot |{\rm ED}|} }\right)^{s'}= \frac{1}{1-\omega(\varepsilon)}~,
\end{eqnarray}
where
\begin{equation}
\label{omega-define}
\omega(\varepsilon) = \frac{ed\varepsilon\cdot |{\rm ED}|}{(1-\varepsilon)^{|{\rm exRec}| - (d-1)\cdot |{\rm ED}|} }~.
\end{equation}
We note that eq.~(\ref{prob-bad-global3}) implies that 
\begin{eqnarray}
\label{prob-bad-global4}
{\rm Prob}({\cal I}_r~ {\rm bad ~| ~ global ~acceptance})
~\le ~\left(\varepsilon^2/\varepsilon_0\right)^r~, 
\end{eqnarray}
where $\varepsilon_0$ is defined implicitly by the equation
\begin{equation}
\label{epsilon-zero-define}
\varepsilon_0 = \frac{1}{A} \cdot {D(\varepsilon_0)}\Big( 1-\omega(\varepsilon_0)\Big)^d~,
\end{equation}
and where it is understood that the expression for $D(\varepsilon_0)$ is evaluated by choosing the 1-exRec that maximizes $D(\varepsilon_0)$ (the ex-Rec with the largest value of $(d+1)\cdot |{\rm exRec}| - d^2 \cdot|{\rm ED}|$), and that $( 1-\omega(\varepsilon_0))^d $ is also evaluated by choosing the 1-exRec that maximizes its value. 
We obtain:

\medskip
\noindent {\bf Lemma 6. Probability of bad 1-exRecs conditioned on global acceptance}. {\em Suppose that an ideal quantum circuit is simulated using level-1 fault-tolerant error-detecting gadgets satisfying the properties specified in Lemmas 1 and 2, subject to locally correlated stochastic noise with strength $\varepsilon\le \varepsilon_0$. The probability ${\rm Prob}({\cal I}_r~{\rm bad ~| ~ global ~acceptance})$ that all $r$ 1-exRecs in a specified set ${\cal I}_r$ are bad, conditioned on global acceptance by every 1-ED in the circuit, obeys
\begin{eqnarray}
{\rm Prob}({\cal I}_r~ {\rm bad ~| ~ global ~acceptance}) ~\le ~\left(\varepsilon^2/\varepsilon_0\right)^r~,
\end{eqnarray}
where $\varepsilon_0$ is defined by eq.~(\ref{epsilon-zero-define},\ref{D-define},\ref{omega-define}), and
$A$ is the number of pairs of locations contained in the largest 1-exRec.}
\medskip

\noindent We note that $\varepsilon_0 \le 1/A$, and that because the right-hand-side of eq.~(\ref{epsilon-zero-define}) is a decreasing function of $\varepsilon_0$, we can obtain a lower bound on $\varepsilon_0$ by substituting $\varepsilon_0\to 1/A$ in the right-hand-side.

Aside from a slight modification to be discussed in Sec.~\ref{sec:threshold}, arising from the decoding of the $C_1^{\circ k}$ blocks in Knill's scheme, $\varepsilon_0$ can be regarded as an estimate of the accuracy threshold for postselected fault-tolerant quantum computation. Of course, to obtain a numerical lower bound on the threshold, we must determine $A$, $D$, and $\omega$ by analyzing the properties of an explicit set of universal fault-tolerant gadgets. And as discussed in Sec.~\ref{sec:malignant}, we can improve our estimate by counting malignant pairs of locations in the 1-exRecs, rather than all pairs of locations.

\section{The level reduction lemma}
\label{sec:level-reduction}

Suppose we consider an ideal quantum circuit that begins with the preparation of qubits in the standard state $|0\rangle$ and ends with single-qubit measurements in the standard basis $\{|0\rangle,|1\rangle\}$ that read out the final result. We will say such a circuit is ``closed,'' meaning that the circuit ends with measurements. In contrast, we say that a circuit is ``open'' if its output is a quantum state. In Knill's scheme, the purpose of the postselected computation, protected by the concatenated error-detecting code $C_1^{\circ k}$, is to prepare an ancilla state that is protected by the code $C_2$. This postselected circuit is open.

In the case of the level-1 postselected simulation of a closed circuit, subject to locally correlated stochastic noise with strength $\varepsilon$, Lemmas 1, 2, and 6 tell us how to reduce the circuit to an equivalent level-0 circuit (equivalent in the sense that both circuits produce exactly the same probability distribution for the final measurement outcomes). By Lemma 1, we can convert each good 1-preparation to an ideal 0-preparation followed by an ideal 1-encoder, and we can convert each good 1-measurement to an ideal 0-measurement preceded by an ideal 1-decoder. Using Lemmas 1, 2, and 6, we can move the 1-encoders forward one step at a time, converting each good* 1-location to an ideal 0-gate, and each bad 1-location to a faulty level-0 gate, until the 1-encoders finally meet the 1-decoders and annihilate them. We refer to this procedure that replaces a noisy level-1 circuit by its equivalent level-0 circuit as ``level reduction.''

Furthermore, the equivalent level-0 circuit is also subject to locally correlated stochastic noise (assuming the 1-ED gadgets have depth at least two), but with a reduced strength $\varepsilon^{(1)}= \varepsilon^2/\varepsilon_0$. Lemma 6 actually establishes that the noise in the reduced circuit is local stochastic noise with strength $\varepsilon^{(1)}$, but as discussed in Sec.~\ref{sec:prob-bad-exRec} we can also infer that syndrome information is unable to propagate from one good carved 1-exRec to a following one. In addition, as can be seen from Fig.~\ref{fig:weak-correct2}, messages that are passed from one level-0 gate to another must penetrate a very good 1-ED in order to successfully propagate through two successive good carved 1-exRecs; this will not be possible if the 1-ED has depth two or more (as discussed in Sec.~\ref{sec:more-local-noise}). We conclude that in the reduced circuit the adversaries are unable to send syndrome information through two consecutive good gates, the defining property of the locally correlated stochastic noise model. 

Because the reduced circuit is also subject to locally correlated stochastic noise, the level reduction can be repeated. One level reduction step, applied to a closed level-$k$ postselected computation subject to locally correlated stochastic noise with strength $\varepsilon$, yields a closed level-$(k{-}1)$ postselected computation subject to locally correlated stochastic noise with strength $\varepsilon^{(1)}=\varepsilon_0\left( \varepsilon/\varepsilon_0\right)^2$. Repeating the level reduction $k$ times, we finally obtain a level-0 circuit with noise strength
\begin{equation}
\varepsilon^{(k)}= \varepsilon_0\left(\varepsilon/\varepsilon_0\right)^{2^k}~.
\end{equation}
Thus, for $\varepsilon < \varepsilon_0$, the noise strength drops steeply as the level increases, and the postselected computation becomes highly reliable.

An open level-$k$ postselected circuit ends, not with measurements, but rather with the recursive decoding of each $C_1^{\circ k}$ block to a single qubit. The decoder 1-Ga maps the 1-block to a single output qubit, while also extracting the error syndrome as in eq.~(\ref{one-decoder-action}). The decoder 1-Rec consists of the decoder 1-Ga only --- no 1-ED is inserted between the preceding 1-Ga and the decoder 1-Ga. Thus the 1-exRec that contains this preceding 1-Ga has a trailing 1-decoder rather than a trailing 1-ED. Nevertheless, if this 1-exRec contains one fault it is correct in the usual sense, because a very good 1-decoder, like a very good 1-ED, will detect a single error in its input 1-block. The $k$-decoder consists of a level-$(k{-}1)$ simulation of the decoder 1-Rec, followed by a level-$(k{-}2)$ simulation of the decoder 1-Rec, etc., culminating in a final decoder 1-Rec. Therefore, the level-$k$ decoder is not really fault tolerant --- a single fault in the last step can cause the decoder to fail.

When we apply a level reduction step to the level-$k$ postselected circuit, the ideal 1-encoders, as they plow forward through the circuit, convert the level-$k$ recursive decoder to a level-$(k{-}1)$ recursive decoder, followed by one last 0-location acting on each output qubit. This last 0-location arises when the ideal 1-encoder meets the final noisy 1-decoder in the level-$k$ decoder. Ideally, this final 0-location should be the identity gate. But just one fault in the final 1-decoder could cause the 1-decoder to fail, and therefore this final 0-location has a fault with a probability bounded above by $D\varepsilon$, where $D$ is the number of locations in the 1-decoder. 

In summary, we have proved:

\medskip
\noindent {\bf Lemma 7. Level reduction for an open circuit}. {\em Suppose that an ideal open quantum circuit is simulated using level-$k$ fault-tolerant error-detecting gadgets, subject to locally correlated stochastic noise with strength $\varepsilon\le \varepsilon_0$, where the level-1 gadgets satisfy the properties specified in Lemmas 1 and 2, and the 1-ED has depth two or more. Then if no errors are detected by any 1-EDs in the circuit, there is an equivalent level-$(k{-}1)$ circuit, subject to locally correlated stochastic noise with strength $\varepsilon^{(1)}=\varepsilon_0\left(\varepsilon/\varepsilon_0\right)^2$, that produces the same output state. In addition, the equivalent circuit contains one final time step in which faults occur independently on the output qubits with probability $ \le D\varepsilon$. Here $D$ is the number of locations in the  1-decoder, and $\varepsilon_0$ is defined as in Lemma 6.}
\medskip 

\section{Threshold theorem}
\label{sec:threshold}

In a level-$k$ postselected simulation, the computation is aborted if an error is detected at any level. Therefore, we can apply Lemma 7 $k$ times in succession. For the first level reduction step, we only require that no errors are detected in the level-1 blocks. In the next step we again require only a trivial syndrome at level 1 in the effective level-$(k{-}1)$-simulation, but this is equivalent to requiring a trivial level-2 syndrome in the original level-$k$ simulation, and so on. 

We conclude that if no errors are detected at any level by any $k$-ED, then, after applying level reduction $k$ times to the open circuit, the equivalent circuit (apart from the step arising from the $k$-fold reduction of the final $k$-decoders) is subject to locally correlated stochastic noise with strength $\varepsilon^{(k)}=\varepsilon_0\left(\varepsilon/\varepsilon_0\right)^{2^k}$. Each of the final $k$-decoders is reduced to a single 0-location acting on an output qubit, ideally the identity, with fault probability bounded above by 
\begin{equation}
\label{epsilon-decode}
\varepsilon_{\rm dec}^{(k)} = D\sum_{j=0}^{k-1}\varepsilon^{(j)}= D\varepsilon_0 \sum_{j=0}^{k-1}\left(\varepsilon/\varepsilon_0\right)^{2^j}=D\varepsilon +O(\varepsilon^2)~.
\end{equation} 

If, following Knill, we simulate the encoding circuits for the $C_2$ ancillas using the level-$k$ postselected computation, then for $\varepsilon < \varepsilon_0$ and $k$ sufficiently large the errors arising from faults in the encoding circuit become negligible, and we only need to worry about the faults during decoding. These faults produce errors independently distributed among the qubits in the ancilla, with error rate $\varepsilon_{\rm dec}^{(k)}$. If this error rate is low enough, the errors will be corrected with high probability by the code $C_2$. 

In Knill's scheme (see Fig.~\ref{fig:knill}), the code $C_2$ protects a simulated computation in which every gate is teleported. Let's recall what it means to ``teleport a gate'' \cite{gottesman-chuang}. To be concrete we will consider the case of a single-qubit gate, but the same idea also applies to multi-qubit gates. To teleport the gate $V$, first the entangled two-qubit ancilla state $(I_A\otimes V_B)|\phi\rangle_{AB}$ is prepared, where $|\phi\rangle = (|00\rangle+|11\rangle)/\sqrt{2}$ is a Bell state. Then a Bell measurement is performed on the ancilla qubit $A$ and the input data qubit $C$, obtaining the outcome $(\sigma_C\otimes I_A)|\phi\rangle_{CA}$, where $\sigma\in\{I,X,Y,Z\}$ is a known Pauli operator. If the input state of the data is $|\psi\rangle_C$, this procedure prepares qubit $B$, which may be regarded as the output data qubit, in the state $V\sigma^{-1}|\psi\rangle_B$.

Now, if $V$ is a Clifford group gate, then the output state is
\begin{equation}
V\sigma^{-1}|\psi\rangle = (V\sigma^{-1}V^{-1})V|\psi\rangle~,
\end{equation}
where $(\sigma')^{-1} =V\sigma^{-1}V^{-1}$ is a known Pauli operator. To ensure that the output state is $V|\psi\rangle$ as desired we could apply $\sigma'$ to the output to complete the teleportation step. However, it is not actually necessary to apply $\sigma'$; rather it suffices to keep a classical record indicating that the deviation of the data from its ideal state can be repaired by applying $\sigma'$. In a computation in which all gates are Clifford gates, this record can be continuously updated, and ultimately used to interpret the results of Pauli operator measurements at the end of the computation. Note that, because when a Clifford gate is teleported the outcome of the Bell measurement is not needed to determine how later gates are implemented, it is not necessary to wait for the Bell measurement to be completed before preceding to the next step of the computation.

The destructive measurement of encoded Pauli operators can also be realized by ``teleportation.'' To measure the Pauli operator $Z$, we prepare an ancilla qubit in the state $|0\rangle$ (the eigenstate of $Z$ with eigenvalue ${+}1$), and then perform a joint Bell measurement on the ancilla qubit and the input qubit, obtaining the outcome $(\sigma\otimes I)|\phi\rangle$. If $\sigma\in\{I,Z\}$, then the outcome of the measurement is $Z=1$, while if $\sigma\in\{X,Y\}$, then the outcome is $Z={-}1$. Similarly, to measure the Pauli operator $X$, we prepare an ancilla qubit in the state $|+\rangle$ (the eigenstate of $X$ with eigenvalue ${+}1$), and then perform Bell measurement on the ancilla qubit and input qubit. If $\sigma\in\{I,X\}$, then the measurement outcome is $X=1$, and if $\sigma\in\{Y,Z\}$, then the outcome is $X=-1$.

Of course, to achieve universal quantum computation we will need a non-Clifford gate in our gate set. But, as will be discussed in Sec.~\ref{sec:FTUniversal}, our non-Clifford gate $U$ can be chosen such that if $\sigma$ is any Pauli operator, then  $U\sigma^{-1}U^{-1}$ is a Clifford-group operator in our gate set. Therefore, each time $U$ is teleported, the Bell measurement outcome (together with the record of outcomes of Bell measurements performed to implement previous Clifford gates) determines a known Clifford gate that will complete the teleportation of $U$. Thus, when $U$ is teleported, we {\em do} wait for the Bell measurement to be completed; the measurement outcome determines which Clifford gate comes next and therefore which ancilla state will be used in the next teleportation step.

We are to simulate an ideal circuit of teleported gates by using ancilla states encoded in $C_2$ code blocks, and by performing Bell measurements on $C_2$ code blocks. The circuit that prepares each ancilla state has been protected by the error-detecting code $C_1^{\circ k}$; after applying level reduction $k$ times to this ancilla preparation circuit, we obtain a noisy $C_2$ ancilla encoder, which has two parts. The first part, the {\em ancilla preparation gadget}, has a negligible probability of being bad if $\varepsilon < \varepsilon_0$ and $k$ is sufficiently large, and if good it outputs the ideal $C_2$ ancilla state. The second part, the {\em independent noise gadget}, applies independent noise to the qubits in the $C_2$ code block, with error probability per qubit bounded above by $\varepsilon_{\rm dec}^{(k)}\le \varepsilon_{\rm dec}^{(\infty)}$.

Using a suitable notion of level reduction, we can reduce the noisy $C_2$ simulation to a level-0 circuit, where each good teleported gate in the $C_2$ simulation is mapped to an ideal level-0 teleported gate, and each bad teleported gate in the $C_2$ simulation is mapped to a faulty level-0 teleported gate. An exRec in the $C_2$ simulation consists of two ancilla preparation gadgets, the two independent noise gadgets that follow the ancilla preparation gadgets, and a Bell measurement. (If a non-Clifford gate occured in the previous step, the exRec also includes two storage steps, while one of the ancillas waits for the outcome of the previous Bell measurement.) We say that the exRec is good if both of the ancilla preparation gadgets are good, and if the errors arising from independent noise gadgets, the noisy Bell measurement, and the storage steps (if there are any) occur at a correctable set of positions in the $C_2$ code block. If the exRec is good, it is also correct; that is, the logical Pauli operator inferred from the Bell measurement agrees with what it would have been if all operations were ideal.

Formally, if the exRec is correct, we can move an ideal decoder acting on the output of the $C_2$ simulation of gate teleportation forward so that the ideal decoder acts on the input, thereby transforming the simulated gate teleportation to the ideal level-0 gate teleportation, as illustrated here:

\setlength{\unitlength}{0.90pt}
\vspace{.5cm}
\begin{center}
\begin{picture}(358,106)
\put(0,72){\framebox(36,24){\shortstack{\small ancilla\\ \small prep.}}}
\put(18,96){\line(0,1){10}}
\put(36,84){\line(1,0){10}}
\put(46,72){\framebox(36,24){\shortstack{\small indep.\\ \small noise}}}
\put(82,84){\line(1,0){56}}
\put(100,84){\circle*{4}}
\put(118,84){\circle*{4}}
\put(46,0){\framebox(36,24){\shortstack{\small ancilla\\ \small prep.}}}
\put(64,24){\line(0,1){24}}
\put(82,12){\line(1,0){10}}
\put(92,0){\framebox(36,24){\shortstack{\small ideal\\ \small decoder}}}
\put(128,12){\line(1,0){10}}
\put(64,48){\line(1,0){28}}
\put(92,36){\framebox(36,24){\shortstack{\small indep.\\ \small noise}}}
\put(128,48){\line(1,0){10}}
\put(138,36){\framebox(36,60){\shortstack{\small Bell\\ \small meas.}}}
\put(154,96){\line(0,1){10}}
\put(158,96){\line(0,1){10}}
\put(174,64){\line(1,0){10}}
\put(174,68){\line(1,0){10}}
\put(194,50){\makebox(20,12){=}}
\put(220,72){\framebox(36,24){\shortstack{\small ancilla\\ \small prep.}}}
\put(238,96){\line(0,1){10}}
\put(256,84){\line(1,0){10}}
\put(266,72){\framebox(36,24){\shortstack{\small ideal\\ \small decoder}}}
\put(302,84){\line(1,0){10}}
\put(266,0){\framebox(36,36){\shortstack{\small ideal\\ \small ancilla\\ \small 0-prep.}}}
\put(284,36){\line(0,1){12}}
\put(302,12){\line(1,0){10}}
\put(284,48){\line(1,0){28}}
\put(312,36){\framebox(36,60){\shortstack{\small ideal\\ \small Bell\\ \small 0-meas.}}}
\put(328,96){\line(0,1){10}}
\put(332,96){\line(0,1){10}}
\put(348,64){\line(1,0){10}}
\put(348,68){\line(1,0){10}}
\end{picture}
\end{center}
\vspace{.5cm}

\noindent In the diagram, the double-line input and output to the Bell measurement indicates the classically recorded logical Pauli operator which is updated when Clifford-group gates are teleported. The circles on one input wire of the Bell measurement indicate the two storage steps that must be included if the previous gate is non-Clifford, in which case the outcome of the previous Bell measurement is needed to determine what ancilla state to use for the current gate. 

To complete our new proof of the threshold theorem, we are to show that the probability that the $C_2$ simulation of a teleported gate is bad is bounded above by $\varepsilon_{\rm th}$, the previously established accuracy threshold. The encoded Bell measurement is done transversally; that is, a Bell measurement on $C_2$ blocks $A$ and $B$ is realized by first performing Bell measurements on all pairs of qubits, where one qubit in each pair is from block $A$ and the other qubit is at the same position in block $B$. Then the logical Bell measurement outcome is extracted by classical post-processing. This classical decoding will be successful if the Bell measurements with errors occur at a correctable set of positions. 

Because the noise that follows the ancilla preparation gadget is independent, and because there is no subsequent communication between Bell pairs in the rest of the circuit, the errors on different Bell pairs are uncorrelated (neglecting the very small probability that the ancilla preparation gadgets are bad). The probability of error in each two-qubit Bell measurement can be bounded above by
\begin{equation}
\label{epsilon-bell-infinity}
\varepsilon_{\rm Bell}^{(k)} ~=~ 2\varepsilon_{\rm dec}^{(k)} + 5\varepsilon~.
\end{equation}
The factor 2 in front of $\varepsilon_{\rm dec}^{(k)}$ arises because either of the two independent noise gadgets could act at a particular position in the block, and the factor 5 in front of $\varepsilon$ counts the three 0-locations in the Bell measurement circuit (one {\sc cnot} gate and two single-qubit measurements), plus (in the case where the preceding gate is non-Clifford) the two storage 0-locations while the data waits for the outcome of the previous Bell measurement. 

Teleportation of an $m$-qubit gate requires $m$ independent Bell measurements, any of which might fail. Therefore, if the probability of a failed ancilla preparation is negligible, the probability of failure for an $m$-qubit teleported gate is no larger than $m\varepsilon_{\rm Bell}^{(k)}$. Suppose that each gate in the universal set that is protected by the code $C_2$ is a one-qubit or two-qubit gate. Then the code $C_2$ should be chosen so that independent errors occuring at rate $\varepsilon_{\rm Bell}^{(k)}$ can be corrected, such that the probability of a $C_2$ encoded error is below $\frac{1}{2}\varepsilon_{\rm th}$, where $\varepsilon_{\rm th}$ is the value of the quantum accuracy threshold that can be achieved using concatenated error-correcting codes. That is, if the probability of an encoded error is less than $\frac{1}{2}\varepsilon_{\rm th}$ when the error rate in the $C_2$ block is $\varepsilon'$, then a lower bound on the accuracy threshold is found by solving for $\varepsilon$ in the equation
\begin{equation}
\label{epsilon-bell-infinity-threshold}
\varepsilon_{\rm Bell}^{(\infty)} ~\equiv ~ 2\varepsilon_{\rm dec}^{(\infty)} + 5\varepsilon= \varepsilon'~.
\end{equation}

Therefore we have proved:

\medskip
\noindent {\bf Theorem 1. Accuracy threshold for postselected quantum computation}. {\em Suppose that fault tolerant level-1 gadgets can be constructed that satisfy the properties specified in Lemmas 1 and 2, where the 1-ED has depth two or more, and consider a noisy quantum circuit subject to locally correlated stochastic noise with strength $\varepsilon$. Then for any constant $\delta > 0$ the noisy circuit can simulate an ideal quantum circuit of size $L$ with error $\delta$, with an overhead cost polylogarithmic in $L$, provided that $\varepsilon < \tilde \varepsilon_0$. Here $\tilde \varepsilon_0$ is the value of $\varepsilon$ that solves eq.~(\ref{epsilon-bell-infinity-threshold}), where $\varepsilon_{\rm dec}^{(\infty)}$ is defined by eq.~(\ref{epsilon-decode}),and $\varepsilon_0$ is defined as in Lemma 6.}
\medskip

\noindent
Theorem 1 follows because we have shown that, under the specified conditions, Knill's postselected scheme can simulate quantum gates of adequate fidelity for the previously proved threshold theorem to apply. See for example \cite{AGP} for further explanation of this previously established theorem, including in particular the definition of the error $\delta$ and an estimate of the overhead. 

We will describe in Sec.~\ref{sec:malignant} a further refinement of Theorem 1 that will yield an improved lower bound on the accuracy threshold. Let us nevertheless pause to evaluate the lower bound provided by Theorem 1. For the fault tolerant circuits that are constructed in Sec.~\ref{sec:gadgets}, the threshold estimate is dominated by the two-qubit {\sc cnot} gate (the only multi-qubit gate in our universal gate set). The 1-ED contains $|{\rm ED}|$=28 0-locations and the {\sc cnot} 1-Ga contains  $|{\rm Ga}|$=4 0-locations. Therefore the {\sc cnot} 1-exRec contains $4\times 28+4= 116$ 0-locations, and the number of pairs of 0-locations contained in the {\sc cnot} 1-exRec is $A={116\choose 2}=6670$. Plugging these numbers and $d=4$ into eq.~(\ref{epsilon-zero-define}), we find
\begin{eqnarray}
\label{epsilon0-threshold-number}
\varepsilon_0 = 1.410 \times 10^{-4} ~,
\end{eqnarray}
which is only about 6\% below the naive estimate $\varepsilon_0=1/A=1.499 \times 10^{-4}$.

The decoder has $D=3$ locations and  since, for example,
\begin{equation}
\sum_{k=0}^\infty (.999)^{2^k} \approx 9.633
\end{equation}
we may choose $\varepsilon= 1.40 \times 10^{-4}$ and still have 
\begin{equation}
\varepsilon'= \varepsilon_{\rm Bell}^{(\infty)}~ \le (6\times 9.633 +5)\varepsilon < 63 ~\varepsilon < .009~.
\end{equation}
We may choose the code $C_2$ to be any code that can correct independent errors occuring at the rate $\varepsilon'$, with probability of a decoding failure smaller than $\varepsilon_{\rm th}$. To be concrete, suppose that $C_2$ is the level-2 concatenated 5-qubit code, decoded recursively. Since the five-qubit code can correct one error, two errors in a block of five are needed to cause failure; therefore for each inner block of five qubits, the probability $\varepsilon^{(1)}$ of a decoding failure is bounded above by $\varepsilon^{(1)}\le {5\choose 2}(\varepsilon')^2=10(\varepsilon')^2$. For the outer block of five qubits, the probability $\varepsilon^{(2)}$ of a decoding failure is bounded above by $\varepsilon^{(2)}\le 10(\varepsilon^{(1)})^2=10^3(\varepsilon')^4\approx 6.6 \times 10^{-6}$. This quantity is an upper bound on the probability of badness for each single-qubit gate in the simulated $C_2$ computation. To teleport a two-qubit gate, there are two Bell measurements, either of which might fail, so that probability of badness is bounded above by $2(6.6 \times 10^{-6})\le 1.4\times 10^{-5}$, which is still safely below $\varepsilon_{\rm th}$. We conclude that $1.40\times 10^{-4}$ is a lower bound on the accuracy threshold that can be achieved using error detection and postselection.

However, this lower bound on the accuracy threshold is weaker than the lower bound $1.9 \times 10^{-4}$ that has already been established in \cite{aliferis-cross} using concatenated quantum error-correcting codes. To demonstrate that Knill's postselection method really improves the threshold we need to strengthen Theorem 1.

\section{Malignant pairs of locations}
\label{sec:malignant}

The threshold estimate formulated in Sec.~\ref{sec:threshold} was based on counting all pairs of 0-locations in the largest 1-exRec, along with a correction factor to take into account correlations among the faults in the circuit. This estimate is too pessimistic because there are many pairs of 0-locations that are {\em benign} --- the 1-exRec is correct even if arbitrary faults occur at that pair of 0-locations. As in \cite{AGP}, we can improve the threshold estimate by counting the pairs of 0-locations that are {\em malignant} (i.e., not benign).

Recall there are several different notions of correctness, where the applicable notion depends on the context in the circuit. For example, for a two-qubit gate, we may consider AA-correctness, AB-correctness, BB-correctness, weak AA-correctness, weak AB-correctness, and weak BB-correctness; the malignant pairs of 0-locations can be identified in all six cases. Let us say that a pair of 0-locations is malignant if any of these notions of correctness are violated for some choice of faults at this pair of 0-locations.

If a 1-exRec is incorrect, then either faults occur at a malignant pair of 0-locations, or else faults occur at three or more 0-locations (where no two of these locations constitute a malignant pair). This means that Lemma 3 can be strengthened, with eq.~(\ref{eye-are-bad}) replaced by  
\begin{equation}
{\rm Prob}({\cal I}_r~{\rm bad}) \le \left(\tilde A\varepsilon^2 +B\varepsilon^3\right)^r~;
\end{equation}
here $\tilde A$ is now the largest number of malignant pairs of 0-locations for any 1-exRec, and $B={|{\rm exRec}|\choose 3}$ is the number of ways to choose three 0-locations in the largest 1-exRec. We may also make the replacement $A\varepsilon^2\to \tilde A\varepsilon^2 + B\varepsilon^3$ in eq.~(\ref{prob-bad-global3}) to improve Theorem 1.

There are a variety of ways to further refine this estimate, some of which we will exploit to estimate the threshold in Sec.~\ref{sec:estimate}. One useful observation is that a 1-exRec is sure to be correct if faults occur only in the trailing 1-EDs. Therefore, when we enumerate the sets of faulty 0-locations that cause the 1-exRec to fail we may exclude the case where all the faulty 0-locations are contained in the trailing 1-EDs, and so we may make the replacement (for an $m$-qubit gate)
\begin{equation}
B\to \tilde B={|{\rm exRec}|\choose 3} - {m |{\rm ED}|\choose 3}~.
\end{equation}

Furthermore, when we count the sealed clusters, we may distinguish the case where the 1-exRec contains exactly two faults from the case where the 1-exRec contains more than two faults. For an $m$-qubit gate, if there are only two faults then at least $2m-2$ 1-EDs in the 1-exRec are very good. Thus for any fixed fault path with exactly two faults in a specified 1-exRec, we can restrict the sum over the sealed clusters containing that 1-exRec to those such that $2m-2$ specified 1-EDs are at the cluster's edge. Therefore the factor $\gamma(\varepsilon)^d$ in eq.~(\ref{prob-bad-global3}) can be replaced by $\gamma(\varepsilon)^2$, and also the size of the interior of the smallest sealed cluster becomes
\begin{equation}
|{\cal K}_{\rm smallest}| \le 3\cdot |{\rm Ga}| + 2 \cdot |{\rm ED}|= 3\cdot |{\rm exRec}| - 10 \cdot|{\rm ED}|~.
\end{equation}
Adding together the contributions due to 1-exRecs with exactly two faults and due to 1-exRecs with three or more faults, the expression for ${\rm Prob}({\cal I}_r~ {\rm bad ~| ~ global ~acceptance})$ becomes
\begin{eqnarray}
\label{prob-bad-global-malignant}
{\rm Prob}({\cal I}_r~ {\rm bad ~| ~ global ~acceptance})
~\le ~\left(\frac{\tilde A\gamma(\varepsilon)^2 \varepsilon^2}{C(\varepsilon)} + \frac{\tilde B\gamma(\varepsilon)^d \varepsilon^3}{D(\varepsilon)}\right)^r 
~,
\end{eqnarray}
where
\begin{equation}
\label{C-define}
C(\varepsilon)= (1-\varepsilon)^{3\cdot |{\rm exRec}| - 10 \cdot|{\rm ED}|}~.
\end{equation}
The equation that implicitly defines the threshold estimate $\varepsilon_0$ becomes
\begin{equation}
\varepsilon_0^{-1}= \frac{\tilde A\gamma(\varepsilon_0)^2 }{C(\varepsilon_0)} + \frac{\tilde B\gamma(\varepsilon_0)^d \varepsilon_0}{D(\varepsilon_0)}~.
\end{equation}
Some further ideas for improving the estimate will be explained in Sec.~\ref{sec:estimate}.

For the proof of the threshold theorem it is necessary to establish that the locally correlated stochastic noise model is preserved by level reduction. One possible approach is to design our gadgets such that even when good 1-exRecs are permitted to contain faults at a benign pair of locations (rather than just a single fault), it is still true that information about the previous fault history is unable to propagate through a good* 1-exRec to a good* 1-exRec that immediately follows. In fact, the definitions of AA, AB, and BB-correctness  already ensure that a correct 1-exRec clears the syndrome --- the input to each of the trailing 1-EDs is a codeword. Similarly, in the case of a weakly correct 1-exRec, the input to the trailing 1-ED that is shared with the following good* 1-exRec is a codeword. We should check in addition that all gadgets are constructed such that no level-0 messages can pass through two consecutive good* 1-exRecs. However, in this paper we follow a somewhat different approach to derive a lower bound on the accuracy threshold; we will reformulate the conditions satisfied by the noise model, as explained in Sec.~\ref{sec:seal-cluster-Knill}.

%---------------------------------------------------------------------------------------------%
\section{Postselected quantum computation: the gadgets}
\label{sec:gadgets}

Before proceeding to give more details of our threshold analysis, we will describe in this section the fault-tolerant gadgets that we will use. Then, in Sec.~\ref{sec:estimate}, we will derive our improved lower bound on the accuracy threshold.

Following Knill \cite{knill_detect}, the error-detecting code $C_1$ used in our fault-tolerant constructions has parameters [[$n,k,d$]]=[[4,2,2]]; that is, it encodes $k=2$ logical qubits in $n=4$ physical qubits and has distance $d=2$. Using the notation $X\equiv \sigma_{\rm x}$ and $Z\equiv \sigma_{\rm z}$ for the Pauli matrices, the generators of the code's stabilizer are the operators $X^{\otimes 4}$ and $Z^{\otimes 4}$. These generators have eigenvalues $\pm 1$, and measuring both generators determines a ``syndrome'' that can take any one of $2^2=4$ values; the code space has the trivial syndrome $X^{\otimes 4}=Z^{\otimes 4}=1$. The logical $X$ and $Z$ operators can be taken to be $X_L=XXII$ and $Z_L=ZIZI$ acting on the first logical qubit and $X_T=XIXI$ and $Z_T=ZZII$ acting on the second logical qubit (where $I$ is the identity operator and we have omitted the tensor product symbol). 

It is helpful to recognize that the [[4,2,2]] code can be regarded as the distance-2 ``Bacon-Shor code'' described in \cite{bacon}. The 16-dimensional Hilbert space $\mathcal{H}$ of the four qubits in the code block can be partitioned into four four-dimensional subspaces, each labeled by a value of the syndrome. Each of these subspaces --- and, in particular, the code space corresponding to the trivial syndrome --- can be decomposed as a tensor product $\mathcal{H}_L\otimes\mathcal{H}_T$ of two logical qubits; hence
\begin{equation}
\label{eq:gad.1.1}
\mathcal{H} = \bigoplus_{{\rm eigen}(X^{\otimes 4}, Z^{\otimes 4})} \left( \mathcal{H}_L \otimes \mathcal{H}_T \right)  \; .
\end{equation}
Either of the two logical qubits could be chosen as the logical qubit for our computation --- by convention, we choose the first logical qubit. The second logical qubit then acts as a ``gauge'' qubit \cite{bacon,poulin} (or ``spectator'' qubit \cite{knill_detect}) whose state does not affect the computation. In principle, we could just disregard the gauge qubit in our subsequent discussion. But as we will discuss in Sec.~\ref{sec:FTED}, we can increase the effectiveness of error detection if we keep the gauge qubit in a known state throughout our computation. 

The remainder of this section is organized as follows: In Sec.~\ref{sec:FTED} we present 1-ED gadgets that implement fault-tolerant error detection using the [[4,2,2]] code, and in Sec.~\ref{sec:seal-cluster-Knill} we explain how these gadgets have suitable properties for our proof of the threshold theorem to be applicable. In Sec.~\ref{sec:FTClifford} we discuss 1-Ga gadgets for implementing the logical gates $\{${\sc cnot}$, H, S \}$ that generate the Clifford group. Then in Sec.~\ref{sec:FTUniversal} we discuss how to implement a logical non-Clifford single-qubit rotation that completes our universal gate set. Finally, in Sec.~\ref{sec:FTPrepMeas}, we discuss how to construct 1-Ga's that prepare eigenstates of the logical Pauli operators and 1-Ga's that measure the logical Pauli operators.

%---------------------------------------------------------------------------------------------%
\subsection{Fault-tolerant error detection}
\label{sec:FTED}

\begin{figure}[tbh]
\begin{center}
\epsfig{file=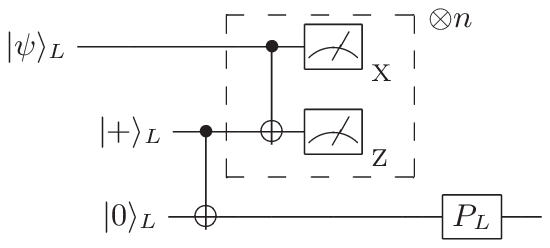,width=5.5cm} \vspace{0.2cm}
\end{center}
\fcaption{\label{fig:1.1} A schematic of Knill's 1-ED gadget applied to a code with $n$ qubits in the code block. Two ancilla blocks are prepared in a logical Bell state, $|\Phi_0\rangle_L \propto |00\rangle_L + |11\rangle_L$, and then transversal {\sc cnot} gates are applied as shown. Finally, all qubits in the data and the first ancilla block are measured in the eigenbases of the Pauli operators indicated at the lower right of the measurement symbols. Parity checks applied to the measurement outcomes detect errors in the incoming data block or the first ancilla block. If there are no errors, the measurement outcomes determine the logical Pauli operator ($P_L$) needed to complete the {\em logical teleportation} of the state $|\psi\rangle_L$ of the input code block. }
\end{figure}

To implement fault-tolerant error detection we will use Knill's 1-ED gadget shown in Fig.~\ref{fig:1.1}: First, two ancilla blocks are encoded in the logical Bell state $|\Phi_0\rangle_L \propto |00\rangle_L + |11\rangle_L$ and, next, a transversal {\sc cnot} gate is applied with the data block as control and the first ancilla block as target. Then, all qubits in the data block are measured in the $X$ eigenbasis and all qubits in the first ancilla block are measured in the $Z$ eigenbasis. In the case of the [[4,2,2]] code, the parity of the four $X$ measurement results determines $X^{\otimes 4}_{\rm data}\otimes X^{\otimes 4}_{\rm anc}$, which has the value $+1$ if there are no $Z$ errors in the data or ancilla block and the value $-1$ if there is a single $Z$ error in the two blocks. Similarly, the parity of the four $Z$ measurement results determines $Z^{\otimes 4}_{\rm data}\otimes Z^{\otimes 4}_{\rm anc}$, which has the value $+1$ if there are no $X$ errors in the data or ancilla block and the value $-1$ if there is a single $X$ error in the two blocks.

If there are no errors in either block, then the parity of the first two $X$ measurement results determines the eigenvalue of $X_L\otimes X_L$ acting on the data and first ancilla blocks, and the parity of the first and third $Z$ measurement results determines the eigenvalue of $Z_L\otimes Z_L$ acting on the data and first ancilla blocks. These two eigenvalues determine the logical Pauli operator that needs to be applied to the second ancilla block to complete the teleportation of the logical state of the input data block. 
 
\begin{figure}[tb]
\begin{center}
\epsfig{file=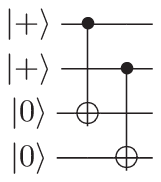,width=1.6cm} \hskip 1.5cm \epsfig{file=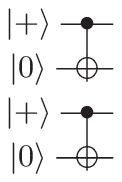,width=1.2cm}
\put(-5.1,1.6){(a)} \put(-1.9,1.6){(b)} \vspace{0.2cm}
\end{center}
\fcaption{\label{fig:1.2} Encoding circuits for the states (a) $|0\rangle_L|+\rangle_T$, and (b) $|+\rangle_L|0\rangle_T$. Note that a {\sc cnot} gate acting on two qubits initialized in the state $|+\rangle | 0\rangle$ creates a Bell state $|\Phi_0\rangle \propto |00\rangle + |11\rangle$.}
\end{figure}

In the case of the [[4,2,2]] code, the preparation of the logical Bell state can be simplified by choosing the state of the gauge qubit appropriately. We note that the code state $|0\rangle_L |+\rangle_T$ is the simultaneous +1 eigenstate of $Z_L=ZIZI$ and $X_T=XIXI$; using the code's stabilizer generators $ZZZZ$ and $XXXX$ we infer that $|0\rangle_L |+\rangle_T$ is also a +1 eigenstate of $IZIZ$ and $IXIX$. It follows, then, that
\begin{equation}
|0\rangle_L |+\rangle_T= |\Phi_0\rangle_{13}|\Phi_0\rangle_{24}
\end{equation}
is a product of a Bell state of the first and third qubits in the block and a Bell state of the second and fourth qubits. By similar reasoning we see that 
\begin{equation}
|+\rangle_L |0\rangle_T= |\Phi_0\rangle_{12} |\Phi_0\rangle_{34}~.
\end{equation}
Circuits for preparing $|0\rangle_L|+\rangle_T$ and $|+\rangle_L|0\rangle_T$ are shown in Fig.~\ref{fig:1.2}. These 1-preparation gadgets satisfy Property Ga2 as formulated in Sec.~\ref{sec:goodness} because a Bell state can have no more than one error --- since $(I\otimes O)|\Phi_0\rangle = (O^T \otimes I )|\Phi_0\rangle $ for any matrix $O$ (where $O^T$ denotes the transpose of $O$), an error acting on the second qubit of a Bell state is equivalent to an error acting on the first qubit.

\setlength{\unitlength}{1cm} 

\begin{figure}[tb]
\begin{center}
\epsfig{file=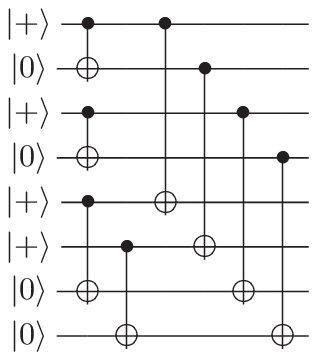,width=3cm} \hskip 1.7cm \epsfig{file=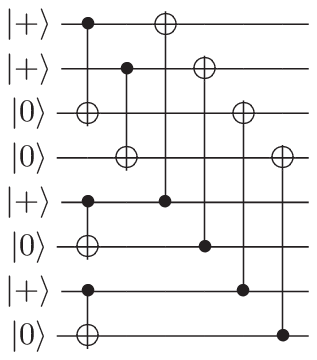,width=3cm}
\put(-8.5,3.1){(a)} \put(-3.7,3.1){(b)} \vspace{0.2cm}
\end{center} 
\fcaption{\label{fig:1.3} Two encoding circuits for the logical Bell state $|\Phi_0\rangle_{L}$. In (a), the first block is prepared in the state $|+\rangle_L|0\rangle_T$ and the second block is prepared in the state $|0\rangle_L|+\rangle_T$. Then, the transversal {\sc cnot} gate creates a logical Bell state while preserving the state $|0\rangle_{T}|+\rangle_{T}$ of the gauge qubits. In (b), the two code blocks are interchanged, so that the gauge qubits are prepared in the state $|+\rangle_{T}|0\rangle_{T}$.}
\end{figure}

Furthermore, since the [[4,2,2]] code is a CSS code, the logical {\sc cnot} is transversal (it can be implemented by bitwise {\sc cnot} gates). This implies that in order to prepare two blocks in a logical Bell state we may use either of the circuits in Fig.~\ref{fig:1.3}. The two circuits are related by interchanging the two code blocks; therefore both circuits prepare the logical state $|\Phi_0\rangle_L$ (which is invariant under interchange), but they prepare different states of the gauge qubits. These Bell pair 1-preparation gadgets satisfy Property Ga2 of Sec.~\ref{sec:goodness} because both the preparation of encoded blocks and the transversal {\sc cnot} gate are fault tolerant.

\begin{figure}[tb]
\begin{center}
\epsfig{file=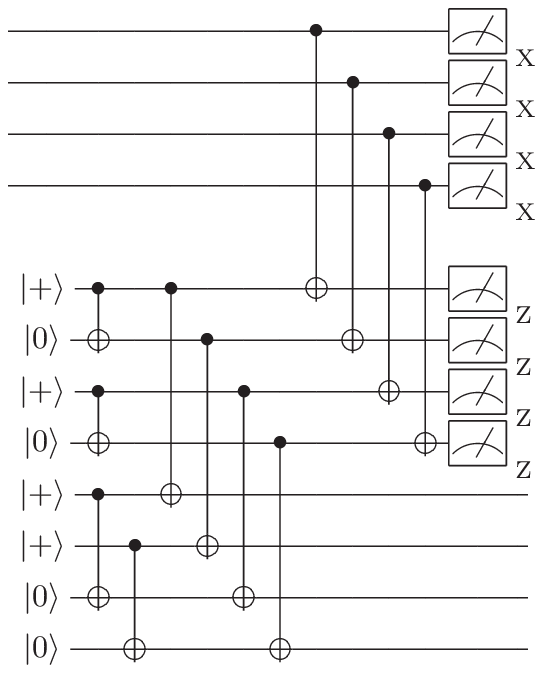,width=4.5cm} \hskip 2cm \epsfig{file=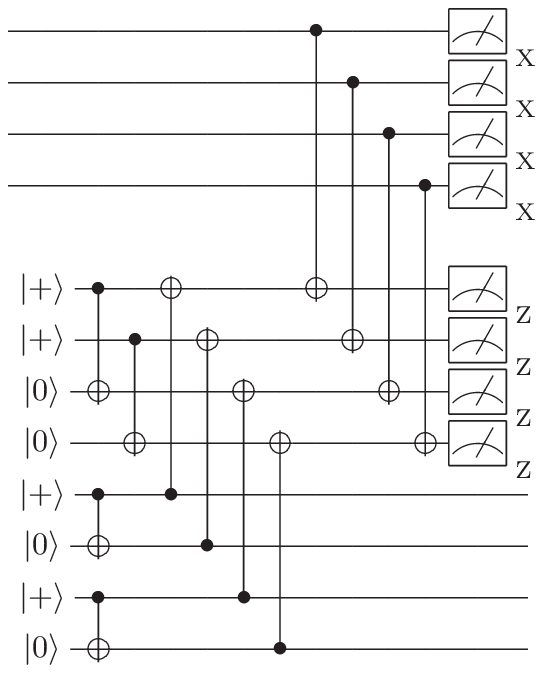,width=4.5cm} 
\put(-11.9,5.1){(a)} \put(-5.3,5.1){(b)} \vspace{0.2cm} 
\vspace{0.2cm}
\end{center}
\fcaption{\label{fig:1.4} Two alternative versions of Knill's 1-ED gadget for the [[4,2,2]] code. (a) In ``zero 1-ED'' (1-zED), the gauge qubit of the first ancilla block is prepared in the state $|0\rangle_{T}$. If the input data block has its gauge qubit in the ``matching'' state $|0\rangle_T$, then the eigenvalue of the operator $Z_T{\otimes}Z_{T}$ acting on the data and the first ancilla block can be extracted during error detection. (b) In ``plus 1-ED'' (1-pED), the gauge qubit of the first ancilla block is prepared in the state $|+\rangle_{T}$. If the input data block has a gauge qubit in the ``matching'' state $|+\rangle_T$, then the eigenvalue of the operator $X_T{\otimes}X_{T}$ acting on the data and the first ancilla block can be extracted during error detection.  }
\end{figure}

Depending on which one of the two circuits in Fig.~\ref{fig:1.3} we use to prepare the logical Bell state, we obtain one of the two 1-ED gadgets shown in Fig.~\ref{fig:1.4}. The 1-ED circuit in Fig.~\ref{fig:1.4}(a) prepares the gauge qubit in the first ancilla block in the state $|0\rangle_{T}$; we therefore refer to this type of error detection as {\em zero} error detection, denoted 1-zED. The 1-ED circuit in Fig.~\ref{fig:1.4}(b) prepares the gauge qubit in the first ancilla block in the state $|+\rangle_{T}$; we therefore refer to this type of error detection as {\em plus} error detection, denoted 1-pED. 

As suggested by Knill \cite{knill_detect}, we perform 1-zED if the gauge qubit of the incoming data block has been prepared in the state $|0\rangle_{T}$; then we can perform an additional parity check on the measurement outcomes to evaluate the eigenvalue of $Z_T{\otimes}Z_{T}$ acting on the data and the first ancilla block. This scheme allows us to detect, for example, the Pauli error $XIIX$ acting on the data, which commutes with the code stabilizer but anticommutes with both $Z_L$ and $Z_T$. Similarly, we perform 1-pED if the gauge qubit of the incoming data block has been prepared in the state $|+\rangle_T$, which enables us to extract the eigenvalue of $X_T\otimes X_T$ acting on the data and the first ancilla block, thus detecting the Pauli error $ZIIZ$ acting on the data. 

The output from 1-zED has the gauge qubit prepared in the state $|+\rangle_T$ and the output from 1-pED has the gauge qubit prepared in the state $|0\rangle_T$. Therefore, we {\em alternate} between the 1-zED and 1-pED gadgets throughout our computation.  Half of the error detection steps have an enhanced ability to detect $X$ errors and the other half have an enhanced ability to detect $Z$ errors. 

%---------------------------------------------------------------------------------------------%
\subsection{Knill's ED gadget and the locally correlated stochastic noise model}
\label{sec:seal-cluster-Knill}

\begin{figure}
\begin{center}
\setlength{\unitlength}{1pt}
%\vspace{0.5cm}
\begin{picture}(106,72)
\put(0,3){\framebox(36,18){1-prep}}
\put(36,12){\line(1,0){40}}
\put(0,27){\framebox(36,18){1-prep}}
\put(36,36){\line(1,0){40}}
\put(76,6){\makebox(20,12){{\em out}}}
\put(76,24){\framebox(30,48){1-BM}}
\put(56,60){\line(1,0){20}}
\put(36,54){\makebox(20,12){{\em in}}}
\put(56,36){\circle*{4}}
\put(56,12){\circle{8}}
\put(56,8){\line(0,1){28}}
\end{picture}
\vspace{0.3cm}
\end{center}
\fcaption{Schematic view of Knill's 1-ED gadget. In our modified noise model, a good level-0 Bell measurement that is preceded by a good 0-Ga will prevent messages from passing from the input to the output; therefore a very good 1-ED blocks communication.}
\label{fig:ED-w-BM}  
\end{figure}

For our proof of the threshold theorem, we need to verify that two properties of the level-1 gadgets are satisfied by the noise model. First, we require that communication through the circuit is blocked by a very good 1-ED (one with no faults). This property is used in our analysis of sealed clusters in Sec.~\ref{sec:local-to-global} and Sec.~\ref{sec:malignant}. Second, we must check that under level reduction the noise model is mapped to a noise model of the same form, but with a renormalized noise strength.

We have seen that the locally correlated stochastic noise model formulated in Sec.~\ref{sec:noise} has both properties, if a (possibly truncated) good* 1-exRec is allowed to contain no more than one fault. But the stability of the noise model under level reduction must be revisited if faults at benign pairs of locations are allowed. We will continue to suppose that correlations among faults can be established only due to messages passed by the local adversaries who reside at each 0-Ga in the circuit. However, the condition invoked in our previous analysis, that no message can pass through two consecutive good 0-Ga's, needs to be modified. 

As will be explained in Sec.~\ref{sec:coexRecs} , after one or more level reduction steps we will regard each Bell measurement and each Bell state preparation as a single 0-Ga in the circuit --- using such ``contracted'' gadgets will enable us to improve the threshold estimate. With this convention, Knill's 1-ED contains only 12 0-Ga's: 4 Bell state preparations, 4 {\sc cnot} 0-Ga's, and 4 Bell measurement 0-Ga's. Our noise model must be chosen so that the desired properties of the noise model are compatible with this procedure for enumerating 0-Ga's.

\begin{figure}
\begin{center}
\setlength{\unitlength}{1pt}
%\vspace{0.5cm}
\begin{picture}(156,144)
\put(0,27){\framebox(36,18){1-prep}}
\put(36,36){\line(1,0){40}}
\put(76,0){\framebox(30,48){1-BM}}
\put(56,12){\line(1,0){20}}
\put(36,6){\makebox(20,12){{\em in}}}
\put(56,36){\circle*{4}}
\put(56,60){\circle{8}}
\put(56,36){\line(0,1){28}}
\put(0,51){\framebox(36,18){1-prep}}
\put(36,60){\line(1,0){90}}
\put(126,48){\framebox(30,48){1-BM}}
\put(0,75){\framebox(36,18){1-prep}}
\put(36,84){\line(1,0){90}}
\put(56,108){\circle*{4}}
\put(56,84){\circle{8}}
\put(56,80){\line(0,1){28}}
\put(0,99){\framebox(36,18){1-prep}}
\put(36,108){\line(1,0){40}}
\put(76,96){\framebox(30,48){1-BM}}
\put(56,132){\line(1,0){20}}
\put(36,126){\makebox(20,12){{\em in}}}
\end{picture}
\vspace{0.3cm}
\end{center}
\fcaption{The Bell measurement 1-exRec, consisting of two leading 1-EDs followed by a Bell measurement 1-Ga.}
\label{fig:BM1-Rec}  
\end{figure}

Schematically, the 1-ED of Fig.~\ref{fig:1.4} has the form shown in Fig.~\ref{fig:ED-w-BM}, where each 1-preparation contains two Bell state preparations performed in parallel, the {\sc cnot} 1-Ga contains four parallel {\sc cnot} gates, and the level-1 Bell measurement contains four parallel Bell measurements (or 0-BMs). If the 1-ED is very good then its output is in the code space after postselection, in accord with property ED1 of Sec.~\ref{sec:goodness}. Since no syndrome information propagates through the very good 1-ED, only the messages passed by the local adversaries can establish correlations between its input and its output. 

Since the 1-ED under examination may have resulted from several level reduction steps applied to a higher level 1-ED, we must recognize that postselection could be hidden inside a good 0-BM. Because of this hidden postselection, a good 0-BM might establish correlations between the two messages it receives. To exclude this possibility, let us assume, as the defining feature of our modified noise model, that {\em a good level-0 Bell measurement blocks communication, unless it is immediately preceded by bad level-0 gates acting on both of its input qubits}. Under this assumption, the four good 0-BMs inside a very good 1-ED (each of which is preceded by a good {\sc cnot} gate acting on one of its inputs) prevent correlations from being established between the input and output messages passed through the 1-ED; thus a very good 1-ED blocks communication as we desire.

It remains to show that this condition is stable under level reduction. At level 1, the BM 1-Rec, shown in Fig.~\ref{fig:BM1-Rec}, consists of a BM 1-Ga preceded by two leading 1-EDs (the 1-exRec, which coincides with the 1-Rec, contains no trailing 1-EDs and is never truncated). If the 1-BM is good, and if (at least) one of the two immediately preceding 1-exRecs is also good*, then the input to the 1-ED that is shared by the 1-BM Rec and the preceding good* 1-exRec is guaranteed to be a codeword. Thus there is no nontrivial syndrome information carried by this input block that could become correlated with the syndrome of the other block. Therefore, unless both of the immediately preceding 1-exRecs are bad, only the level-0 communication can establish correlations between the two input blocks.

Furthermore, a good BM 1-Rec contains at most two faults, and therefore since a very good 1-ED blocks communication, communication through a good BM 1-Rec is possible only if each leading 1-ED contains exactly one fault. It follows that the BM 1-Ga must have no faults, and the four good 0-BMs contained in this 1-BM will block the level-0 messages unless one of the 0-BMs is immediately preceded by two bad 0-Ga's. Thus we see that communication is surely blocked unless the two faulty gates in the good BM 1-Rec are {\sc cnot} gates that both precede the same 0-BM. 

Now consider the leading 1-ED in the BM 1-Rec that is shared with (one of) the good* 1-exRecs that immediately precedes the BM 1-Rec. We have already seen that, if communication through the BM 1-Rec is possible, then this 1-ED must have exactly one faulty {\sc cnot} 0-Ga that is contained in the {\sc cnot} 1-Ga. Therefore the 1-BM contained in the 1-ED has no faults, and hence blocks the level-0 communication, {\em unless} one of the 0-BMs contained in the 1-BM is preceded by faulty 0-Ga's acting on both of its input qubits; one of these is the faulty {\sc cnot} 0-Ga contained in the 1-ED, and the other must be a faulty 0-Ga that acts at the same position in the code block and is contained in the 1-Ga that immediately precedes the 1-ED.

However, this pair of fault locations in the preceding 1-exRec --- one fault in the 1-{\sc cnot} of the trailing 1-ED and another fault in the 1-Ga that acts at the same position in the code block --- is surely a malignant pair: the output of the 1-Ga may have one error caused by the fault in the 1-Ga, and this error may escape detection because of the fault in the 1-ED. Therefore, if the 1-exRec is indeed good*, then the 1-BM in the 1-ED  is guaranteed to block the messages passed by the local adversaries. We conclude that, under level reduction, the good BM 1-Rec is mapped to a good 0-BM that blocks communication, unless both of the immediately preceding 0-Ga's are bad; therefore, the noise model is stable under level reduction as desired.

%---------------------------------------------------------------------------------------------%
\subsection{Fault-tolerant Clifford-group computation}
\label{sec:FTClifford}
In the [[4,2,2]] code, as for any CSS code, the logical {\sc cnot} is transversal. The logical Hadamard is also transversal, up to a permutation of the second and third qubits in the code block. To see this, note that a bitwise Hadamard transformation interchanges $XXXX$ and $ZZZZ$, thereby preserving the code's stabilizer group. It also maps $X_L=XXII$ to $ZZII$ and $Z_L=ZIZI$ to $XIXI$; after swapping the second and third qubits in the code block, then, $X_L$ and $Z_L$ are interchanged. This is the logical Hadamard transformation.

Thus, for the [[4,2,2]] code, there are fault-tolerant realizations of all the operations in the set
\begin{equation}
\mathcal{G}_{\rm CSS} = \{{\rm CNOT}, H, \mathcal{P}_{|0\rangle}, \mathcal{P}_{|+\rangle}, \mathcal{M}_X, \mathcal{M}_Z \}~,
\end{equation}
where $\mathcal{P}_{|\phi\rangle}$ denotes the preparation of the single-qubit state $|\phi\rangle$ and $\mathcal{M}_A$ denotes the measurement of the single-qubit operator $A$. We refer to these operations as the {\em CSS operations}. Together with the operations in $\mathcal{G}_{\rm CSS}$, the phase gate 
\begin{equation}
S\,{\equiv}\,\exp(-i {\pi \over 4}\sigma_{\rm z})
\end{equation}
suffices to generate the Clifford group. Though no direct transversal implementation of the logical $S$ gate is possible for the [[4,2,2]] code, we may use the simulation circuit in Fig.~\ref{fig:1.5}. The simulation of the logical $S$ gate then reduces to logical CSS operations and the preparation of an ancilla block in the logical state $|{+}i\rangle_L \propto |0\rangle_L + i |1\rangle_L$. 
\begin{figure}[h]
\begin{center}
\parbox{1cm}{\Qcircuit @C=0.7ex @R=1.5ex @!R { 
   & \push{|\psi\rangle \hspace{0.1cm}} & \qw & \targ      & \qw & \meterz \cwx[1]  \\
   & \push{|{+}i\rangle \hspace{0.1cm}} & \qw & \ctrl{-1}  & \qw & \gate{Y} & \qw & \qw & \rstick{S|\psi\rangle} 
                                }} \vspace{0.3cm}  
\end{center}
\fcaption{\label{fig:1.5} Circuit simulating the $S$ gate using the ancilla state $|{+}i\rangle$, the $+1$ eigenstate of $Y$.}
\end{figure}

\subsubsection{Distillation threshold}
Although it is not obvious how to devise a fault-tolerant circuit that directly prepares highly accurate copies of the logical state $|{+}i\rangle_L$, we can ``distill'' $|{+}i\rangle_L$ states of progressively higher fidelity provided that the error probability of the initial copies is below a threshold value. For example, the distillation protocol indicated in Fig.~\ref{fig:1.6} takes as input two noisy copies of $|{+}i\rangle$, each with error probability $\varepsilon_{\rm anc}$, and (after postselection) outputs a single cleaner copy with error probability of order $\varepsilon_{\rm anc}^2$. Repeating the distillation protocol reduces the error probability further. Note that the operations used in the distillation protocol are CSS operations (the {\sc cphase} gate can be built from {\sc cnot} and Hadamard gates).

\begin{figure}[h]
\begin{center}
\parbox{1cm}{\Qcircuit @C=0.7ex @R=2.1ex @!R { 
   & \push{|{+}i\rangle \hspace{0.1cm}} & \qw & \control \qw & \qw & \targ      & \qw & \qw & \qw & \rstick{|{+}i\rangle}  \\
   & \push{|{+}i\rangle \hspace{0.1cm}} & \qw & \ctrl{-1}    & \qw & \ctrl{-1}  & \qw & \meterx & \cw & \rstick{+1} 
                                }} \vspace{0.3cm}  
\end{center}
\fcaption{\label{fig:1.6} A distillation protocol for the $|{+}i\rangle$ state. Two copies are prepared and then interact via a {\sc cphase} gate followed by a {\sc cnot} gate. If the initial copies have no errors and the operations are executed without faults, the $X$ measurement on the second copy yields the outcome $+1$. If a different outcome is observed, the output state is rejected.}
\end{figure}

To be more concrete, consider a state of a single qubit that is close to the state $|{+}i\rangle$. By flipping a coin and applying $Y$ to the state with probability 1/2, we induce decoherence in the basis of $Y$ eigenstates; thus the resulting state is a mixture of $|{+}i\rangle$ with probability $1-\varepsilon_{\rm anc}$ and $|{-}i\rangle_{\rm anc}$ with probability $\varepsilon_{\rm anc}$. We say that $\varepsilon_{\rm anc}$ is the ``error probability.'' 

This procedure, which uses randomness (i.e. a toss of a fair coin) to induce decoherence in a particular basis, has been called {\em twirling} \cite{bdsw}. For postselected circuits, the analysis of a protocol that involves twirling can be problematic --- the adversary might arrange for the probability of acceptance to depend on the outcome of the coin toss, which in effect biases the coin \cite{reichardt-thesis}. Fortunately, the effectiveness of the distillation protocol is not impaired if twirling is omitted. We will explain this point in Sec.~\ref{sec:no-twirling} below.

Suppose for the moment that all operations in the distillation protocol are executed ideally without faults. Then, after one iteration of the distillation protocol in Fig.~\ref{fig:1.6}, the error probability $\varepsilon^{(1)}_{\rm anc}$ for the output state {\em conditioned on acceptance} (i.e., conditioned on getting the ${+}1$ measurement outcome) satisfies the upper bound
\begin{equation}
\label{eq:gad.1.2}
\varepsilon^{(1)}_{\rm anc}
 = {\varepsilon_{\rm anc}^2 \over \varepsilon_{\rm anc}^2 + \left(1-\varepsilon_{\rm anc}\right)^2} \leq 2\varepsilon_{\rm anc}^2
\; ,
\end{equation}
since both input states must have an error for an error in the output state to go undetected\footnote{To obtain eq.$\,$(\ref{eq:gad.1.2}), we used $\eta^2 + (1-\eta)^2 \geq 1/2$, $\forall \eta$. Since in fact $\eta_1 \eta_2 + (1-\eta_1)(1-\eta_2) \geq 1/2$, $\forall \eta_1,\eta_2\le 1/2$, eq.$\,$(\ref{eq:gad.1.2}) applies even when the two ancillas have unequal error probabilities, where $\varepsilon_{\rm anc}$ is the larger error probability.}. 

If we repeat the distillation protocol, the error probability $\varepsilon_{\rm anc}^{(2)}$ after two rounds, conditioned on acceptance, is bounded above by $2 (\varepsilon_{\rm anc}^{(1)})^2$. (The input states in the second round were prepared independently in the first round, so their errors are uncorrelated.) Similarly, after $m$ rounds, 
\begin{equation}
\varepsilon_{\rm anc}^{(m)}\leq \frac{1}{2}(2\varepsilon_{\rm anc} )^{2^m}~,
\end{equation}
which decreases doubly exponentially as long as the initial error probability $\varepsilon_{\rm anc}$ is below the {\em distillation threshold} $\varepsilon_{0}^{\rm dist, |{+}i\rangle} \equiv 1/2$.

Of course, in practice the CSS operations used in the distillation protocol will not be ideal. However, we will show that there is an accuracy threshold $\varepsilon_0^{\rm CSS}$ for CSS operations: if the fault rate is below $\varepsilon_0^{\rm CSS}$ then we can perform logical CSS operations with arbitrarily good accuracy using a concatenated code. Furthermore, we can prepare encoded $|{+}i\rangle$ states with error rate below $\varepsilon_{0}^{\rm dist, |{+}i\rangle}$ using the method of {\em teleportation into the code block}. Therefore, encoded $|{+}i\rangle$ states with a very low error rate can be prepared using the distillation protocol, and highly reliable Clifford group computation is possible, if the fault rate is below an accuracy threshold $\varepsilon_0^{\rm Clif}$ for Clifford operations. Further details will be discussed in Sec.~\ref{sec:ThrClifford}.

\subsubsection{Distillation without twirling}
\label{sec:no-twirling}
In the above discussion of the distillation threshold, we have included in the distillation protocol a ``twirling'' step to induce decoherence in the $Y$-eigenstate basis. This step allows us to describe the deviation from the ideal state in terms of an error probability that can be updated in each round of the protocol. 

In fact, our conclusions concerning the accuracy of the logical $S$ gate apply just as well if the twirling step is omitted. First note that $Y$ gates applied to one or both of the input qubits can be propagated through the circuit shown in Fig.~\ref{fig:1.6}. One finds that these $Y$ gates have no effect on the $X$ measurement, and that for either measurement outcome a $Y$ gate applied with probability $1/2$ to the output qubit is equivalent to $Y$ gates applied with probability $1/2$ to each input qubit (the density operator of the output qubit is the same either way). Invoking this observation for a protocol with many rounds, we conclude that twirling the output qubit after the distillation protocol is complete is equivalent to twirling all of the input qubits before the protocol begins.

We may therefore imagine that the twirl is applied to the input ancilla qubit for the circuit shown in Fig.~\ref{fig:1.5}. By propagating the $Y$ gate through the circuit, we find that a $Y$ gate applied to the input ancilla qubit has the same effect as a $Y$ gate applied to the output qubit, but also flips the outcome of the $Z$ measurement. Taking into account that a $Y$ gate acting on the output qubit is conditioned on the $Z$ measurement outcome, we find that after averaging over the measurement outcomes the density operator of the output qubit is the same, whether or not $Y$ is applied to the input. We conclude, finally, that the output state of the simulated $S$ gate is the same, whether or not twirling is included in the distillation protocol.

%---------------------------------------------------------------------------------------------%
\subsection{Fault-tolerant universal quantum computation}
\label{sec:FTUniversal}

Now that we have seen how to realize fault-tolerant Clifford group computation, we can complete our universal set of quantum gates by adding the non-Clifford gate $T_{\rm y}\equiv \exp(-i{\pi \over 8}\sigma_{\rm y})$. The gate $T_{\rm y}$ can be simulated, as shown in Fig.~\ref{fig:1.7}, using Clifford operations and the ability to prepare the ``magic'' state $|H\rangle \equiv \cos({\pi \over 8})|0\rangle + \sin({\pi \over 8})|1\rangle$, the eigenstate with eigenvalue ${+}1$ of the Hadamard transformation $H$. Therefore, we can achieve universal fault-tolerant quantum computation if we can prepare encoded copies of $|H\rangle$ with a low error rate.

\begin{figure}[htb]
\begin{center} 
\parbox{1cm}{\Qcircuit @C=0.7ex @R=1.8ex @!R { 
   & \push{|\psi\rangle \hspace{0.1cm}} & \qw & \control \qw & \qw & \targ      & \qw & \gate{T_{\rm y}^2} & \qw & \qw & \rstick{T_{\rm y}|\psi\rangle}  \\
   & \push{|H\rangle \hspace{0.1cm}}    & \qw & \ctrl{-1}    & \qw & \ctrl{-1}  & \qw & \meterx \cwx[-1] 
                                }} \vspace{0.3cm}  
\end{center}
\fcaption{\label{fig:1.7} Circuit simulation of the gate $T_{\rm y}$ using the ancilla $|H\rangle$, the ${+}1$ eigenstate of the Hadamard. Note that $T_{\rm y}^2=(SH)S(SH)^\dagger$ is a Clifford group operation.}
\end{figure}  

Just as for the state $|{+}i\rangle$ discussed in Sec.~\ref{sec:FTClifford}, highly accurate copies of $|H\rangle$ can be distilled from noisy copies \cite{bravyi}. Assuming that the Clifford group operations used in the protocol are ideal, distillation will succeed if the initial noisy copies have an error rate below the threshold value $\varepsilon_{0}^{\rm dist, |H\rangle}$. For a protocol that includes ``twirling,'' a lower bound of $14\%$ on the distillation threshold was established by Bravyi and Kitaev in \cite{bravyi}; however, as explained in Sec.~\ref{sec:FTClifford}, including twirling can complicate the analysis of a postselected simulation. Fortunately, we can establish a lower bound on the distillation threshold for a protocol that does not include twirling. Our estimate of the distillation threshold is lower than the estimate found in \cite{bravyi}, but it will suffice for our purposes.

We consider the same protocol as in \cite{bravyi}, which maps 15 noisy input copies to one output copy of the desired state. We suppose that each input is either a perfect copy $\rho_{\rm ideal}$ of the desired state, with probability at least $1-\varepsilon$, or a damaged state $\rho_{\rm junk}$ with probability at most $\varepsilon$ (where the damaged state $\rho_{\rm junk}$ is arbitrary). The Bravyi-Kitaev protocol is based on a distance-3 code with length $n=15$ and $k=1$ encoded qubits. If one or two of the 15 input copies are damaged, then the protocol (which uses only Clifford group operations) is certain to either detect the damage or to project the damaged copies to ideal copies. If errors are detected the input is rejected, and if no errors are detected the code block is decoded to a single output qubit. Thus if the input is accepted then the output is ideal, unless the input copies have (three or more) errors at positions in the code block that can escape detection.

For the 15-qubit code (which is based on the 15-bit classical Hamming code), there are 35 ways to choose a ``bad'' set of three qubits such that errors occuring on those qubits can escape detection, and there are 945 ways to choose a set of four qubits such that no bad set of three is contained in the set of four. Since the probability of acceptance is at least $(1-\varepsilon)^{15}$, we find by applying Bayes's rule that the probability of error in the output qubit, conditioned on acceptance, can be bounded above as
\begin{eqnarray}
\varepsilon_{\rm out} ~\le ~\frac{35 \varepsilon^3 + 945 \varepsilon^4}{(1-\varepsilon)^{15}}~;
\end{eqnarray}
therefore $\varepsilon_{\rm out} < \varepsilon$ for 
\begin{equation}
\varepsilon < .0630~.
\end{equation}
Since the output qubit is either ideal with probability at least $1-\varepsilon_{\rm out}$ or damaged with probability at most $\varepsilon_{\rm out}$, subsequent rounds of the protocol can be analyzed in the same way. Thus we find that, even for a protocol without twirling, the distillation threshold obeys $\varepsilon_{0}^{\rm dist, |H\rangle} > .0630$ --- distillation succeeds if the input copies have a probability of error below $6.3\%$.

If the fault rate is below $\varepsilon_0^{\rm Clif}$, so that Clifford operations can be accurately simulated, and encoded magic states with an error probability below 6.3\% can be prepared by teleporting into the code block, then we can apply an encoded version of the distillation protocol to prepare highly accurate encoded magic states. Details will be discussed in Sec.~\ref{sec:ThrUniversal}.

%---------------------------------------------------------------------------------------------%
\subsection{Fault-tolerant preparation and measurement}
\label{sec:FTPrepMeas}

\begin{figure}[tb]
\begin{center}
\epsfig{file=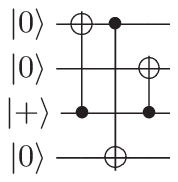,width=1.6cm} \hskip 1.5cm \epsfig{file=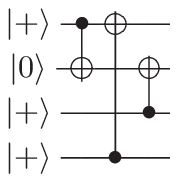,width=1.6cm}
\put(-5.5,1.3){(a)} \put(-2.4,1.3){(b)} \vspace{0.2cm}
\end{center}
\fcaption{\label{fig:1.8} (a) An encoding circuit for $|0\rangle_L |0\rangle_T$. No single fault in this circuit can lead to a $X_L$ error that escapes detection. A fault in either of the last two {\sc cnot} gates may cause the error $X_L\otimes X_T = IXXI$; however this error will be detected by the subsequent 1-zED in the preparation 1-exRec if no other fault occurs. (b) An encoding circuit for $|+\rangle_L |+\rangle_T$. Similarly, no single fault in this circuit can lead to a $Z_L$ error that escapes detection. A fault in either of the last two {\sc cnot} gates may cause the error $Z_L\otimes Z_T = IZZI$, which, if no other fault occurs, will be detected by the subsequent 1-pED in the preparation 1-exRec. }
\end{figure}

Finally, let us discuss how to construct gadgets for preparing the encoded Pauli operator eigenstates $|0\rangle$ and $|+\rangle$ and for measuring the logical Pauli operators $Z$ and $X$. We have already described the fault-tolerant circuits in Fig.~\ref{fig:1.2} that prepare $|0\rangle_L|+\rangle_T$ and $|+\rangle_L|0\rangle_T$. However we might in some cases need to prepare the states $|0\rangle_L|0\rangle_T$ and $|+\rangle_L|+\rangle_T$ such that the logical and gauge qubits have the same value. For this purpose we can use the circuits shown in Fig.~\ref{fig:1.8}.

For the $|0\rangle_L|0\rangle_T$ encoder in Fig.~\ref{fig:1.8}(a), a single fault in either of the last two {\sc cnot} gates can cause the error $IXXI=X_L\otimes X_T$ which can be interpreted as a logical $X_L$ error accompanied by an $X_T$ error on the gauge qubit. If this faulty encoding circuit is followed by a perfect 1-zED gadget (shown in Fig.~\ref{fig:1.4}(a)), then the outcome of the $X_T$ measurement that is included in the 1-zED will be ${-}1$ instead of ${+}1$, and the state will be rejected. Similarly, a single fault in the $|+\rangle_L|+\rangle_T$ encoder shown in Fig.~\ref{fig:1.8}(b) might cause the error $IZZI=Z_L\otimes Z_T$, but this error would be detected by the $Z_T$ measurement in the following 1-pED gadget.

Level-1 measurement gadgets are also simple to construct for the [[4,2,2]] code. A measurement of the logical Pauli operator $Z_L$ can be executed by measuring all the qubits in the code block and then processing the classical measurement outcomes. First, error detection is performed by computing the parity of the measurement outcomes, the eigenvalue of the stabilizer generator $ZZZZ$. If the state of the gauge qubit is known to be $|0\rangle_T$, then an additional parity check can be done, the evaluation of the eigenvalue of the operator $Z_T=ZZII$. Finally, if no errors are detected, the eigenvalue of the logical operator $Z_L=ZIZI$ is computed by taking the parity of the first and third measurement outcomes. 
Measurement of the logical Pauli operator $X_L$ can be executed similarly by measuring all qubits in the $X$ basis and processing the measurement outcomes.

%---------------------------------------------------------------------------------------------%
\section{Postselected quantum computation: threshold analysis}
\label{sec:estimate}

%---------------------------------------------------------------------------------------------%
\subsection{Accuracy threshold for postselected Clifford-group computation} 
\label{sec:post-clifford}

We will see that, once we have established an accuracy threshold $\varepsilon_0^{\rm CSS}$ for the CSS operations, we can appeal to the purification protocols described in Sec.~\ref{sec:FTClifford} and Sec.~\ref{sec:FTUniversal} to show that the accuracy threshold $\tilde\varepsilon_0$ for universal postselected quantum computation is quite close to $ \varepsilon_0^{\rm CSS}$. Therefore, the key step in our threshold estimate is an analysis of the logical {\sc cnot} gate, the only two-qubit gate in our CSS set for the [[4,2,2]] code. 

\begin{figure}[tb]
\begin{center}
\epsfig{file=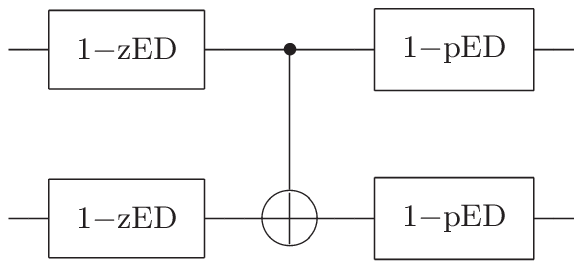,width=4.5cm} \vspace{0.2cm}
\end{center}
\fcaption{\label{fig:1.9} The {\sc cnot} 1-exRec, which consists of the logical {\sc cnot} gate preceded and followed by error-detection gadgets. Here 1-zED and 1-pED denote the 1-ED gadgets shown in Fig.~\ref{fig:1.4}(a) and (b), respectively. In the case shown, the gauge qubit of each input code block has been prepared in the state $|0\rangle_T$, matching the gauge qubit in the first ancilla block of the 1-zED. }
\end{figure}

The {\sc cnot} 1-exRec consists of the transversal {\sc cnot} 1-Ga, preceded by leading 1-EDs acting on each input block and followed by trailing 1-EDs acting on each output block. If the input gauge qubits have been prepared in the state $|0\rangle_T$ (the case shown in Fig.~\ref{fig:1.9}), then the leading error detections are 1-zEDs and the trailing error detections are 1-pEDs, while if the input gauge qubits have been prepared in the state $|+\rangle_T$, then the leading error detections are 1-pEDs and the trailing error detections are 1-zEDs. Either way, the transversal {\sc cnot} acts trivially on the gauge qubits --- the {\sc cnot} preserves $|+\rangle_T|+\rangle_T$ in the first case and $|0\rangle_T|0\rangle_T$ in the second case.

\subsubsection{Decoding circuits}
For the threshold estimate, we are to count the number of malignant pairs of locations inside the {\sc cnot} 1-exRec. Recall that a pair of locations is declared malignant if, for some choice of the Pauli errors at the two locations, at least one of our notions of correctness fails: AA-correctness, AB-correctness, or BB-correctness. (For reasons to be explained in Sec.~\ref{sec:ThrCSS}, we will do a separate count of the malignant pairs of locations that cause {\em weak} correctness to fail.) Each notion of correctness is defined in terms of an ideal 1-*decoder that extracts the syndrome and maps a state in the code space to a single qubit.

\begin{figure}[tb]
\begin{center}
\epsfig{file=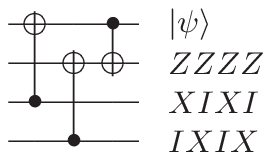,width=2.8cm} \hskip 1.5cm \epsfig{file=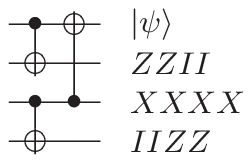,width=2.5cm}
\put(-7.6,1.3){(a)} \put(-3.3,1.3){(b)} 
\vspace{0.2cm}
\end{center}
\fcaption{\label{fig:1.10} (a) The decoding circuit when the state of the gauge qubit is $|+\rangle_T$. This decoder maps the state  $|\psi\rangle_L$ of the encoded qubit to the first qubit in the block, maps the stabilizer generator $ZZZZ$ to $Z$ acting on the second qubit, maps $X_T=XIXI$ to $X$ acting on the third qubit, and maps  $X_T\cdot(XXXX)=IXIX$ to $X$ acting on the fourth qubit. (b) The decoding circuit when the state of the gauge qubit is $|0\rangle_T$. This decoder maps the state  $|\psi\rangle_L$ of the encoded qubit to the first qubit in the block, maps $Z_T=ZZII$ to $Z$ acting on the second qubit, maps the stabilizer generator $XXXX$ to $X$ acting on the third qubit, and maps  $Z_T\cdot(ZZZZ)=IIZZ$ to $Z$ acting on the fourth qubit. }
\end{figure}

For the [[4,2,2]] code there are two 1-*decoders that are shown in Fig.~\ref{fig:1.10}, one for each input state of the gauge qubit ($|+\rangle_T$ or $|0\rangle_T$).  The circuit shown in Fig.~\ref{fig:1.10}(a) is used when the state of the gauge qubit is $|+\rangle_T$; it acts by conjugation according to
\begin{eqnarray}
&& ZZZZ \mapsto IZII\nonumber\\
&& XIXI \mapsto IIXI\nonumber\\
&& IXIX \mapsto IIIX~.
\end{eqnarray}
Therefore, after decoding, a $Z$ measurement of the second qubit extracts the eigenvalue of the stabilizer generator $ZZZZ$, an $X$ measurement of the third qubit extracts the eigenvalue of $X_T=XIXI$, and an $X$ measurement of the fourth qubit extracts the eigenvalue of $X_T\cdot (XXXX)=IXIX$. The circuit shown in Fig.~\ref{fig:1.10}(b) is used when the state of the gauge qubit is $|0\rangle_T$; it acts by conjugation according to
\begin{eqnarray}
&& ZZII \mapsto IZII\nonumber\\
&& XXXX \mapsto IIXI\nonumber\\
&& IIZZ \mapsto IIIZ~.
\end{eqnarray}
Therefore, after decoding, a $Z$ measurement of the second qubit extracts the eigenvalue of the $Z_T=ZZII$, an $X$ measurement of the third qubit extracts the eigenvalue of the stabilizer generator $XXXX$, and a $Z$ measurement of the fourth qubit extracts the eigenvalue of $Z_T\cdot (ZZZZ)=IIZZ$. For each 1-*decoder there is a corresponding 1-*encoder: the time-reversed circuit.

%---------------------------------------------------------------------------------------------%
\subsubsection{Location counting and contracted exRecs}
\label{sec:coexRecs}

Both 1-ED gadgets in Fig.~\ref{fig:1.4} contain %$|{\rm Bell}|=16$ 
16 locations for preparing the two ancilla blocks in a logical Bell state and $12$ locations for the transversal Bell measurements, a total of $|{\rm ED}|=28$ locations. The {\sc cnot} 1-exRec in Fig.~\ref{fig:1.9} therefore contains $|{\rm exRec}|=4\times 28+4= 116$ locations.

We note that in a 1-ED gadget, four {\sc cnot} gates are immediately preceded by preparation steps applied to both qubits, and four {\sc cnot} gates are immediately followed by measurements of both qubits. If we were to follow our usual recursive procedure for constructing a level-$(k{+}1)$ simulation from a level-$k$ simulation (as in Fig.~\ref{fig:level-1-sim}), then each 1-preparation inside the 2-ED would be followed by a 1-ED and each 1-measurement inside the 2-ED would be preceded by a 1-ED. 

\begin{figure}[tb]
\begin{center}
\epsfig{file=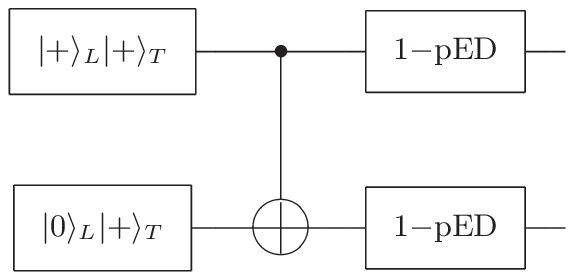,width=4.55cm} \hskip 1.5cm \epsfig{file=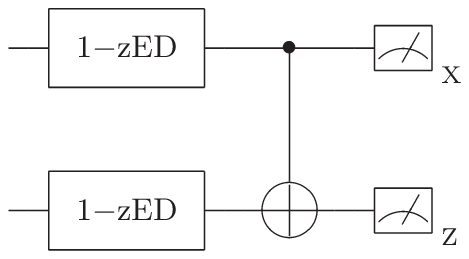,width=4cm}
\put(-11,1.8){(a)} \put(-4.8,1.8){(b)} \vspace{0.2cm}
\vspace{0.2cm}
\end{center}
\fcaption{\label{fig:1.11} Contracted 1-exRecs. (a) The Bell-state preparation 1-exRec. The preparation circuit for $|+\rangle_L |+\rangle_T$ is shown in Fig.~\ref{fig:1.8}(b); $|0\rangle_L |+\rangle_T$ is prepared as in Fig.~\ref{fig:1.2}. (b) The Bell-measurement 1-exRec. The indicated measurements of logical Pauli operators are performed transversally. }
\end{figure}

But we will achieve a mild improvement in our threshold estimate by modifying the simulation \cite{aliferis-terhal}. Inside the 2-ED, we omit the 1-EDs between the 1-preparations and the following 1-{\sc cnot}, and we regard the combination of the 1-preparations and the 1-{\sc cnot} as a single level-1 gadget: the Bell-state preparation gadget. Similarly, we omit the 1-EDs between the 1-{\sc cnot} and the following measurements, and we regard the combination of the 1-{\sc cnot} and the measurements as a single level-1 gadget: the Bell-measurement gadget. The 1-exRecs for the Bell preparation and Bell measurement are shown in Fig.~\ref{fig:1.11}. We will refer to each of these 1-exRecs as a ``contracted $1$-exRec'' or {\em $1$-conexRec}, because it contains fewer 0-locations than the {\sc cnot} 1-exRec. The level-1 combined gadgets are still fault tolerant even though the 1-EDs in the middle are removed; a 1-conexRec that contains just one fault is correct. Furthermore, a 1-conexRec contains fewer malignant pairs of locations than a {\sc cnot} 1-exRec, and is therefore less likely to fail.

The advantage of combining gadgets is that it reduces the number of level-1 locations inside the {\sc cnot} 2-exRec, and also reduces the number of malignant pairs of level-1 locations inside the 2-exRec. Thus, after one level reduction step the number of 0-locations inside a {\sc cnot} 1-exRec is diminished. In effect, the 8 preparations and 8 measurements in each 1-ED are removed, eliminating $4\times 16=64$ locations and leaving $116-64=52$ locations in the 1-exRec.

Note that we count the conexRec as a single location only at level $k=2$ and above. We could use the same strategy at level $1$ as well, but that would not improve the threshold. Inside a 1-exRec, each composite Bell measurement and Bell preparation gadget contains three elementary 0-Ga's; therefore, in our noise model it is about three times as likely to fail as an elementary 0-Ga.

%---------------------------------------------------------------------------------------------%
\subsubsection{The threshold for postselected CSS operations}
\label{sec:ThrCSS}

To estimate an accuracy threshold for postselected computation with CSS gates, we may apply eq.~(\ref{prob-bad-global-malignant}) to the {\sc cnot} 1-exRec, using $|{\rm exRec}|=116$, $|{\rm ED}|=28$, and $d{=}2m{=}4$. The number $\tilde A$ of malignant pairs of locations in the {\sc cnot} 1-exRec can be determined by a computer-assisted combinatorial analysis: For every pair of locations inside the {\sc cnot} 1-exRec, we consider all possible combinations of Pauli operators acting at those two locations, and for each combination we determine whether the 1-exRec is correct. If the 1-exRec is correct for any choice of the Pauli operators, then, since the Pauli operators are an operator basis, it will be correct for arbitrary faults at the specified locations; the pair of locations is benign. Otherwise, there is a choice of Pauli operators for which the 1-exRec is incorrect, and the pair of locations is malignant. Our computer-assisted analysis found in the {\sc cnot} 1-exRec $\tilde{A}=1,306$ malignant pairs, for which either correctness or weak correctness fails.

The threshold estimate can be improved using a few tricks. For example, the upper bound $M_d(s) \leq (ed)^{s-1}$ on the number of clusters, derived in Lemma 5, can be tightened by counting the clusters exactly for small $s$: $M_d(2)=d$ and $M_d(3)={d\choose 2} + d(d-1)= {3\over 2}d(d-1)$. Thus we improve the estimate of $\gamma(\varepsilon)$ in eq.~(\ref{gamma-evaluate}):
\begin{equation}
\label{eq:gad.1.7}
\gamma(\varepsilon) \leq {1\over 1-\omega(\varepsilon)} - \left( 1 - {1\over e} \right)\omega(\varepsilon) - \left(1 - {3(d-1) \over 2 e^2 d}\right)\omega(\varepsilon)^2 \; .
\end{equation}

Another useful observation relates to the distinction between correctness and weak correctness. It turns out that there are many pairs of locations in the {\sc cnot} 1-exRec for which weak correctness fails but correctness does not fail. When we consider a full untruncated 1-exRec rather than a truncated 1-exRec, the number of malignant pairs of locations is only $\hat A=722$ rather than $\tilde A= 1,306$. Furthermore, as we will explain in Sec.~\ref{sec:non-successive}, an incorrect  {\sc cnot} $k$-exRec, for $k>1$, contains either a pair of bad $(k{-}1)$-exRecs, both of which are {\em not} truncated, or it contains three or more bad $(k{-}1)$-exRecs, which may or may not be truncated. This means that after one level reduction step, we can use the correctness criterion rather than weak correctness for the purpose of counting the number of malignant pairs in subsequent steps, which reduces the number of malignant pairs substantially. We therefore formulate separate recursion relations for untruncated and truncated $k$-exRecs to track how the probability of failure evolves as $k$ increases.

And, as we have already discussed in Sec.~\ref{sec:coexRecs}, after one level reduction step we can reduce the number of locations inside the {\sc cnot} 1-exRec (from 116 to 52) and inside the 1-ED (from 28 to 12) by combining gadgets to form 1-conexRecs.  Repeating the counting of malignant pairs of locations for the case where a contracted gadget counts as a single location, we find $\tilde A'=550$ pairs for 1-exRecs that may or may not be truncated, and $\hat A'= 336$ for untruncated 1-exRecs. Let us use $\varepsilon^{(k)}$ to denote the failure probability for a 1-conexRec or 1-exRec that may or may not be truncated, and $\tilde \varepsilon^{(k)}$ to denote the failure probability for an untruncated 1-conexRec or 1-exRec. Then for $k>1$ the recursion relation for $\varepsilon^{(k)}$ becomes
\begin{equation}
\label{eq:gad.1.9}
\varepsilon^{(k)} \leq \frac{\tilde{A}'\left( \gamma'(\varepsilon^{(k-1)}) \right)^2 \left(\tilde{\varepsilon}^{(k-1)}\right)^2}{C'(\varepsilon^{(k-1)})} + \frac{\tilde{B}'\left(\gamma'(\varepsilon^{(k-1)})\right)^4 \left(\varepsilon^{(k-1)} \right)^3}{D'(\varepsilon^{(k-1)})} \; .
\end{equation}
Here $\tilde B'$ and the functions $\gamma'$, $C'$, $D'$  are similar to $\tilde B$, $\gamma$, $C$, and $D$ in Sec.~\ref{sec:malignant}, except that we make the replacements
\begin{equation}
|{\rm exRec}|\rightarrow |{\rm exRec}'|=52~,\quad   |{\rm ED}|\rightarrow |{\rm ED}'|=12
\end{equation}
to take into account the reduced size of the contracted gadgets, and we use the improved version of $\gamma(\varepsilon)$ from eq.~(\ref{eq:gad.1.7}). 
Note that $\left(\tilde{\varepsilon}^{(k-1)}\right)^2$ rather than $\left( {\varepsilon}^{(k-1)}\right)^2$ appears in the numerator of the first term on the right-hand side, reflecting our observation that the exRecs that constitute a malignant pair must be untruncated. The recursion relation for $\tilde \varepsilon^{(k)}$ is similar, except for the replacement $\tilde A'= 550 \rightarrow \hat A'= 336$.

\begin{figure}[t]
\begin{center}
\hspace{1.8cm} \epsfig{file=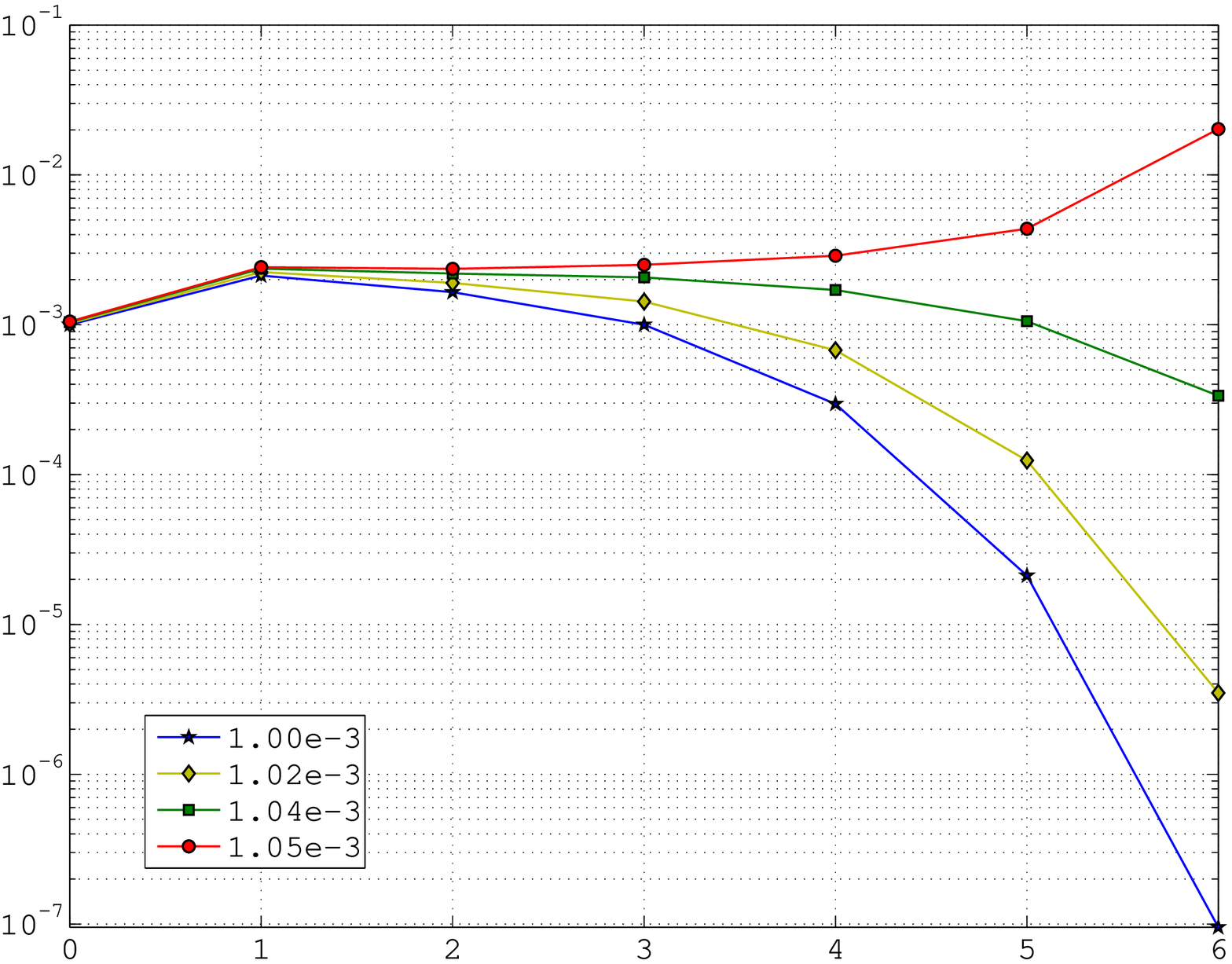,width=11.5cm}
\put(-12.85,4.15){\parbox{2cm}{\footnotesize Noise strength, $\varepsilon^{(k)}$, for level-$k$ postselected CSS operations}}
\put(-7,0.3){\parbox{3cm}{\footnotesize Concatenation level $k$}}
\end{center}
\fcaption{\label{fig:results} The effective strength $\varepsilon^{(k)}$ of locally correlated stochastic noise for level-$k$ postselected CSS operations, as a function of $k$, for various values of the physical level-0 error probability $\varepsilon$ (shown in the legend). The plot demonstrates that the threshold error probability is above $1.04\times 10^{-3}$. } 
\end{figure}

Given a bound $\varepsilon$ on the physical error probability at level-0, we can iterate our recursion equations to find an upper bound on the effective noise strength for level-$k$ postselected CSS operations, as a function of the concatenation level $k$. Fig.~\ref{fig:results} shows the result of this iteration; we identify
\begin{equation}
\label{eq:threshold}
\varepsilon < \varepsilon^{\rm CSS}_{0} \; ; \; {\rm with} \; \varepsilon^{\rm CSS}_{0} \geq 1.04\times 10^{-3}
\end{equation}  
\noindent as the accuracy threshold condition for postselected CSS computation. This analysis applies to both the {\sc cnot} 1-exRec shown in Fig.~\ref{fig:1.9}, and also to the  {\sc cnot} 1-exRec with the 1-zED and 1-pED gadgets interchanged. 

%---------------------------------------------------------------------------------------------%
\subsubsection{The threshold for postselected Clifford-group operations}
\label{sec:ThrClifford}

As discussed in Sec.~\ref{sec:FTClifford}, we can now add the phase gate $S$ to $\mathcal{G}_{\rm CSS}$ to obtain a gate set that generates the Clifford group. Because the $S$ gate is not used inside ED gadgets, it is not necessary to simulate $S$ at each level of concatenation. Rather we need to simulate $S$ only at the highest level of concatenation, and for that purpose it suffices to prepare the ancilla state $|{+}i\rangle$ encoded at the highest level.
If the fault rate $\varepsilon$ is below the threshold  $\varepsilon_{0}^{\rm CSS}$ for CSS computation, so that arbitrarily accurate CSS operations can be simulated, then we can distill very clean encoded $|{+}i\rangle$ states from noisy encoded states, provided the error rate $\varepsilon_{\rm anc}$ in the noisy encoded states is below the distillation threshold $\varepsilon_{0}^{\rm dist, |{+}i\rangle} = 1/2$. 

\begin{figure}[tb]
\begin{center}
\epsfig{file=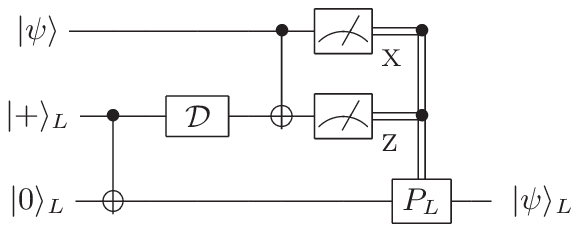,width=5.2cm}
\vspace{0.2cm}
\end{center}
\fcaption{\label{fig:1.12} Teleporting the single-qubit state $|\psi\rangle$ into the code block. First, two ancilla blocks are prepared in a logical Bell state, then the first ancilla block is decoded with the decoding circuit $\mathcal{D}$. Finally, a Bell measurement is performed on an input qubit and the output qubit from the decoder. After the logical Pauli  correction ($P_L$) is applied, the state of the output code block is the encoded state $|\psi\rangle_L$ as desired. }
\end{figure}

A general method for mapping a single-qubit state to the code space of a quantum code is to ``teleport into the code block'' \cite{knill_detect} using the circuit shown in Fig.~\ref{fig:1.12}. First a {\em logical} Bell state is prepared, then the first ancilla block is decoded to a single qubit, and finally a Bell measurement is performed on the decoded qubit and the input qubit. After applying the {\em logical} Pauli operator that completes the teleportation protocol, the state $|\psi\rangle$ of the input qubit has been mapped to the corresponding encoded state $|\psi\rangle_L$.

We can use this teleportation method to prepare the (noisy) encoded $|{+}i\rangle$ states that are to be distilled. Let $k$ be the number of levels of concatenation that are necessary so that the postselected CSS operations of the purification protocol are sufficiently accurate. Then, the input states for the distillation protocol are to be encoded using the [[4,2,2]] code $C_1$ concatenated $k$ times with itself, denoted $C_1^{\circ k}$. To assess the accuracy of the circuit in Fig.~\ref{fig:1.12} we regard it as a level-$k$ open circuit, and we perform level reduction $k$ times to obtain an equivalent level-0 circuit, as described in Sec.~\ref{sec:level-reduction} and Sec.~\ref{sec:threshold}. The probability $\varepsilon_{\rm anc}$ that this reduced circuit contains a fault satisfies the upper bound
\begin{equation}
\label{eq:gad.1.11}
\varepsilon_{\rm anc}~ \leq ~3\varepsilon^{(k)} + \varepsilon_{\rm dec}^{(k)}+4\varepsilon \; .
\end{equation}

\noindent Here $3\varepsilon^{(k)}$ is an upper bound on the  error probability for the preparation of the postselected level-$k$ encoded Bell state (which involves three CSS operations --- two level-$k$ state preparations and one level-$k$ {\sc cnot} gate), $\varepsilon_{\rm dec}^{(k)}$ is an upper bound on the probability of an undetected error arising during decoding, and $4\varepsilon$ is an upper bound on the probability of an  error during the preparation of the input single-qubit state $|{+}i\rangle$ or during the Bell measurement (one {\sc cnot} gate and two single-qubit measurements). We emphasize that $\varepsilon_{\rm anc}$ in eq.~(\ref{eq:gad.1.11}) is the probability that the output code block has an encoded error {\em conditioned on} detecting no errors in all the ED gadgets at all levels of concatenation inside the entire teleportation circuit in Fig.~\ref{fig:1.12}.

\begin{figure}[tb]
\begin{center}
\hspace{1.8cm} \epsfig{file=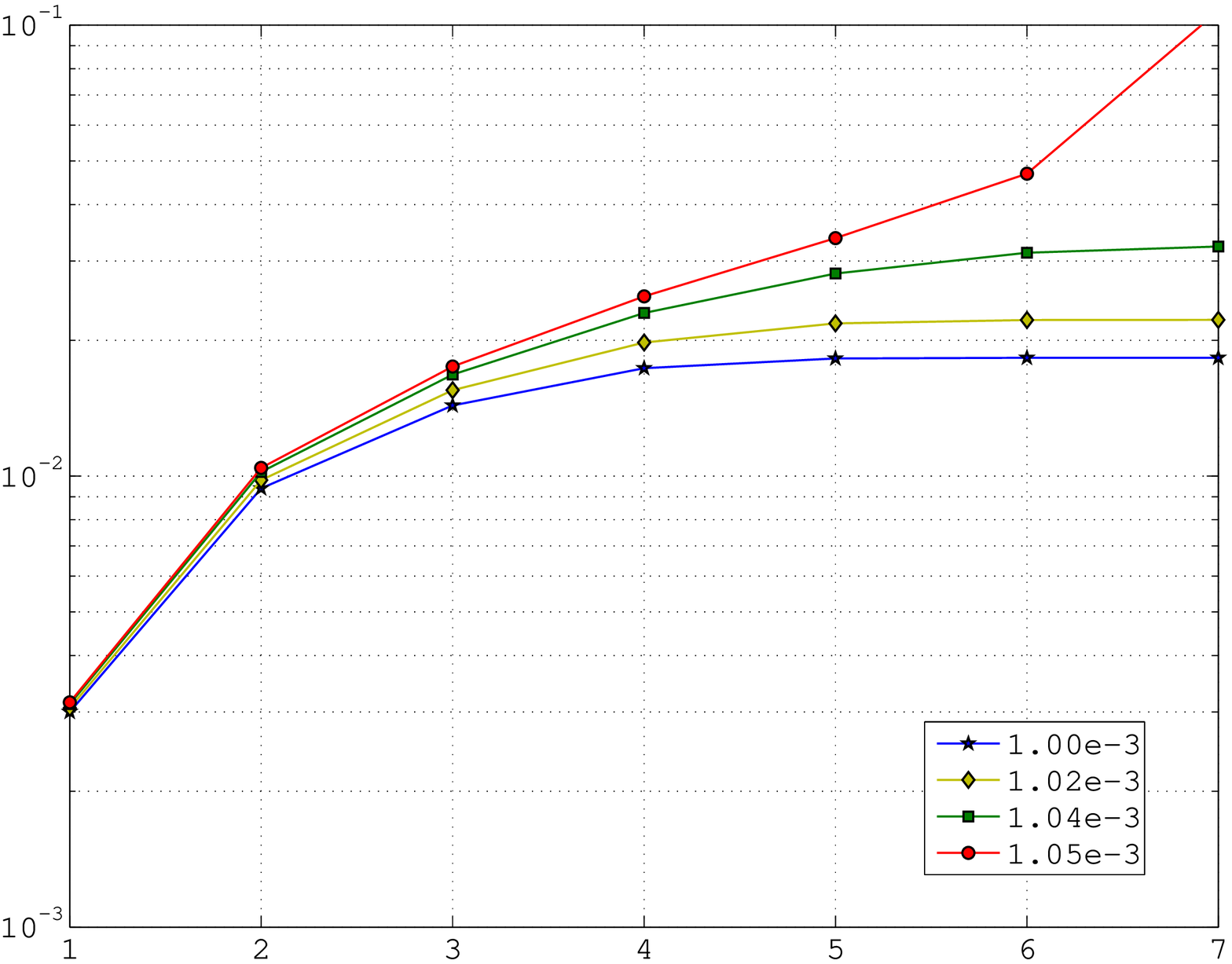,width=11.5cm}
\put(-13,6){\parbox{2.3cm}{\footnotesize Conditional error probability, $\varepsilon_{\rm dec}^{(k)}$, due to decoding}}
\put(-7,0.3){\parbox{3cm}{\footnotesize Concatenation level $k$}}
\end{center}
\fcaption{\label{fig:decoding} Decoding error probability $\varepsilon_{\rm dec}^{(k)}$, conditioned on acceptance by all error detections at all levels, as a function of the concatenation level $k,$ for various values of the physical level-0 error probability $\varepsilon$ (shown in the legend). For $\varepsilon < 1.04\times 10^{-3}$, $\varepsilon_{\rm dec}^{(k)}$ converges to an asymptotic value as $k\to\infty$. The values of $\varepsilon_{\rm dec}^{(k)}$ for $k=7$ shown here are very close to the asymptotic values; this was verified by considering larger values of $k$  that are not shown in the plot. } 
\end{figure}

As discussed in Sec.~\ref{sec:threshold}, the probability of an undetected decoding error satisfies the upper bound
\begin{equation}
\label{eq:gad.1.12}
\varepsilon_{\rm dec}^{(k)} \leq D \sum\limits_{j=0}^{k-1} \varepsilon^{(j)} \; ;
\end{equation}

\noindent here $D=3$ is the number of (CSS) locations in the decoding circuits shown in Fig.~\ref{fig:1.10}. In Fig.~\ref{fig:decoding} we have plotted the decoding error probability $\varepsilon_{\rm dec}^{(k)}$ as a function of the concatenation level $k$. When the physical level-0 fault rate $\varepsilon$ is below $1.04\times 10^{-3}$, we find $\varepsilon_{\rm dec}^{(k)}<3.3\%$ for all $k$. If $\varepsilon$ is also below the threshold $\varepsilon_0^{\rm CSS}\approx 1.04\times 10^{-3}$ for postselected CSS operations, then $\varepsilon^{(k)}$ becomes arbitrarily small for $k$ sufficiently large, and eq.~(\ref{eq:gad.1.11}) then implies that $\varepsilon_{\rm anc}$ is less than $\varepsilon_{\rm dec}^{(k)}+4\varepsilon < .038$, which is definitely below the distillation threshold  $\varepsilon_{0}^{\rm dist, |{+}i\rangle} = 1/2$. Therefore state distillation succeeds, and our estimated accuracy threshold $\varepsilon^{\rm Clif}_{0}$ for postselected Clifford-group computation is very nearly the same as the threshold for postselected CSS computation:
\begin{equation}
\label{eq:threshold-2}
\varepsilon^{\rm Clif}_{0} \geq 1.04\times 10^{-3} \; .
\end{equation}  
(Actually $\varepsilon^{\rm Clif}_{0}$ is slightly below $\varepsilon_0^{\rm CSS}$, but the difference is quite small.)

To ensure that our upper bound on the probability of a decoding error is rigorous for all $k$, we bound $\varepsilon_{\rm dec}^{(k)} $ up to a specified finite value of $k$ by iterating our recursion relations by hand, and then bound the remaining terms that contribute to $\varepsilon_{\rm dec}^{(\infty)} $ using an analytic argument. The remainder we wish to bound is
\begin{equation}
\label{eq:add-1}
\Delta \varepsilon_{\rm dec}^{(k,\infty)}\equiv  \varepsilon_{\rm dec}^{(\infty)}-\varepsilon_{\rm dec}^{(k)}\leq D \sum\limits_{j=k}^{\infty} \varepsilon^{(j)} \;.
\end{equation}
From eq.~(\ref{eq:gad.1.9}), using $D'(\varepsilon) \le C'(\varepsilon)$ and $\tilde\varepsilon^{(k)} \le \varepsilon^{(k)}$, we find that for $j\geq k$ 
\begin{equation}
\label{eq:add-2}
\varepsilon^{(j)}\leq \Gamma^{-1} \left( \Gamma \varepsilon^{(k)} \right)^{2^{j-k}} \; ,  
\end{equation}
\noindent where 
\begin{equation}
\label{eq:add-3}
\Gamma \equiv {\tilde{A}' \left( \gamma' \left( \varepsilon^{(k)} \right) \right)^2 + \tilde{B}' \left( \gamma' \left( \varepsilon^{(k)} \right) \right)^4 \varepsilon^{(k)} \over D' \left( \varepsilon^{(k)} \right) } \; .
\end{equation}

\noindent Inserting Eq.~(\ref{eq:add-2}) into Eq.~(\ref{eq:add-1}) and shifting the lower limit of the summation, we obtain
\begin{equation}
\label{eq:add-4}
\Delta \varepsilon_{\rm dec}^{(k,\infty)} \leq D\Gamma^{-1} \sum\limits_{m=0}^{\infty} \left( \Gamma \varepsilon^{(k)} \right)^{2^m} 
\leq D\varepsilon^{(k)} \sum\limits_{n=0}^{\infty}  \left( \Gamma \varepsilon^{(k)} \right)^{n} 
\leq  D\varepsilon^{(k)}\left( 1- \Gamma \varepsilon^{(k)}  \right)^{-1} \;.
\end{equation}

\noindent For $\varepsilon = 1.04\times 10^{-4}$ and $k=8$, we find by explicitly iterating the recursion relations that $\varepsilon_{\rm dec}^{(7)}\leq 3.23\times 10^{-2}$ and $\varepsilon^{(8)} \leq 1.64\times 10^{-7}$. Substituting into eq.~(\ref{eq:add-3}), we find $\Gamma \leq 551$; this upper bound is quite close to $\tilde{A}' = 550$ because the corrections that are higher order in $\varepsilon^{(8)} $ are quite small. Thus, for $D=3$, eq.~(\ref{eq:add-4}) yields $\Delta \varepsilon_{\rm dec}^{(8,\infty)} \leq  10^{-6}$ and so we conclude that
\begin{equation}
\label{eq:add-5}
\varepsilon_{\rm dec}^{(\infty)} \leq \varepsilon_{\rm dec}^{(7)} + \Delta\varepsilon_{\rm dec}^{(8,\infty)} \leq .033 
\end{equation}
for $\varepsilon < 1.04 \times 10^{-3}$.
%---------------------------------------------------------------------------------------------%
\subsubsection{A consecutive pair of locations cannot be malignant}
\label{sec:non-successive}

In formulating the recursion relation in Sec.~\ref{sec:ThrCSS}, we used the property that an incorrect  {\sc cnot} 2-exRec contains either a pair of bad 1-exRecs, both of which are {\em not} truncated, or contains three of more bad 1-exRecs, which may or may not be truncated. This property can be justified by two observations: (i) A consecutive pair of locations in a 1-exRec cannot be malignant. (ii) Suppose that a 1-exRec occuring in a final step of a 2-exRec is immediately preceded by good 1-exRecs, but might be followed by a bad 1-exRec that is contained in the following 2-exRec. 
Then we may modify our rules for carving bad clusters so that the 1-exRec is untruncated --- doing so will not lead us to underestimate the probability that any 2-exRec is bad.

Observation (i) is almost obvious. For an encoded error to occur, there must be errors acting on each of two distinct qubits in the same code block. Therefore, a second fault afflicting a qubit that has already been damaged in the previous time step cannot cause an encoded error. Nearly all of the gates in the {\sc cnot} 1-exRec act transversally --- that is, each gate acts on only a single qubit in each code block. The only exceptions are the {\sc cnot} gates in the second time step of the 1-ED gadget shown in Fig.~\ref{fig:1.4}; however, as we argued in Sec.~\ref{sec:FTED}, each of these {\sc cnot} gates prepares a two-qubit Bell state, and therefore in effect a fault in the {\sc cnot} damages only one of the two qubits in the code block. Therefore, two consecutive faults can damage only one qubit in a code block, and cannot produce an encoded error.

There are a few subtleties; one arises when we recall that, to assess whether a 1-exRec satisfies AB-correctness or BB-correctness, we are to consider a decoder-encoder pair (which may have a nontrivial syndrome) inserted between a leading 1-ED and the 1-Ga that follows. What if the two consecutive faulty locations are a {\sc cnot} gate in the last step of the 1-ED (before the decoder-encoder pair) and a following {\sc cnot} gate in the 1-Ga (after the decoder-encoder pair)? If the preceding 1-exRec is actually correct (or weakly correct), then the input to the shared 1-ED is a codeword, and we may argue as before that the consecutive faults cannot produce an encoded error. Otherwise the preceding 1-exRec is incorrect. In that case, though, it is harmless to absorb an encoded operation into the bad (incorrect) 1-exRec, and this operation can be chosen so that the encoder's syndrome (if nontrivial) has its support on the same qubit that is damaged by the following fault in the 1-Ga. Again, it follows that the faulty 1-Ga cannot produce an encoded error.

We should also consider what happens if the two consecutive faulty locations are a gate in the last step of the 1-Ga and a following gate in the first step of the trailing 1-ED. Strictly speaking, this pair of locations is not benign according to the criterion we have used up to now; the output of the 1-Ga might have an error due to the first fault, and the error might escape detection due to the second fault. However, in that case the {\em output} of the trailing 1-ED is sure to be a codeword, and to agree with the output of an ideal 1-exRec. Without changing the result of the simulation, we can absorb the noisy trailing 1-ED into the 1-Ga and insert a following very good 1-ED that may now be regarded as the trailing 1-ED of an augmented 1-exRec. This augmented 1-exRec is correct, and so the pair of locations can be declared benign.

Observation (ii) requires more discussion.  Let us consider estimating the probability of badness for a particular set $\mathcal{I}^{(2)}_{r}$ of $r$ specified 2-exRecs, and let us identify one specific 2-exRec $R_1\in\mathcal{I}^{(2)}_r$, whose badness can be attributed to two bad 1-exRecs at a malignant pair of locations inside $R_1$. Each bad 1-exRec in the malignant pair is untruncated unless it is followed by another bad 1-exRec. Observation (i) has told us that these two bad 1-exRecs cannot be consecutive.  Therefore, if there are only two bad 1-exRecs in $R_1$ and one of these bad 1-exRecs is truncated, then there must a third bad 1-exRec that immediately follows the truncated 1-exRec, contained in a 2-exRec $R_2$ that succeeds $R_1$. We distinguish two cases depending on whether $R_2\in\mathcal{I}_r^{(2)}$ or not.

\begin{description}

\item[Case 1: $R_2\not \in\mathcal{I}^{(2)}_r$.] In this case there is no harm in taking the bad 1-exRec in $R_1$ to be untruncated because the following bad 1-exRec in $R_2$ is not relevant for the purpose of estimating the probability that $\mathcal{I}^{(2)}_r$ is bad. More formally, when we estimate the probability that the specified set $\mathcal{I}^{(2)}_r$ is bad, it suffices to ``carve'' bad clusters by starting at the rear edge of $\mathcal{I}^{(2)}_r$, rather than (as in Sec.~\ref{sec:carving}) at the rear edge of any larger bad clusters that extend into the future of $\mathcal{I}^{(2)}_r$. With this modified procedure the bad 1-exRec in $R_1$ is untruncated.

\item[Case 2: $R_2 \in {\cal I}_r^{(2)}$.] Note that in this case, since we are considering both $R_1$ and $R_2$ to be bad, $R_1$ is a truncated 2-exRec. Thus the bad 1-exRec in $R_2$ acts on a 1-block in the input 2-block to a 2-ED --- by inspecting the error detection circuit in Fig.~\ref{fig:ED-w-BM}, we see that it must be a Bell measurement 1-exRec. We will argue that this bad Bell measurement 1-exRec actually makes no contribution to the badness of $R_2$, and hence that it is harmless to regard the bad 1-exRec in $R_1$ as untruncated.

To see this, we will modify our procedure for analyzing bad clusters by introducing yet another type of good* exRec, which we denote by good$'''$. Note that when we consider a good* exRec that follows a bad exRec, we never need to use A-correctness, since the ${\cal D}{\cal D}^{-1}$ pair is inserted after the exRec's leading 1-ED. Under our previously stated rules for analyzing bad clusters, a bad cluster might contain a consecutive pair of exRecs where the later bad exRec violates A-correctness but not B-correctness. If so, there is really no need to classify the later exRec as bad. Therefore, let us augment our procedure: after classifying the exRecs into bad/good$'$/good$''$ according to the previous rules, we sweep forward through the bad cluster and look for cases where an exRec that follows a bad exRec violates A-correctness but not B-correctness. Such exRecs are reclassified as  good$'''$. In the case of a two-qubit location, a good$'''$ exRec respects all of the properties AB-correctness, BB-correctness, weak AB-correctness, and weak BB-correctness.

With this new criterion for badness, $R_2$ must violate one of the B-correctness properties in order to be bad. Let us say a set of locations is ``B-malignant'' if faults at those locations can violate one of the B-correctness properties. Now we claim that any B-malignant set is still B-malignant if we remove any Bell measurements in the leading EDs. This is true because a change in the {\em logical} state that occurs in the leading ED is irrelevant to B correctness.  A single error caused by a fault in the leading ED {\em is} potentially relevant, because it can propagate forward and combine with errors caused by later faults to produce an undetectable logical error. However, for the ED circuit shown in Fig.~\ref{fig:ED-w-BM} only faults during the ancilla preparation can propagate forward; faults in the Bell measurement cannot, and are therefore inessential to the violation of B-correctness.

Since the bad 1-exRec in $R_2$ that succeeds the bad 1-exRec in $R_1$ is a Bell measurement 1-exRec in the leading 2-ED of $R_2$, it actually makes no contribution to the badness of $R_2$. The badness of $R_2$ is unaffected if we consider the Bell measurement 1-exRec to be good*, and accordingly we may consider the bad 1-exRec in $R_1$ to be untruncated.

\end{description}

To summarize, observations (i) and (ii) mean that when a 2-exRec is bad because it contains a malignant pair of 1-exRecs, we may consider both of the bad 1-exRecs to be untruncated. That is the property we used to derive eq.~(\ref{eq:gad.1.9}).

%---------------------------------------------------------------------------------------------%
\subsection{Accuracy threshold for postselected universal quantum computation} 
\label{sec:ThrUniversal}

We have now shown that if the fault rate $\varepsilon$ satisfies $\varepsilon < \varepsilon_0^{\rm Cliff}\approx 1.04\times 10^{-3}$, then postselected Clifford-group computation can be performed to any desired accuracy. To complete our analysis of the accuracy threshold for universal quantum computation, we must show that the non-Clifford gate $T_{\rm y}$ can also be performed to arbitrary accuracy. As discussed in Sec.~\ref{sec:FTUniversal}, it is sufficient to be able to prepare a highly accurate encoded $|H\rangle$ state, which will be possible if we can prepare a noisy encoded $|H\rangle$ state with an error probability below the distillation threshold $\varepsilon_{0}^{\rm dist, |H\rangle} \geq .063$.

The noisy encoded $|H\rangle$ state can be prepared using ``teleportation into the code block,'' and we can estimate the accuracy of the preparation just as in Sec.~\ref{sec:ThrClifford}. Assuming that we can prepare a single-qubit $|H\rangle$ state with error probability no larger than $\varepsilon$, the analysis is exactly the same as before, and we conclude that the encoded $|H\rangle$ state has error probability less than $.038$ for $\varepsilon < 1.04\times 10^{-3}$. Since this error probability is less than the distillation threshold  $\varepsilon_{0}^{\rm dist, |H\rangle}> .063$, the distillation protocol succeeds and reliable universal postselected quantum computation is possible. Thus we have derived a new lower bound on the accuracy threshold $\varepsilon_0$ for universal postselected quantum computation:
\begin{equation}
\label{eq:threshold-3}
\varepsilon_{0} \geq 1.04\times 10^{-3} \; .
\end{equation}  
By counting malignant pairs of locations and employing the other tricks we have described here, we have improved our previous estimate  eq.~(\ref{epsilon0-threshold-number}) by a factor of about 7.4.

Finally, as in Sec.~\ref{sec:threshold}, we consider decoding the $C_1^{\circ k}$ blocks to complete the preparation of the $C_2$ encoded states. The errors occuring in the Bell measurement of two $C_2$ code blocks can be regarded as independent with error probability $\varepsilon'$ given by eq.~(\ref{epsilon-bell-infinity-threshold}):
\begin{equation}
\label{eq:gad.1.13}
 \varepsilon' = \varepsilon_{\rm Bell}^{(\infty)} ~\leq ~2\times (.033) + 5\times (1.04\times 10^{-3}) ~< ~.072 \;.
\end{equation}  

\noindent If for example we take $C_2$ to be the 5-qubit code concatenated five times, the probability of an encoded error in a Bell measurement is no more than $\frac{1}{10}\left(10\varepsilon'\right)^{2^5}< 2.8\times 10^{-6}$; hence the probability of a fault in a two-qubit gate (which involves two Bell measurements) is no more than $5.6\times 10^{-6}$, which is below the accuracy threshold $\varepsilon_{\rm th} > 1.9\times 10^{-4}$ that can be achieved using quantum error-{\em correcting} codes. We have shown at last, then, that  $1.04\times 10^{-3}$ is a lower bound on the accuracy threshold attainable using error detection and postselection.

%---------------------------------------------------------------------------------------------%
\section{Overhead}
\label{sec:overhead}

The main goal of this paper has been to use Knill's postselection method to establish a rigorous lower bound on the quantum accuracy threshold, improving on previous rigorous estimates of the threshold. We have not done detailed studies of the resource requirements for our fault-tolerant simulations. Since the objective of the postselected part of the computation is not to achieve an arbitrarily small effective fault rate, but rather to reach a fault rate below a fixed target value, the additional resource cost incurred by using Knill's method is a ``mere'' multiplicative constant --- that is, it is a multiplicative factor that does not depend on the size of the quantum circuit to be executed. However, as Knill has emphasized \cite{knill_detect}, this constant is quite large when the noise strength is close to the threshold value, and therefore the value of the constant is highly relevant to assessing the feasibility of the method. 

Knill has also described a different scheme for fault-tolerant quantum computation, the ``Fibonacci scheme,'' which has a more modest overhead cost \cite{knill_detect}. This scheme is based on the observation that a distance-2 code, aside from detecting errors, can also correct errors that occur at known positions in the code block. Thus, in the Fibonacci scheme, ancillas are not discarded when errors are detected; rather the positions of the blocks with detected errors are recorded, and this information is used to attempt to correct the located errors using the code at the next level up. (We call it the Fibonacci scheme because the recursion relation for the probability of failure involves a Fibonacci sequence.) According to Knill's numerics, this more efficient scheme also has a threshold above 1\%. However, the threshold analysis we have reported in this paper applies not to the Fibonacci scheme, but rather to the more wasteful scheme in which ancillas with detected errors are discarded. Finding a rigorous lower bound on the accuracy threshold that is attainable with the  Fibonacci scheme (and thus with a less demanding resource cost) is an interesting open problem.

Here we will just make a few simple and general remarks about the constant overhead factor for the more wasteful version of Knill's postselection method, without attempting any detailed estimates. And let us recall that we are only considering the cost of using error detection to realize gates of adequate fidelity that can then be plugged into a ``traditional'' fault-tolerant scheme based on quantum error-{\em correcting} codes; thus this constant factor multiplies the resource requirements for the ``traditional'' scheme, estimated in, for example, \cite{AGP}.

Furthermore, for our sketchy discussion we will consider only the postselected simulation of the CSS gates. The postselection scheme may also require (off-line) state distillation in order to boost from CSS computation to universal quantum computation, and the number of rounds needed for successful distillation depends on how noisy the initial copies are. We must prepare enough initial copies to support the needed number of rounds of the distillation protocol; however we will not consider this feature of the overhead estimate in any detail. 

Rather let us merely observe that the postselection scheme is costly because ancilla preparation needs to be repeated many times before an ancilla is finally accepted. We may imagine that many ancillas are prepared in parallel, sufficiently many so that at least one preparation is likely to be accepted. If in each attempt the probability of acceptance is $p\ll 1$, then if $N$ attempts are made in parallel the probability that all attempts fail is $(1-p)^N\approx e^{-pN}$. We should choose $N$ large enough so that this probability is small compared to other sources of error in our simulation.

Recall that the ancilla is accepted only if no error is detected at any level. The ancilla will certainly be accepted if there are not any faults in any gates. Let $L$ denote the total size of the preparation circuit for a particular $C_2$ ancilla state, including the complete $C_1^{\circ k}$ decoding step. Then, if the fault rate is $\varepsilon$, the probability $p$ of acceptance obeys
\begin{equation}
p \ge \left(1-\varepsilon\right)^{L}\approx e^{-\varepsilon L}~.
\end{equation}
But $L$ depends on $k$, the number of levels of concatenation. How should $k$ and  $L$ scale with the strength of the noise? 

We need to choose $L$ large enough so that our simulated logical $C_2$ gates have a fault rate below $\varepsilon_{\rm th} = 1.9\times 10^{-4}$ \cite{aliferis-cross}. The simulated gate could have a fault for either one of two reasons. One of the postselected $C_1^{\circ k}$ gates in the $C_2$ preparation circuit might fail. Or a logical $C_2$ error might arise because more errors occur in a $C_2$ block during the decoding of the $C_1^{\circ k}$ block and  the following Bell measurement than $C_2$ can correct. Let us take it for granted that $C_2$ has been chosen so that the latter type of error is sufficiently unlikely, and therefore we only need to worry about the first type. 

Suppose the $C_2$ preparation circuit has $L_2$ gates, each simulated using postselected computation with the code $C_1^{\circ k}$. If we want the failure probability of each gate to be below $10^{-4}/L_2$, we choose the level $k$ of concatentation to be such that 
\begin{equation}
2^k\approx \frac{\log (10^4 \varepsilon_0 L_2)}{\log (\varepsilon_0/\varepsilon)}~,
\end{equation}
(where $\varepsilon_0$ is the threshold for postselected CSS computation) so that the size of the $C_2$ ancilla preparation circuit is
\begin{equation}
\label{L-estimate}
L\approx L_2 \left(\frac{\log (10^4 \varepsilon_0 L_2)}{\log (\varepsilon_0/\varepsilon)}\right)^{\log_2\ell}~.
\end{equation}
Here $\ell$ is the maximal number of locations in any $C_1$ 1-Rec ($\ell=2\times 28+4=60$ and $\log_2\ell\approx 5.91$ for the simulation we have described). 

The number of gates needed for the preparation of each $C_2$ ancilla scales like $NL$, because the preparation is attempted $N$ times and $L$ gates are used each time. $N$ is chosen so that the probability $e^{-pN}$ of failing in all $N$ attempts is also less than $10^{-4}$, or 
\begin{equation}
\label{NL-estimate}
NL ~>~ p^{-1}\cdot \ln(10^4)\cdot L\approx (9.2)\cdot Le^{\varepsilon L}~.
\end{equation}
Combining eq.~(\ref{L-estimate}) with eq.~(\ref{NL-estimate}), we have a crude estimate of the constant overhead factor. 
It is a ``constant'' in the sense that it does not depend on the size of the quantum circuit to be executed, but it diverges as $\varepsilon$ approaches the threshold from below (and it is a discouraging constant when $\varepsilon$ is close to $\varepsilon_0$). To optimize the overhead it is important to keep the $C_2$ ancilla preparation circuits as efficient as possible. Note that these preparation circuits need not be fault tolerant --- we are assuming that a single bad gate in the circuit causes it to fail. 

In principle there is a tradeoff between the overhead cost of the postselected part of the computation and the overhead associated with, e.g., the concatenated Bacon-Shor quantum error-correcting code \cite{aliferis-cross} that protects against errors in the encoded $C_2$ computation. But the overhead cost of the error-detecting postselected computation is so high compared to the overhead cost of the error-correcting simulation that in the optimal construction the noise in the logical $C_2$ gates should be only slightly below the error correction threshold ($\approx 10^{-4}$) that can be attained using quantum error-correcting codes.

\section{Conclusions}
\label{sec:conclusions}

In this paper, following Knill's perceptive suggestion \cite{knill_detect}, we have studied fault-tolerant simulations of quantum circuits based on concatenated error-{\em detecting} codes and postselection. We have proved that there is an accuracy threshold $\tilde \varepsilon_0$ for postselected quantum computation, and we have derived a lower bound on the value of this threshold: $\tilde \varepsilon_0 > 1.04\times 10^{-3}$.

To prove this result we must analyze the failure probability of simulated quantum gates conditioned on acceptance (i.e., given that no errors are detected during the simulation). The analysis shows that if the noise strength $\varepsilon$ is less than $\tilde\varepsilon_0$, then by choosing the level $k$ of the concatenated code sufficiently large we can make the conditional probability of failure per gate arbitrarily small. When $\varepsilon$ is small enough, faults are sufficiently sparse that errors are almost sure to be detected before they can accumulate to cause an encoded error at the code's top level.

If we use error detection and postselection to simulate a large circuit, then we can be confident in the reliability of the simulation only if every simulated gate accepts, and unfortunately the probability of acceptance by every gate declines exponentially with the circuit size $L$. Therefore for $L$ large we would need to repeat the simulation an unreasonable number of times to have a reasonable probability of acceptance. Fortunately, though, postselected simulations are of potential practical interest even for $L$ fixed, in which case the overhead cost is only a (possibly large) multiplicative constant. 

In particular, for $\varepsilon < \tilde \varepsilon_0$ we can use postselected quantum computation (with constant overhead cost) to simulate quantum gates with fault rate below $ 1.9\times 10^{-4}$, a value already shown in \cite{aliferis-cross} to be a lower bound on the quantum accuracy threshold. Thus we can establish that, for $\varepsilon < \tilde\varepsilon_0$, an arbitrarily long quantum circuit can be simulated accurately with {\em polylogarithmic} overhead. In other words, we have improved the best rigorously established lower bound on the accuracy threshold for quantum computation by about a factor of $5.5$, from $1.9\times 10^{-4}$ to $1.04\times 10^{-3}$.

To study the accuracy of postselected circuits, we used some of the same ideas that we developed in \cite{AGP} for the study of simulations using quantum error correction, but some additional new techniques were needed, too. For example, simulations based on error correction or on error detection can be analyzed using the concept of level reduction, but in the case of error detection we needed the carefully constructed rules described in Sec.~\ref{sec:carving} for carving a bad cluster, and also the notion of weak correctness. We also needed to extend the level-reduction concept, in Sec.~\ref{sec:level-reduction}, from closed circuits concluded by measurements to open circuits concluded by (noisy) decoders. Furthermore, bounding the probability of failure of circuit simulations {\em conditioned on acceptance} requires firmer control over the correlations in the noise than we needed in \cite{AGP}, and hence the noise model formulated in Sec.~\ref{sec:more-local-noise} --- {\em locally correlated stochastic noise} --- is more restrictive than the local noise model adopted in \cite{AGP}. The crux of our analysis is in Sec.~\ref{sec:local-to-global}, where we developed the crucial concept of a {\em minimal sealed cluster}, and used it to estimate the failure probability for a specified set of gates conditioned on {\em global} acceptance by every error detection gadget in an entire circuit.

The formulation of the recursion relations in Sec.~\ref{sec:ThrCSS} invoked some tricks, such as contracting gadgets and distinguishing failure probabilities for truncated and untruncated extended rectangles, that complicated the analysis but improved the numerical estimate of the accuracy threshold. Even so, our rigorous lower bound is more than an order of magnitude below Knill's threshold estimates based on numerical simulation. Presumably, further bells and whistles in the analysis could push up our threshold estimate further, but the prospect of closing the gap between the rigorous lower bound and Knill's estimates seems daunting. (Reichardt's quite different methods \cite{reichardt-thesis} also yield a lower bound close to $10^{-3}$.) Our work might be fruitfully extended in several other directions; for example, it would be interesting to perform a rigorous analysis of the Fibonacci simulation method mentioned in Sec.~\ref{sec:overhead}, or of the reliability  of postselected quantum computation in the presence of (sufficiently local) coherent noise. In addition, we have assumed that two-qubit gates can act on any pair of qubits, no matter how distantly separated. How will the threshold estimate based on postselected quantum computation be affected if two-qubit gates act on only nearest neighbors in a particular spatial architecture?

\nonumsection{Acknowledgments}
\noindent
We thank Ben Reichardt for helpful comments. This work has been supported in part by: the Department of Energy (DoE) under Grant No. DE-FG03-92-ER40701, the National Science Foundation (NSF) under Grant No. PHY-0456720, the National Security Agency (NSA) under Army Research Office (ARO) Contract No. W911NF-05-1-0294, the province of Ontario through MRI, NSERC of Canada, and CIAR. Some of our circuit diagrams were drawn using the Q-circuit macro package written by Steve Flammia and Bryan Eastin.

\nonumsection{References}


\noindent

\begin{thebibliography}{10}

\bibitem{shor_ft} P.~W. Shor, ``Fault-tolerant quantum computation,'' in {\it Proceedings, 37th Annual Symposium on Foundations of Computer Science}, pp. 56-65 (Los Alamitos, CA, IEEE Press, 1996), arXiv:quan-ph/9605011.
\bibitem{ben-or} D. Aharonov and M. Ben-Or, ``Fault-tolerant quantum computation with constant error,'' {\it Proc. 29th Ann. ACM Symp. on Theory of Computing}, p. 176 (New York, ACM, 1998), arXiv:quan-ph/9611025; D. Aharonov and M. Ben-Or, ``Fault-tolerant quantum computation with constant error rate,'' arXiv:quan-ph/9906129 (1999).
\bibitem{kitaev_threshold} A. Yu. Kitaev, ``Quantum computations: algorithms and error correction,'' Russian Math. Surveys {\bf 52}, 1191-1249 (1997).
\bibitem{knill} E Knill, R. Laflamme, W.~H. Zurek, ``Resilient quantum computation: error models and thresholds,'' Proc. Roy. Soc. London, Ser. A {\bf 454}, 365 (1998), arXiv:quan-ph/9702058.
\bibitem{jp_threshold} J. Preskill, ``Reliable quantum computers,''  Proc. Roy. Soc. Lond. A {\bf 454}, 385-410 (1998), arXiv:quan-ph/9705031. 
\bibitem{gottesman_threshold} D. Gottesman, ``Stabilizer codes and quantum error correction,'' Caltech Ph.D. thesis (1997), arXiv:quan-ph/9705052.
\bibitem{AGP} P. Aliferis, D. Gottesman, and J. Preskill, ``Quantum accuracy threshold for concatenated distance-3 codes,'' Quant. Inf. Comp. 6, 97-165 (2006), arXiv:quant-ph/0504218.
\bibitem{reichardt-threshold} B.~W. Reichardt, ``Threshold for the distance three Steane quantum code,'' arXiv:quant-ph/0509203 (2005).
\bibitem{aliferis-cross} P. Aliferis and A.~W. Cross, ``Sub-system fault tolerance with the Bacon-Shor code,''  arXiv:quant-ph/0610063 (2006).
\bibitem{knill_detect} E. Knill, ``Quantum computing with realistically noisy devices,'' Nature {\bf 434}, 39--44 (2005), arXiv:quant-ph/0410199 (2004).
\bibitem{steane-software} A. Steane, ``Active stabilisation, quantum computation and quantum state synthesis,'' Phys. Rev. Lett. {\bf 78}, 2252-2255 (1997), arXiv:quant-ph/9611027.
\bibitem{gottesman-chuang} D. Gottesman and I. Chuang, ``Quantum teleportation is a universal computational primitive,'' Nature {\bf 402}, 390 (1999), arXiv:quant-ph/9908010.
\bibitem{reichardt-postselection} B.~W. Reichardt, ``Postselection threshold against biased noise,'' arXiv:quant-ph/0608018 (2006).
\bibitem{reichardt-thesis} B.~W. Reichardt. ``Error-detection-based quantum fault tolerance against discrete Pauli noise,'' U.C. Berkeley Ph.D. thesis (2006), arXiv:quant-ph/0612004.
\bibitem{shor_9} P.~W. Shor, ``Scheme for reducing decoherence in quantum computer memory,'' Phys. Rev. A {\bf 52}, 2493 (1995).
\bibitem{steane_7} A. Steane, ``Error-correcting codes in quantum theory,'' Phys. Rev. Lett. {\bf 77}, 793 (1996).
\bibitem{knill-concatenated} E. Knill and R. Laflamme, ``Concatenated quantum codes,'' arXiv:quant-ph/9608012 (1996).
\bibitem{aliferis-thesis} P. Aliferis, ``Level reduction and the quantum threshold theorem,'' Caltech Ph.D. thesis (2007), arXiv:quant-ph/0703230.
\bibitem{knill-schemes} E. Knill, ``Fault-tolerant postselected quantum computation: schemes,'' arXiv:quant-ph/0402171 (2004).
\bibitem{terhal} B.~M. Terhal and G. Burkard, ``Fault-tolerant quantum computation for local non-Markovian noise,'' Phys. Rev. A {\bf 71}, 012336 (2005), arXiv:quant-ph/0402104.
\bibitem{aharonov-kitaev-preskill} D. Aharonov, A. Kitaev, and J. Preskill, ``Fault-tolerant quantum computation with long-range correlated noise,'' Phys. Rev. Lett. {\bf 96}, 050504 (2006); arXiv:quant-ph/0510231.
\bibitem{bacon} D. Bacon, ``Operator quantum error correcting subsystems for self-correcting quantum memories,'' arXiv:quant-ph/0506023 (2005).
\bibitem{poulin} D. Poulin, ``Stabilizer formalism for operator quantum error correction,'' Phys. Rev. Lett. 95, 230504 (2005). arXiv: quant-ph/0508131.
\bibitem{bdsw} C.~H.~Bennett, D.~P.~DiVincenzo, J.A.~Smolin, and W.~K.~Wootters, ``Mixed state entanglement and quantum error correction,'' Phys. Rev. A {\bf 54}, 3824-3851 (1996), arXiv:quant-ph/9604024.
\bibitem{bravyi} S. Bravyi and A. Kitaev, ``Universal quantum computation with ideal Clifford gates and noisy ancillas,'' Phys. Rev. A {\bf 71}, 022316 (2005), arXiv:quant-ph/0403025.
\bibitem{aliferis-terhal} P. Aliferis and B.~M. Terhal, ``Fault-tolerant quantum computation for local leakage faults,'' Quant. Inf. Comp. 7, 139-156 (2007), arXiv:quant-ph/0511065.
    
%
\end{thebibliography}
\end{document}